\documentclass[]{aa}  
\usepackage{graphicx}
\usepackage[varg]{txfonts}
\usepackage{natbib}
\bibpunct{(}{)}{;}{a}{}{,} 
\begin{document}

\title{VLTI/AMBER spectro-interferometry of the late-type supergiants 
V766~Cen (=HR~5171\,A), $\sigma$~Oph, BM~Sco, and HD~206859
\thanks{Based on observations made with the VLT Interferometer (VLTI)
at Paranal Observatory under program ID 093.D-0014}
}
\titlerunning{Spectro-interferometry of supergiants}
\author{
M.~Wittkowski\inst{1}\and
B.~Arroyo-Torres\inst{2,3,4}\and
J.~M.~Marcaide\inst{2}\and
F.~J.~Abellan\inst{2}\and
A.~Chiavassa\inst{5}\and
J.~C.~Guirado\inst{2,6}
}
\institute{
ESO, Karl-Schwarzschild-Str. 2,
85748 Garching bei M\"unchen, Germany,
\email{mwittkow@eso.org}
\and
Dpt. Astronomia i Astrof\' isica, Universitat de Val\`encia,
C/Dr. Moliner 50, 46100, Burjassot, Spain
\and
Instituto de Astrof\'isica de Andaluc\'ia (IAA-CSIC), 
Glorieta de la Astronom\'ia S/N, 18008, Granada, Spain
\and
Centro Astron\'omico Hispano Alem\'an, Calar Alto, (CSIC-MPG), 
Sierra de los Filabres, E-04550 Gergal, Spain
\and
Laboratoire Lagrange, UMR 7293, Universit\'e de Nice Sophia-Antipolis, CNRS,
Observatoire de la C\^ote d’Azur, BP. 4229, 06304 Nice Cedex 4, France
\and
Observatorio Astron\'omico, Universidad de Val\`encia, 46980 Paterna, 
Val\`encia, Spain
}
\date{Received \dots; accepted \dots}
\abstract{}
{We add four warmer late-type supergiants to our previous
spectro-interferometric studies of red giants and supergiants.}
{We measure the near-continuum angular diameter, derive fundamental
parameters, discuss the evolutionary stage, and study
extended atmospheric atomic and molecular layers.}
{V766~Cen (=HR~5171\,A) is found to be a high-luminosity
($\log L/\mathrm{L}_\odot=$\,5.8\,$\pm$\,0.4) source
of effective temperature 4290\,$\pm$\,760\,K and radius
1490\,$\pm$\,540\,R$_\odot$, located
in the Hertzsprung-Russell (HR) diagram close to both the Hayashi limit and Eddington
limit; this source is consistent with a 40\,M$_\odot$ evolutionary track without
rotation and current mass 27--36\,M$_\odot$. V766~Cen exhibits \ion{Na}{i} in emission
arising from a shell of radius 1.5\,R$_\mathrm{Phot}$ and a
photocenter displacement of about 0.1R$_\mathrm{Phot}$. It shows strong
extended molecular (CO) layers and a dusty circumstellar
background component. The other three sources are found to have
lower luminosities of about $\log L/\mathrm{L}_\odot=$3.4--3.5, corresponding
to 5--9\,M$_\odot$ evolutionary tracks.
They cover effective temperatures of 3900\,K to
5300\,K and radii of  60--120\,R$_\odot$. They do
not show extended molecular layers as observed for higher luminosity
red supergiants of our sample. BM~Sco shows an unusually strong contribution
by an over-resolved circumstellar dust component.
}
{V766~Cen is a red supergiant located close to the Hayashi limit instead
of a yellow hypergiant already evolving back toward warmer effective
temperatures as discussed in the literature. Our observations of the \ion{Na}{i}
line and the extended molecular layers suggest
an optically thick pseudo-photosphere at about 1.5\,R$_\mathrm{Phot}$ 
at the onset of the wind.
The stars $\sigma$~Oph, BM~Sco,
and HD~206859 are more likely high-mass red giants
instead of red supergiants as implied by their luminosity class Ib.
This leaves us with an unsampled locus in the HR diagram corresponding
to luminosities $\log L/\mathrm{L}_\odot\sim$3.8--4.8 or masses
10--13\,M$_\odot$, possibly corresponding to the mass region where
stars explode as (type II-P) supernovae during the red supergiant stage.
With V766~Cen, we now confirm that our previously found relation of increasing strength
of extended molecular layers with increasing luminosities extends to double our previous 
luminosities and up to the Eddington limit. This might further point to steadily increasing
radiative winds with increasing luminosity.
}
\keywords{
Techniques: interferometric --
supergiants --
Stars: atmospheres --
Stars: mass-loss --
Stars: individual: HR5171 A --
Stars: individual: BM Sco
}
\maketitle
\section{Introduction}
Red supergiants (RSGs) are cool evolved massive stars before their transition 
toward Wolf-Rayet (WR) stars and core-collapse supernovae (SNe). Their 
characterization and location in the Hertzsprung-Russell (HR) 
diagram are of importance to calibrate stellar evolutionary models for massive 
stars and to understand their further evolution toward WR stars and 
SNe, a topic that has seen an increased interest  
\citep[e.g.,][]{Dessart2013,Smith2014,Meynet2015}. In addition 
the structure and morphology of the close circumstellar environment and wind 
regions, including the atmospheric molecular layers and dusty envelopes, 
are currently a matter of debate 
\citep[e.g.,][]{Josselin2007,deWit2008,Smith2009,Verhoelst2009,Yoon2010,
Walmswell2012}. 
Knowledge of the circumstellar envelope, along with the knowledge of the 
fundamental parameters, is important to understand the matching of 
SN progenitors to the different types of core-collapse 
SNe \citep[e.g.,][]{Heger2003,Groh2013}.

This paper is conceived as the fourth in a series of previous papers in which 
we have been studying the fundamental parameters and extended 
atmospheric layers
of nearby late-type supergiants based on spectro-imaging with the VLTI/AMBER 
instrument, including those by \citet{Wittkowski2012}, 
\citet{Arroyo2013}, and \citet{Arroyo2015}, 
along with a similar study of lower mass red giants \citep{Arroyo2014}. 
Our sample thus far included the late-K- and M-type RSGs 
VY~CMa (spectral type
M3-4\,II), AH~Sco (M5\,Ia-Iab), UY~Sct (M4\,Ia), 
KW~Sgr (M4\,Ia), V602~Car (M3\,Ia-Iab), HD~95687 (M2\,Ia), 
and HD~183589 (K5\,III), along with five lower mass giants.
We use spectral types as adopted by
Simbad throughout this paper; in some cases different
spectral classifications are available in the literature.
We showed that extended molecular atmospheres with extensions comparable to 
Mira variable asymptotic giant branch (AGB) stars are a common feature of 
luminous RSGs, and that --unlike for Miras-- this phenomenon is not predicted 
by current 3D convection or 1D pulsation models. We found a correlation of 
the contribution by extended atmospheric layers with luminosity, 
which may support a 
scenario of radiative acceleration on Doppler-shifted molecular lines as 
suggested by \citet{Josselin2007}. However, the processes that levitate the 
atmospheres of RSGs to observed extensions currently remain unknown.

In this paper, we add four warmer supergiants of spectral types G5 to K2.5 
to our sample. These sources include the late-type supergiants  
V766~Cen (=HR~5171\,A, Simbad spectral type G8\,Ia), $\sigma$~Oph (K2\,III),  
BM~Sco (K2\,Ib), and HD~206859 (G5\,Ib). V766~Cen (=HR~5171\,A) was classified 
as a yellow hypergiant (YHG) by 
\citet{Humphreys1971}, \citet{vanGenderen1992}, and \citet{deJager1998}, 
showing an intense 
optically thin silicate emission feature \citep{Humphreys1971}. 
The star is known to have a companion (HR~5171\,B) 
at a separation of 9.7\arcsec, which was classified as B0\,Ib. 
\citet{Warren1973} reported on the existence of a high temperature, low-density envelope based on the detection of weak nebular [N II] emission. 
\citet{Chesneau2014} suggested that V766~Cen (=HR~5171\,A) itself has a 
low-mass 
companion that is very close to the primary star, possibly in the common-envelope 
phase. BM~Sco (=HD~160371) is a K2.5 supergiant from the sample of 
\citet{Levesque2005}, who aimed at characterizing the temperature scale 
of galactic RSGs using spectro-photometry. HD~206859 is a G5 
supergiant from the sample of \citet{Belle2009}, who measured effective 
temperatures and linear radii of a  sample of galactic RSGs with 
the Palomar Testbed Interferometer (PTI) in the $H$ and $K$ bands. 
The stars $\sigma$~Oph, BM~Sco, and HD~206859 are included in the sample
of \citet{McDonald2012}, who derived temperatures, luminosities, and
infrared excesses of Hipparcos stars by comparing model atmospheres 
to SEDs.
We aim to study the properties of these individual sources by direct
spectro-interferometry. Altogether, they 
increase our sample significantly. In particular we aim to study whether or 
not the presence of extended molecular layers and the correlation of 
the contribution by extended atmospheric layers
with luminosity reaches the level of warmer and more luminous 
late-type supergiants.
\section{Observations and data reduction}
\begin{table*}
\caption{VLTI/AMBER observations}
\centering
\begin{tabular}{ccccccc}
\hline
\hline
Target (Sp. type)  & Date   & Mode        & Baseline & Projected baseline & PA  & Cal$_{\mathrm{before}}$-Cal$_{\mathrm{after}}$   \\
                   & 2014-  & K- ($\mu$m) &          & m                  & deg &              \\
\hline
V766Cen (G8\,Ia) & 04-07 & 2.1 & D0-H0-I1 & 64.0/31.1/78.4 & 53/127/75 & HIP 72135\\
               & 04-07 & 2.3 & D0-H0-I1 & 63.4/33.3/88.9 & 66/137/89 & HIP 87846-HIP 87846\\
$\sigma$~Oph (K3\,Iab)  & 04-25  & 2.1 & D0-H0-I1 & 57.8/39.4/80.5 & 73/143/100 & HIP 87491 - HIP 88101 \\
                                  & 04-07 & 2.3  & D0-H0-I1 & 55.3/40.0/78.8 & 72/142/101 & HIP 87491 - HIP 88101\\                   
BM~Sco (K2.5\,Iab)  & 04-25 & 2.1 & D0-H0-I1 & 54.5/34.2/60.3 & 37/134/71 & HIP 87846\\
               & 04-07 & 2.3 & D0-H0-I1 & 57.3/36.2/67.4 & 45/136/78 & HIP 87846 - HIP 87846\\
               & 06-09 & 2.3 & A1-G1-K0 & 71.2/81.5/104.5 & 134/40/83 & HIP 87846\\
HD~206859 (G5 Ib)  & 06-20  & 2.1 & A1-G1-K0 & 70.1/80.7/128.9 & 99/37/66 & HIP 109068 \\
                                  & 06-25  & 2.3 & A1-G1-K0 & 69.3/81.2/128.9 & 99/37/65  & HIP 109068\\                                
\hline
\end{tabular}
\tablefoot{Details of our observations. The projected baseline is the 
projected baseline length for the AT VLTI baseline used, and PA is the 
position angle of the baseline (north through east).}
\label{tab:Log_obs}
\end{table*}
We observed the late-type supergiants V766~Cen (=HR~5171\,A),
$\sigma$~Oph, BM~Sco, and HD~206859 with the ESO Very Large 
Telescope Interferometer (VLTI), 
utilizing three of the auxiliary telescopes of 1.8\,m diameter, and the 
AMBER instrument (Astronomical Multi-BEam combineR) with the external 
fringe tracker FINITO \citep{Petrov2007}. We employed the medium-resolution 
mode ($R\sim$~1500) in the K-2.1\,$\mu$m and K-2.3\,$\mu$m bands. We observed 
the data as sequences of cal$_{\mathrm{before}}$-sci-cal$_{\mathrm{after}}$. 
Our calibrators were HIP~72135 (RA 14 45 18.8, Dec -51 18 54, 
spectral type K4\,III, angular diameter 1.76\,mas), HIP~87846 
(RA 17 56 47.4, Dec -44 20 32, K2\,III, 1.92\,mas), 
HIP~87491 (RA 17 52 35.4, Dec +01 18 18, K5\,III, 2.46\,mas), 
HIP~88101 (RA 17 59 36.8, Dec -04 49 16, K5\,III, 2.24\,mas), 
and HIP~109068 (RA 22 05 40.7, Dec +05 03 31, K4\,III, 2.39\,mas). 
The calibrators were selected from the ESO Calibration Selector 
CalVin, in turn based on the catalog by \citet{Lafrasse2010}.

Table~\ref{tab:Log_obs} shows the details of our observations including the 
calibrator used in each case. Some sources are listed with only one 
calibrator, in cases where the data quality of the second calibrator observation
was low and we could not use it. 

It is essential for a good calibration of the interferometric transfer 
function that the performance of the FINITO fringe tracking is comparable 
between the corresponding calibrator and science target observations.
Optical path fluctuations (jitter) produce fringe motions during the
recording of one AMBER frame. These motions reduce the squared visibility
by a factor $e^{-\sigma ^2_{\phi}}$, where $\sigma _{\phi}$ is the fringe
phase standard deviation over the frame recording time (more information in
the AMBER User Manual 
\footnote{http://www.eso.org/sci/facilities/paranal/instruments/amber/doc.html}). 
This attenuation is corrected by the interferometric transfer function 
only if $\sigma _{\phi}$ was comparable between science and calibrator 
observations. We thus inspected the mean FINITO phase rms values to confirm 
that they were comparable. In some cases we deselected individual files for 
which this factor deviated by more than about 20\%.
In some cases we had to discard a complete data set because the FINITO phase
rms values were systematically very different between the science target 
observation and calibrator observations.
  
We obtained raw visibility data from our selected AMBER observations using
version 3.0.8 of the \textit{amdlib} data reduction package 
\citep{Tatulli2007,Chelli2009}.
We appended the visibility data of all scans of the same source taken 
consecutively, and we selected and averaged the resulting  visibilities using 
appropriate criteria. We selected all frames that have flux densities that are
three times higher than the noise and 80\% of the remaining frames with 
best signal-to-noise ratio (S/N)\footnote{see AMBER Data Reduction 
Software User Manual; http://www.jmmc.fr/doc/approved/JMMC-MAN-2720-0001.pdf}.

For the last reduction steps, we used scripts of IDL 
(Interactive Data Language), which we developed. First, we performed the absolute 
wavelength calibration by correlating the AMBER flux spectra  with a reference 
spectrum, that of the star BS~4432 (spectral type K4.5 III, similar to our 
calibrators), from \citet{Lancon2000}.
Afterward, we calibrated the flux and visibility spectra. 
The relative flux calibration was performed with the instrumental 
response, which is estimated by the calibrators and the BS~4432 spectrum. 
The visibility calibration was performed 
as in our previous work \citep{Arroyo2013,Arroyo2014}.

\begin{figure*}
\centering
\includegraphics[width=0.475\hsize]{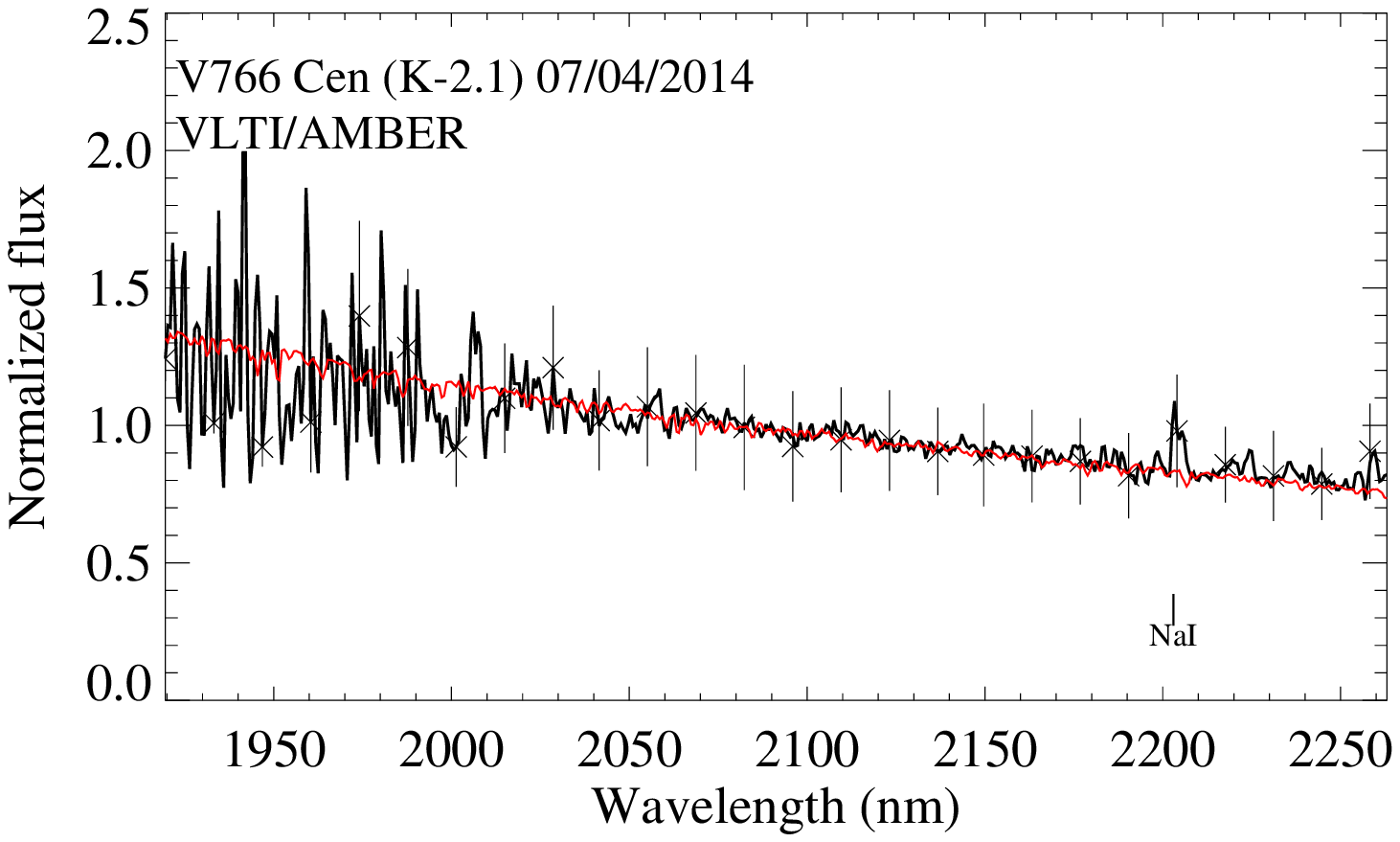}
\includegraphics[width=0.475\hsize]{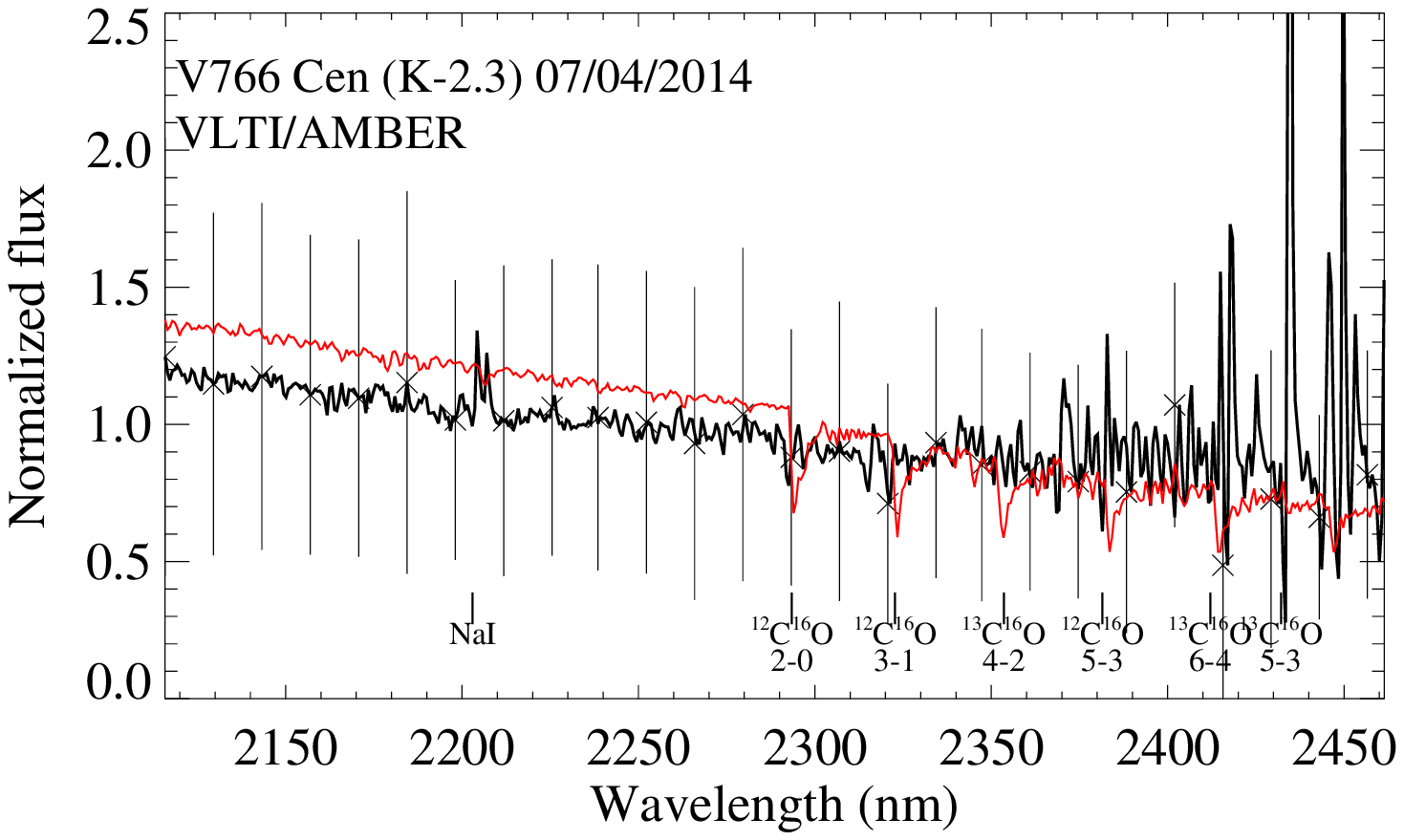}
\includegraphics[width=0.475\hsize]{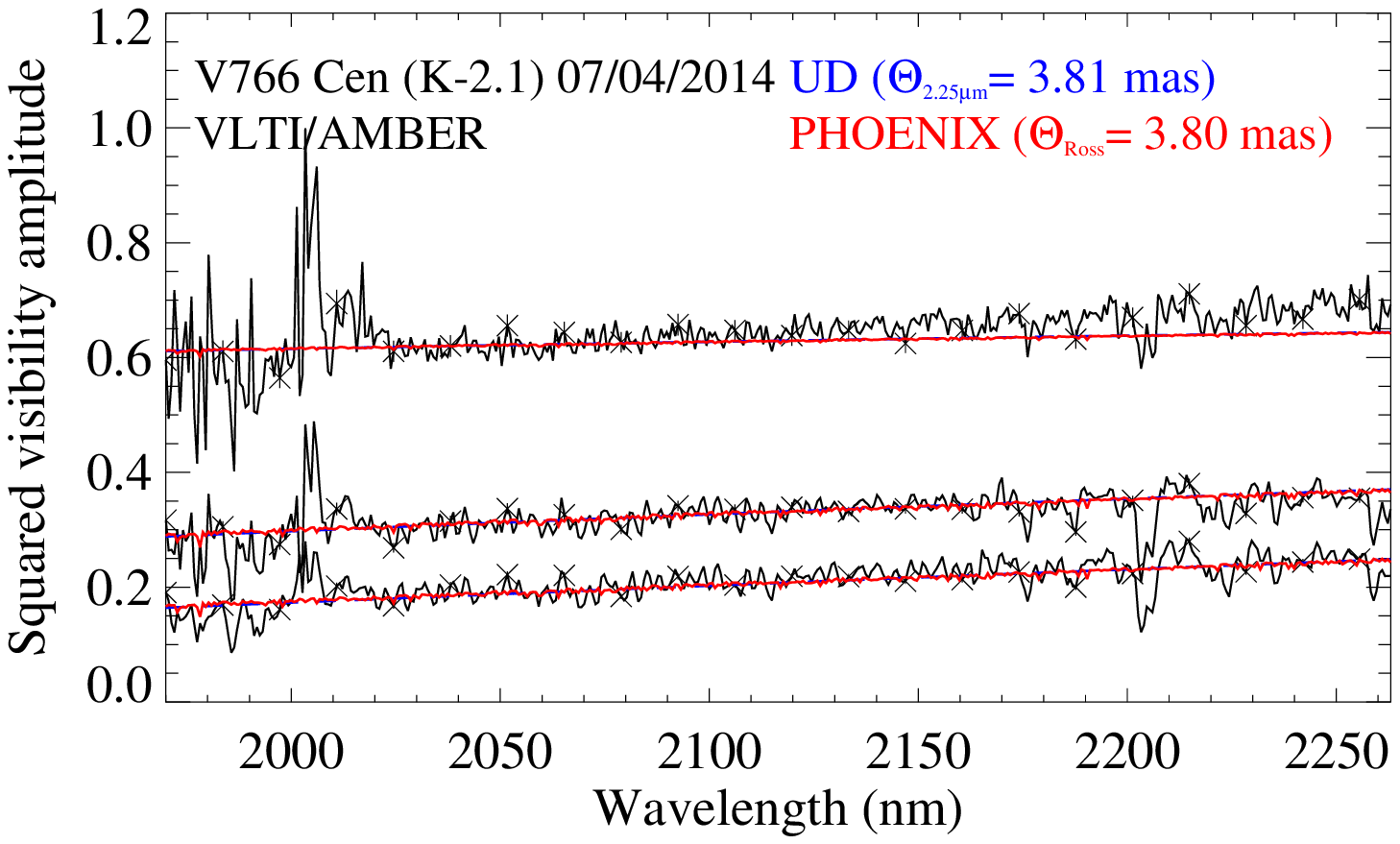}
\includegraphics[width=0.475\hsize]{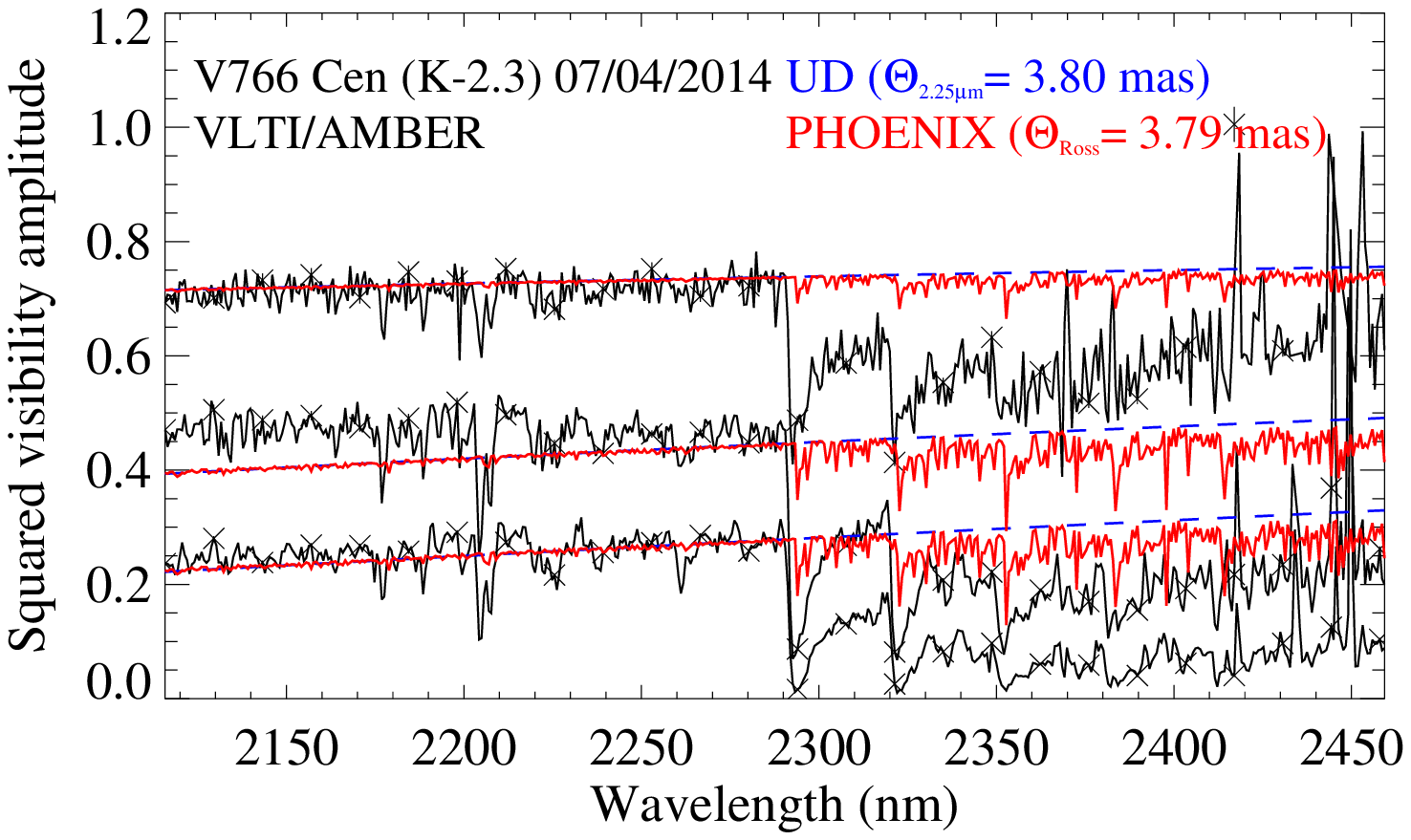}
\includegraphics[width=0.475\hsize]{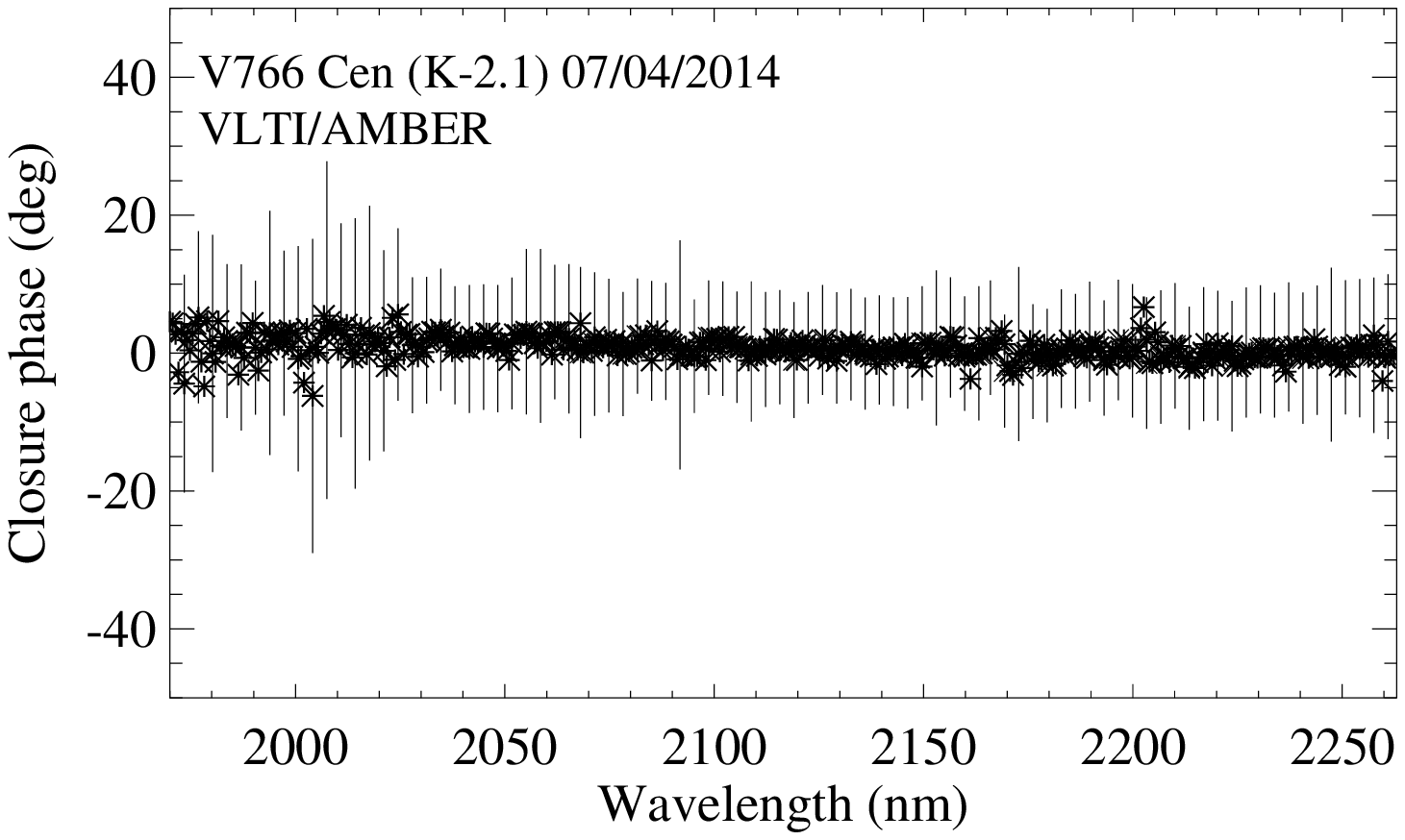}
\includegraphics[width=0.475\hsize]{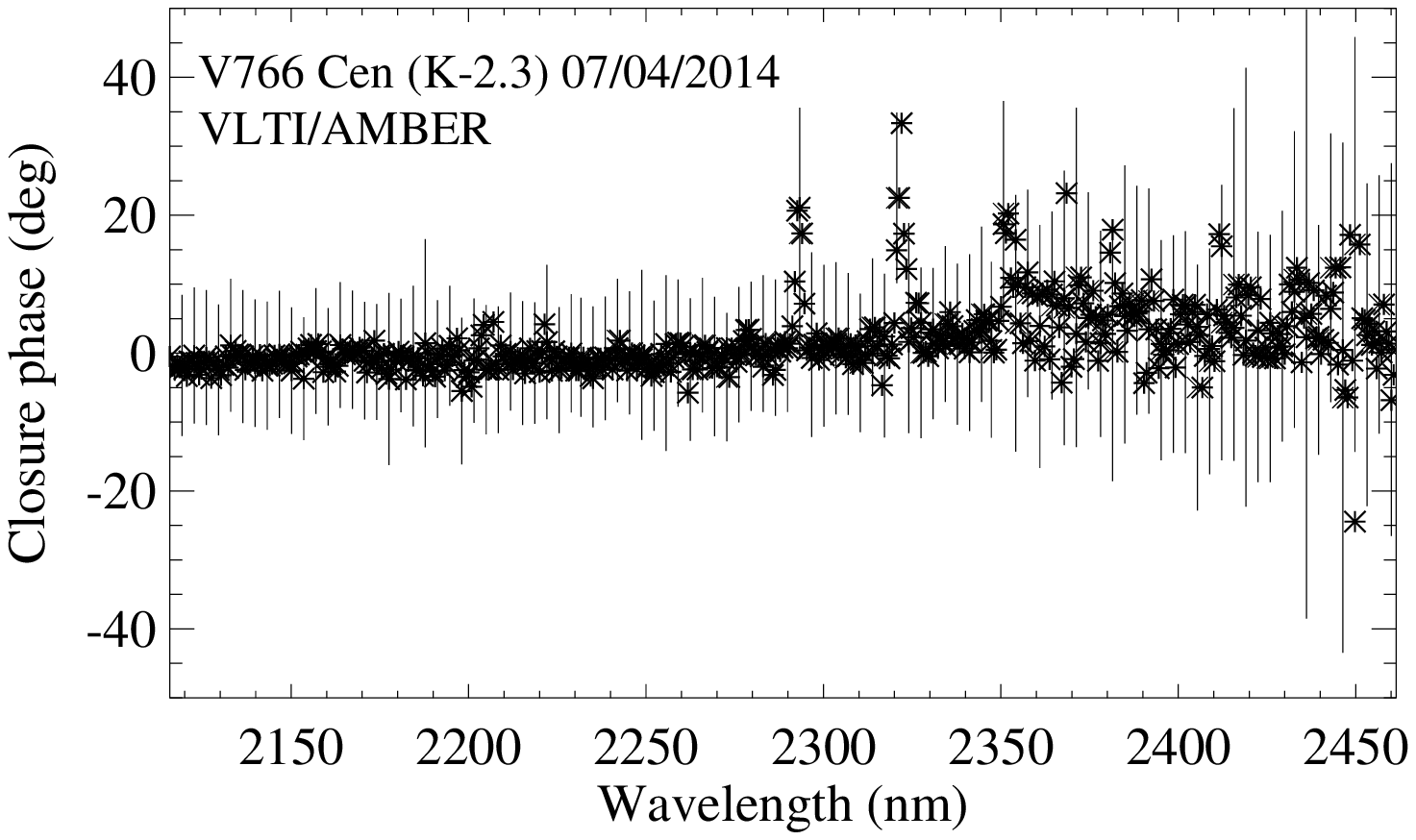}
\caption{Left (from top to bottom): Observed (black) normalized flux, 
squared visibility amplitudes, 
and closure phases
of V766~Cen obtained with the MR-K 2.1\,$\mu$m setting 
on 2014 April 07. Right: Same as left but obtained with the MR-K 2.3\ $\mu$m 
setting on 2014 April 07. 
The black solid lines connect all the observed
data points, while for the sake of clarity, the 'X' symbols and associated
error bars are shown for only every fifth data point.
The blue curves show the best-fit UD model, and 
the red curves the best-fit PHOENIX model prediction.
The visibility plots (middle panels) show three curves, one for each 
of the three baselines of the AMBER triangle.
(cf. Sect.~\protect\ref{sec:modeling}).
}
\label{fig:resul_V766Cen}
\end{figure*}
\begin{figure*}
\centering
\includegraphics[width=0.475\hsize]{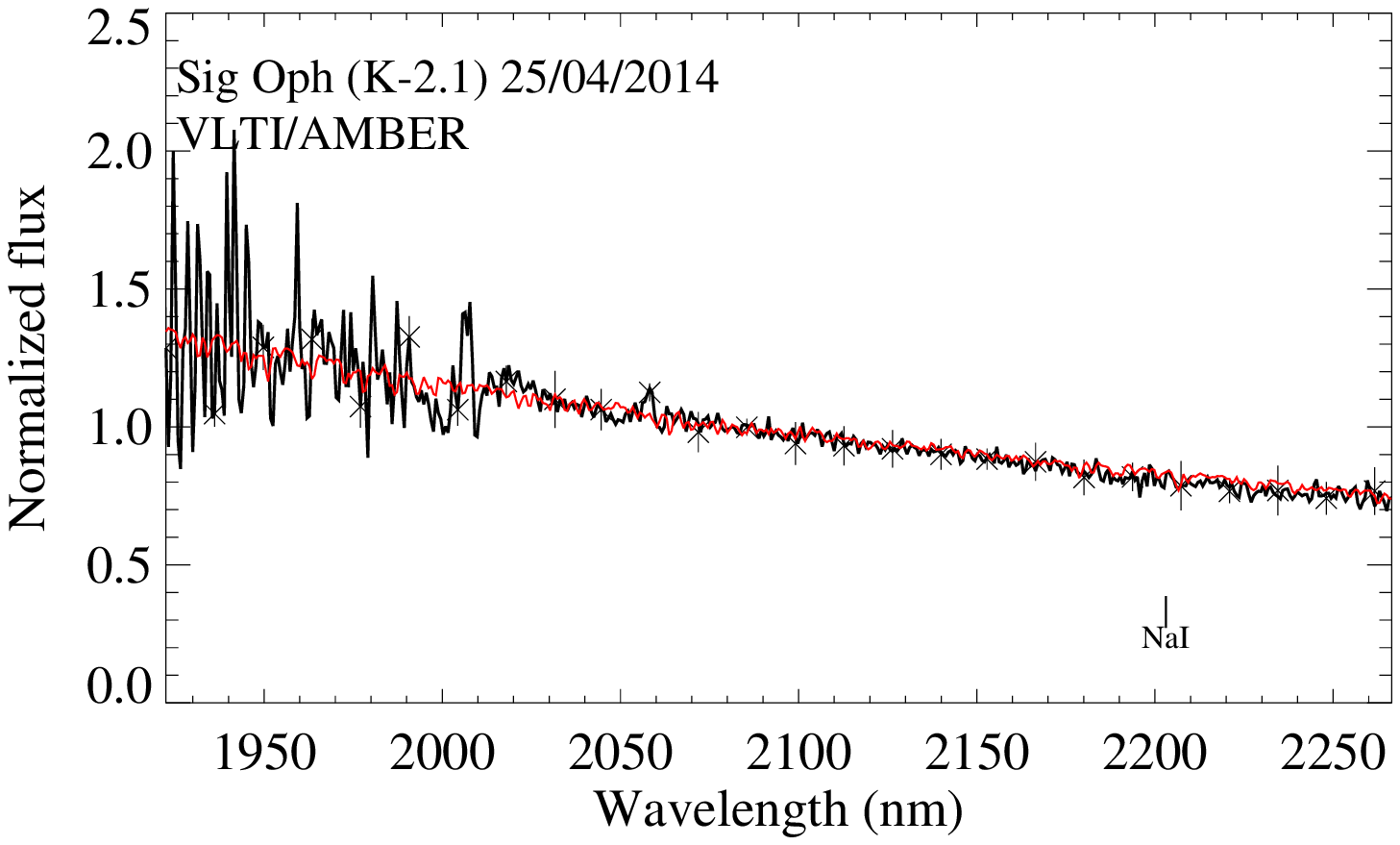}
\includegraphics[width=0.475\hsize]{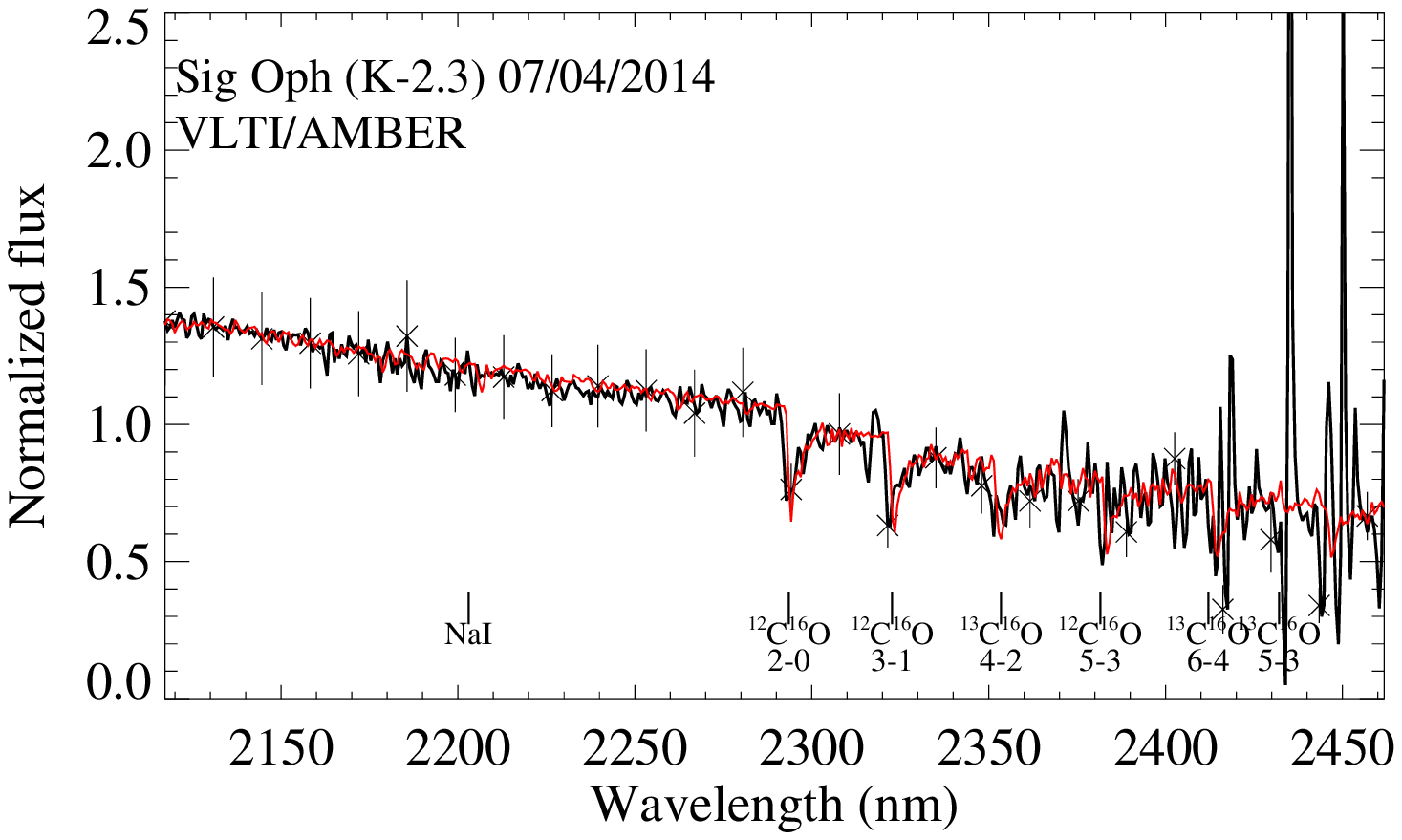}
\includegraphics[width=0.475\hsize]{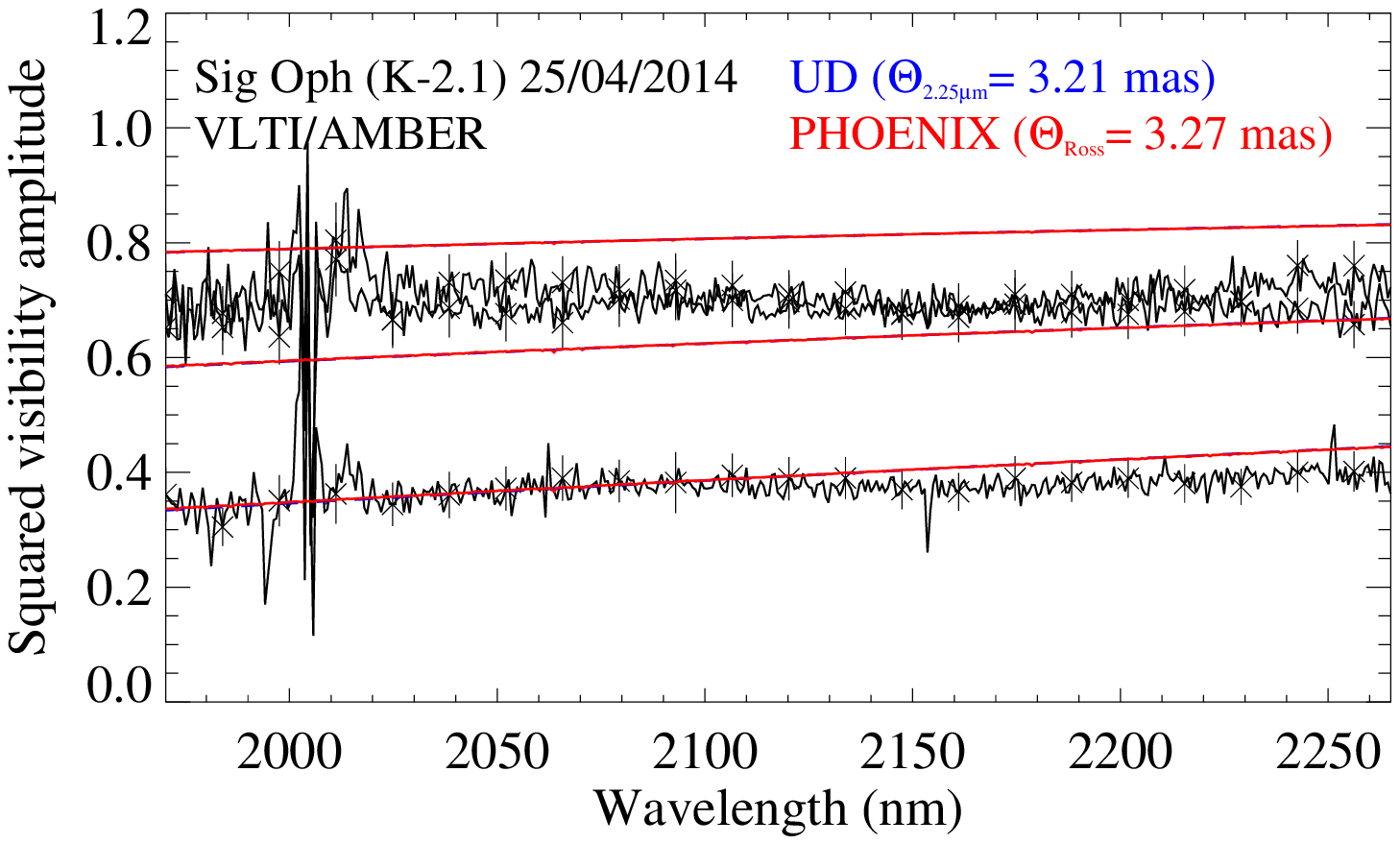}
\includegraphics[width=0.475\hsize]{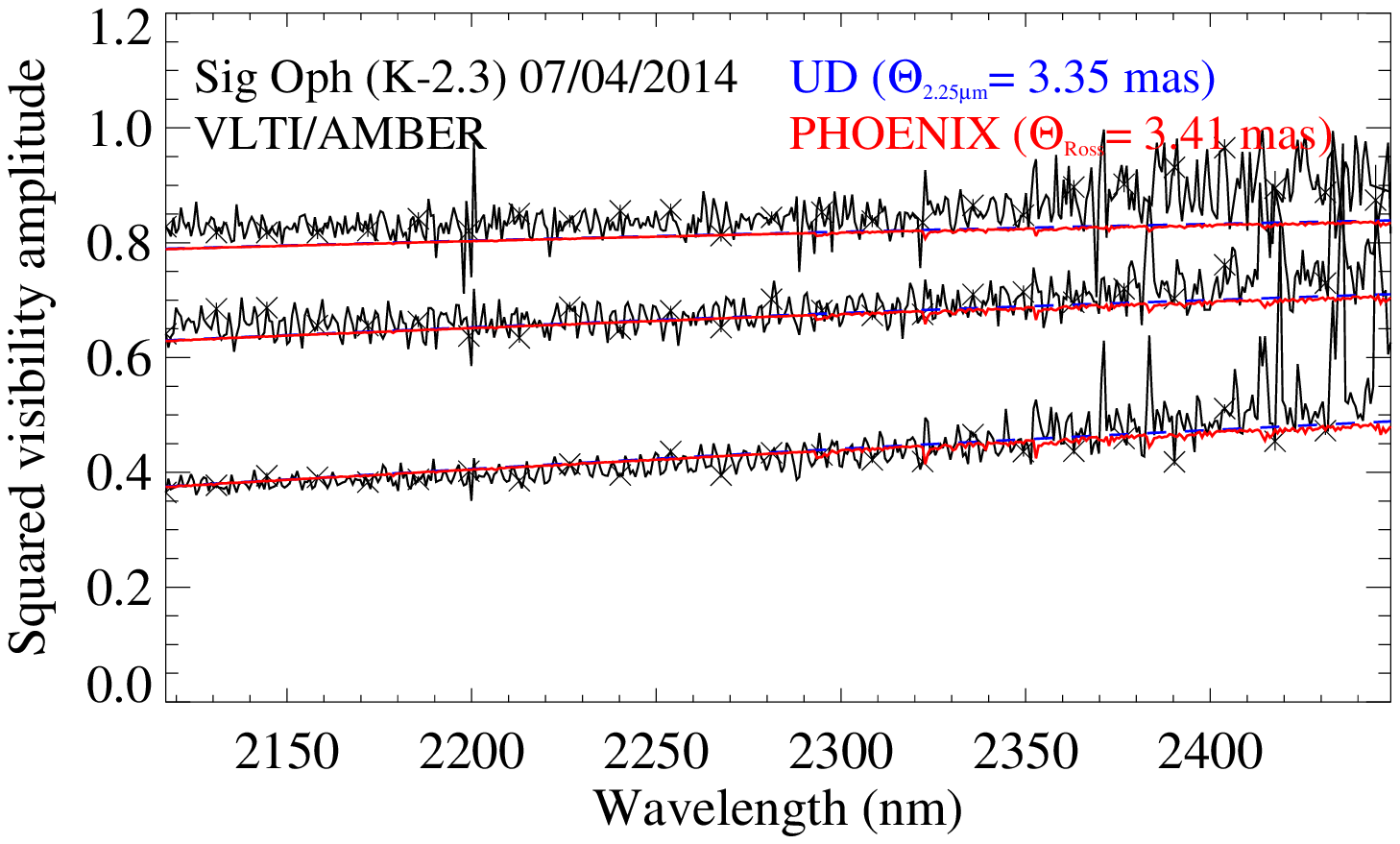}
\includegraphics[width=0.475\hsize]{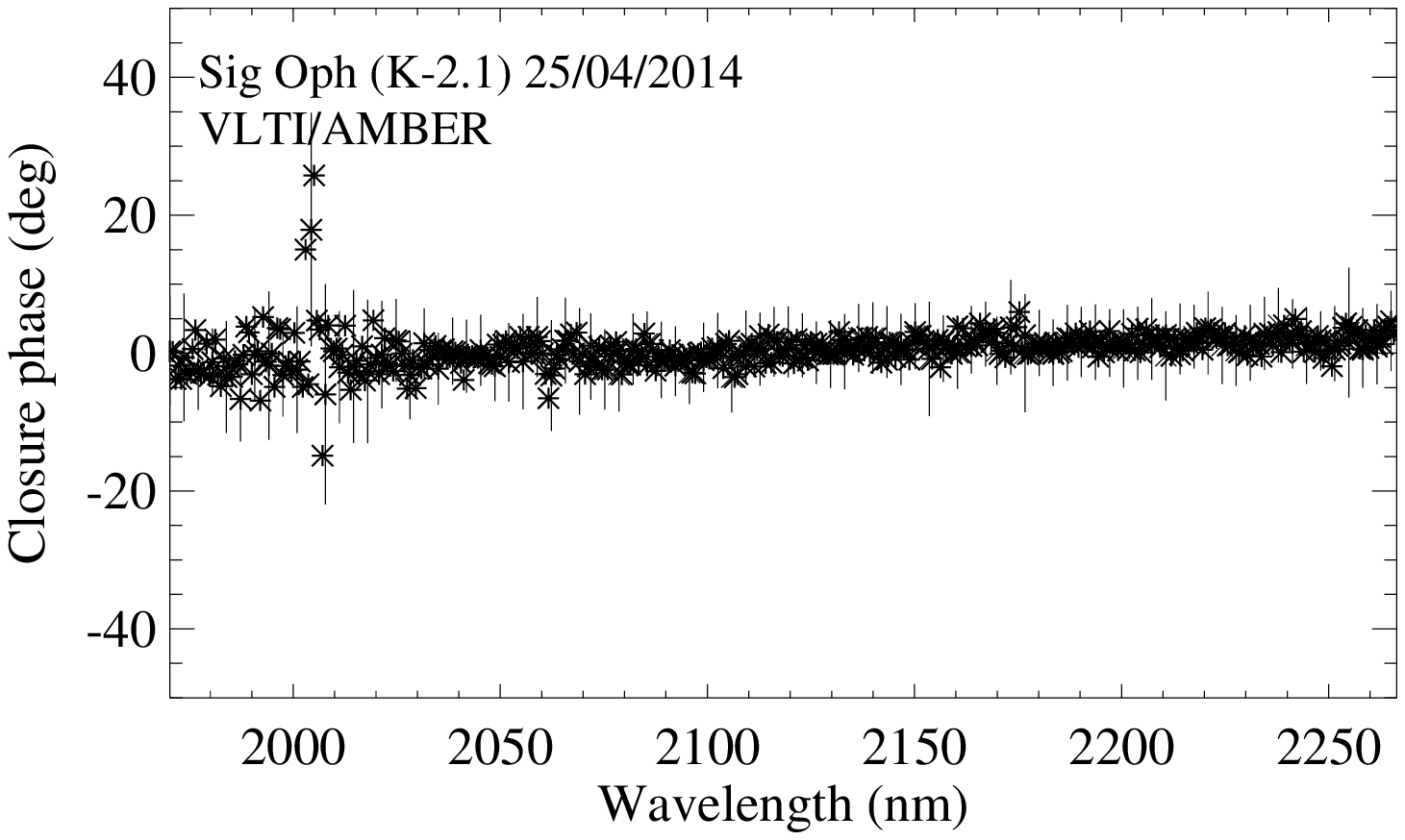}
\includegraphics[width=0.475\hsize]{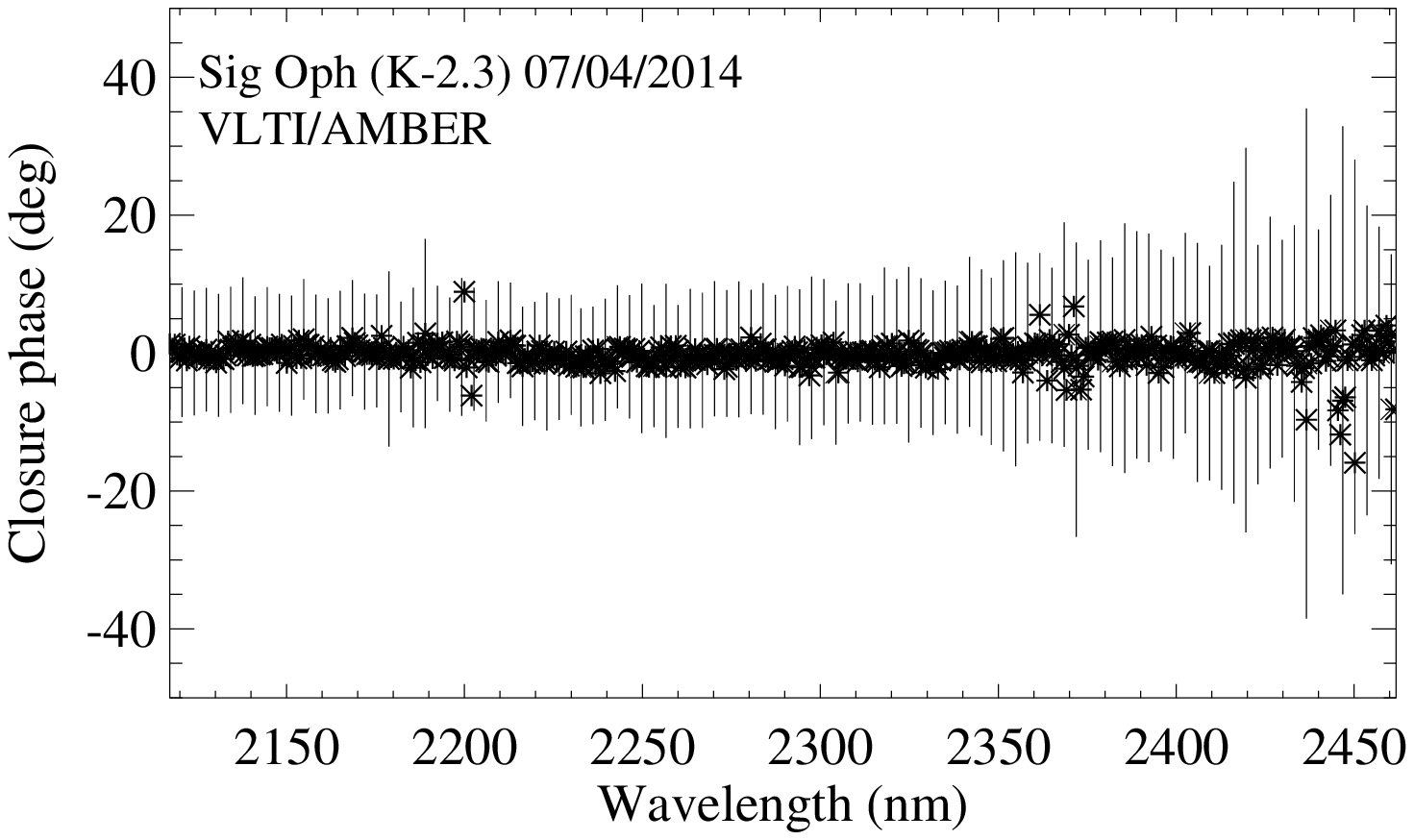}
\caption{As Fig. \ref{fig:resul_V766Cen}, but for data of $\sigma$~Oph 
obtained with the MR-K 2.1\,$\mu$m setting on 2014 April 25 (left) 
and with the MR-K 2.3\,$\mu$m setting on 2014 April 07 (right).}
\label{fig:resul_sigOph}
\end{figure*}
\begin{figure*}
\centering
\includegraphics[width=0.475\hsize]{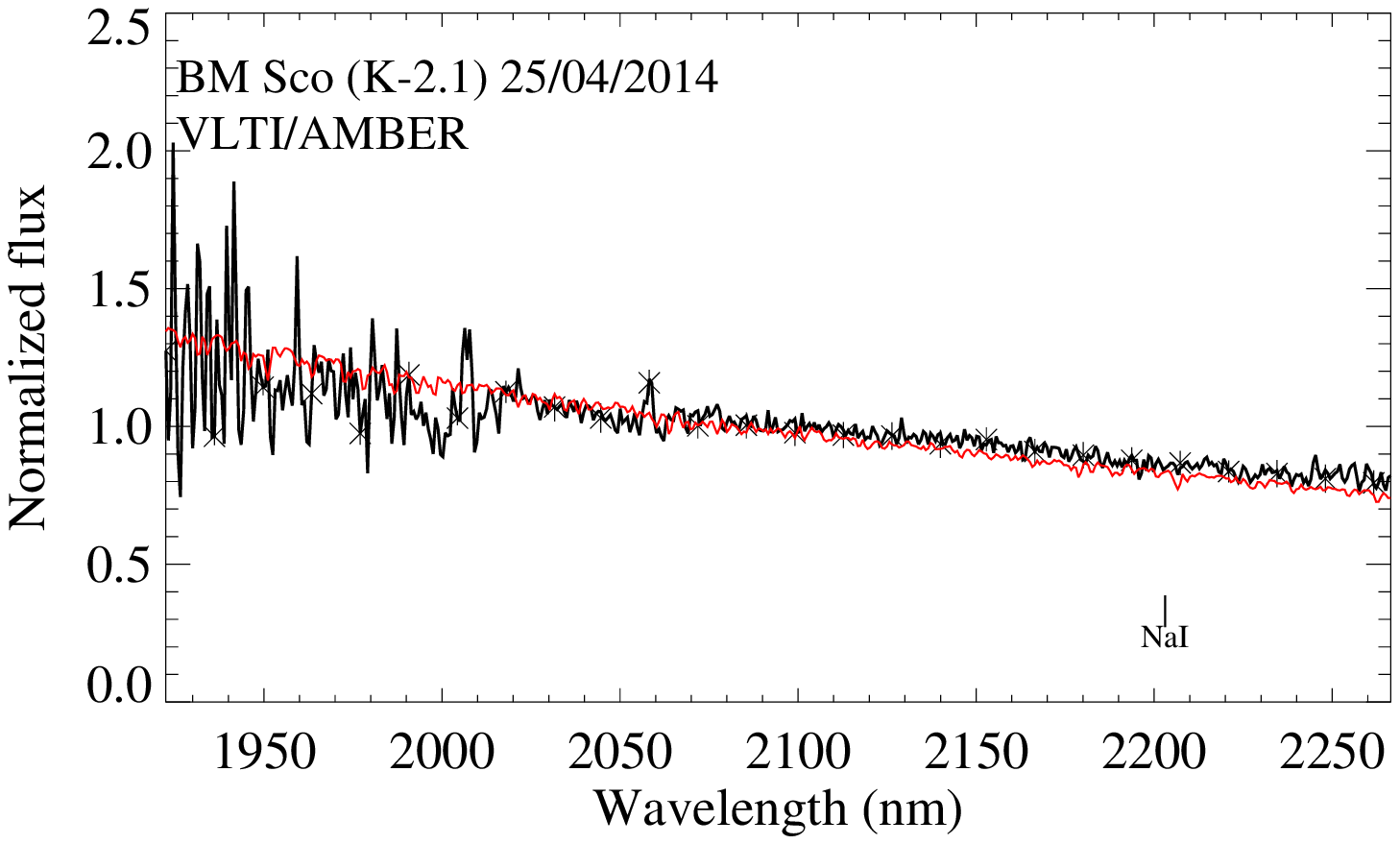}
\includegraphics[width=0.475\hsize]{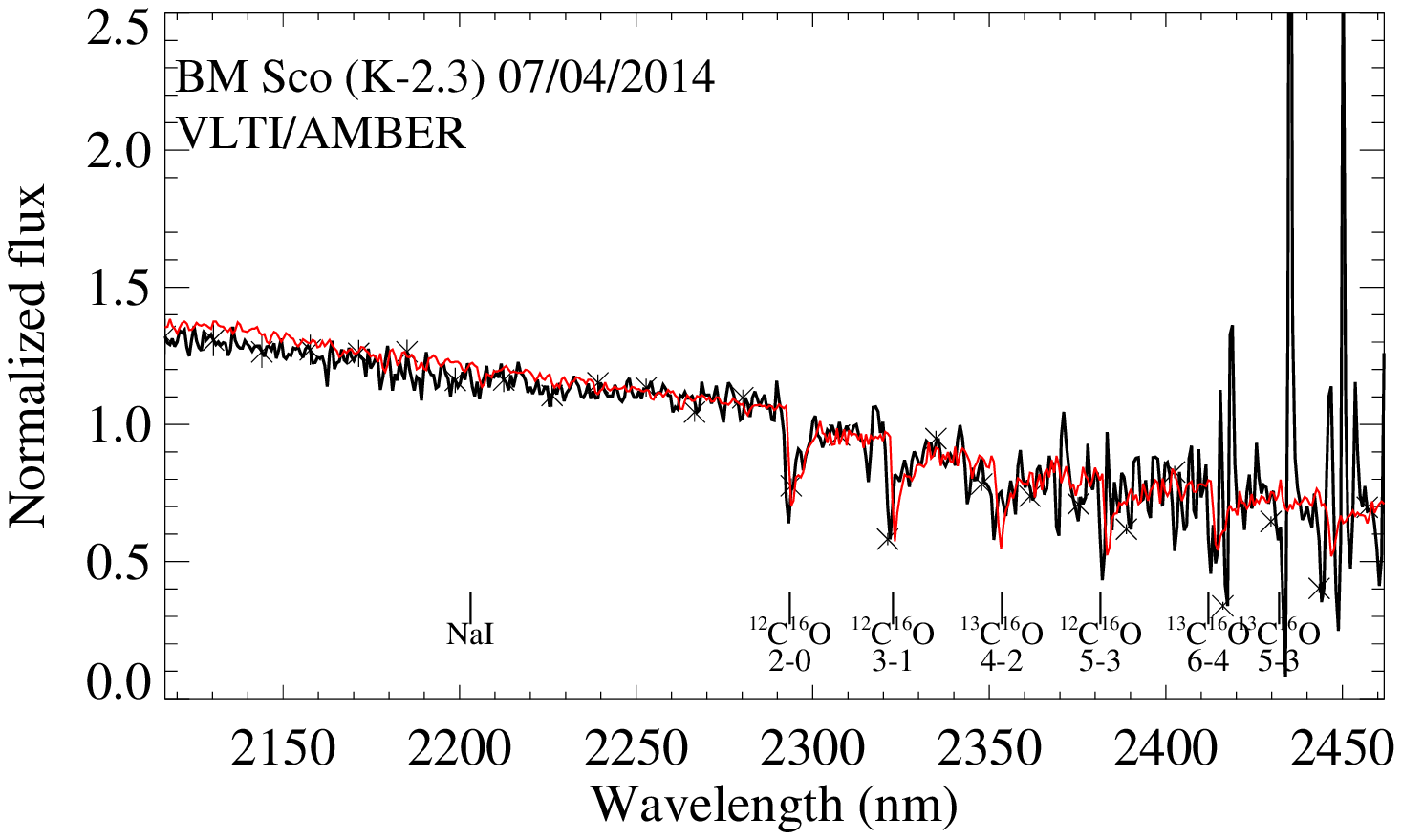}
\includegraphics[width=0.475\hsize]{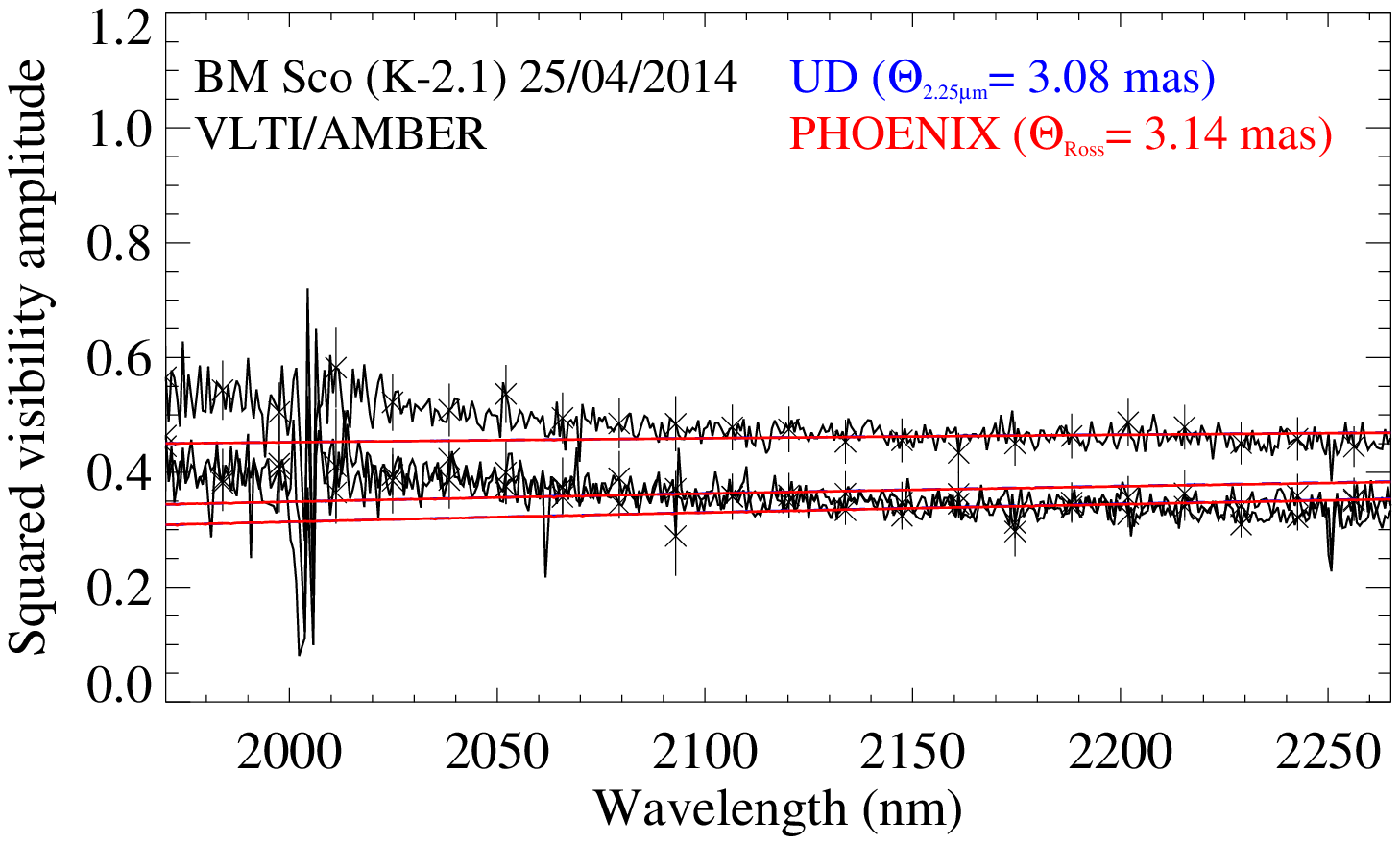}
\includegraphics[width=0.475\hsize]{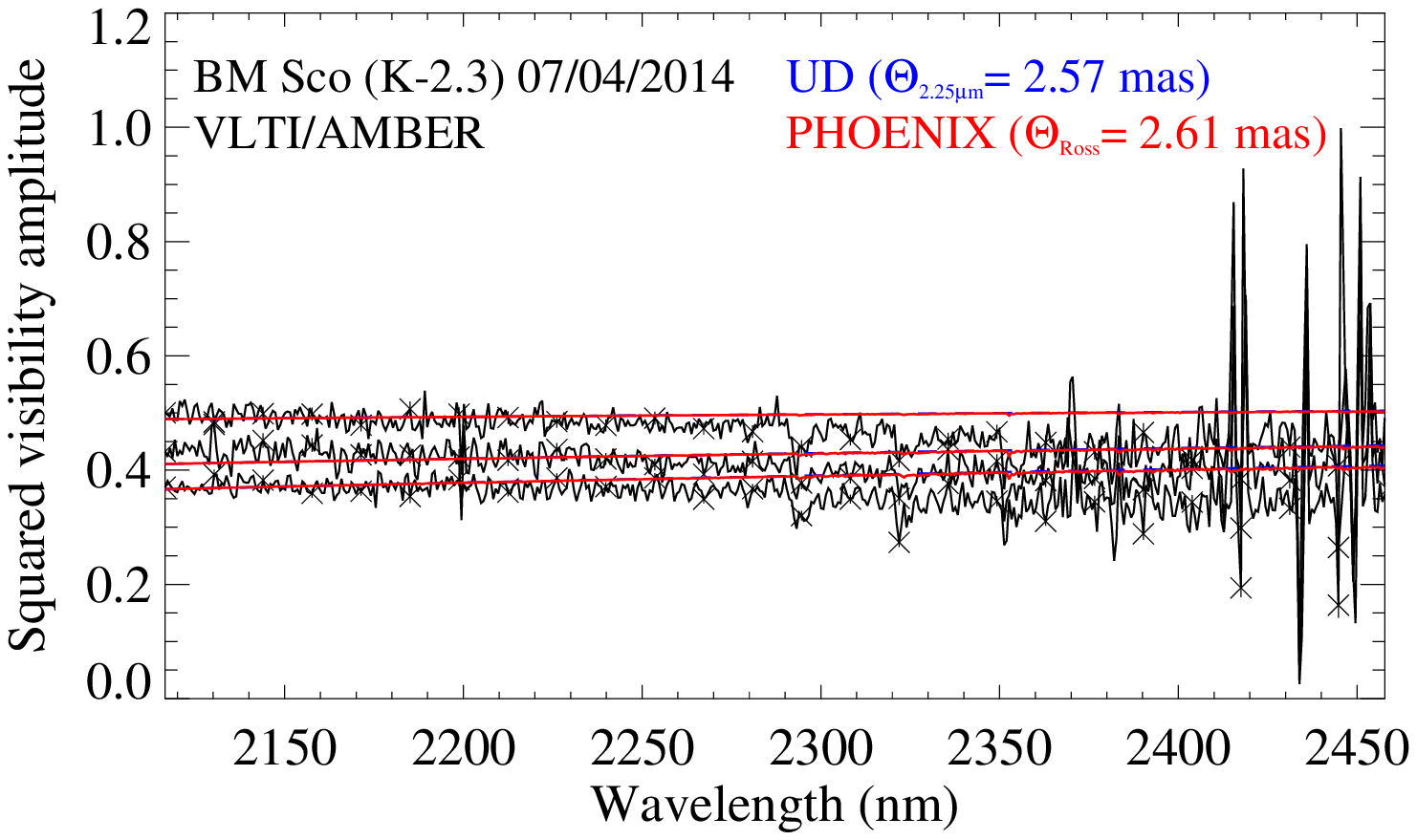}
\includegraphics[width=0.475\hsize]{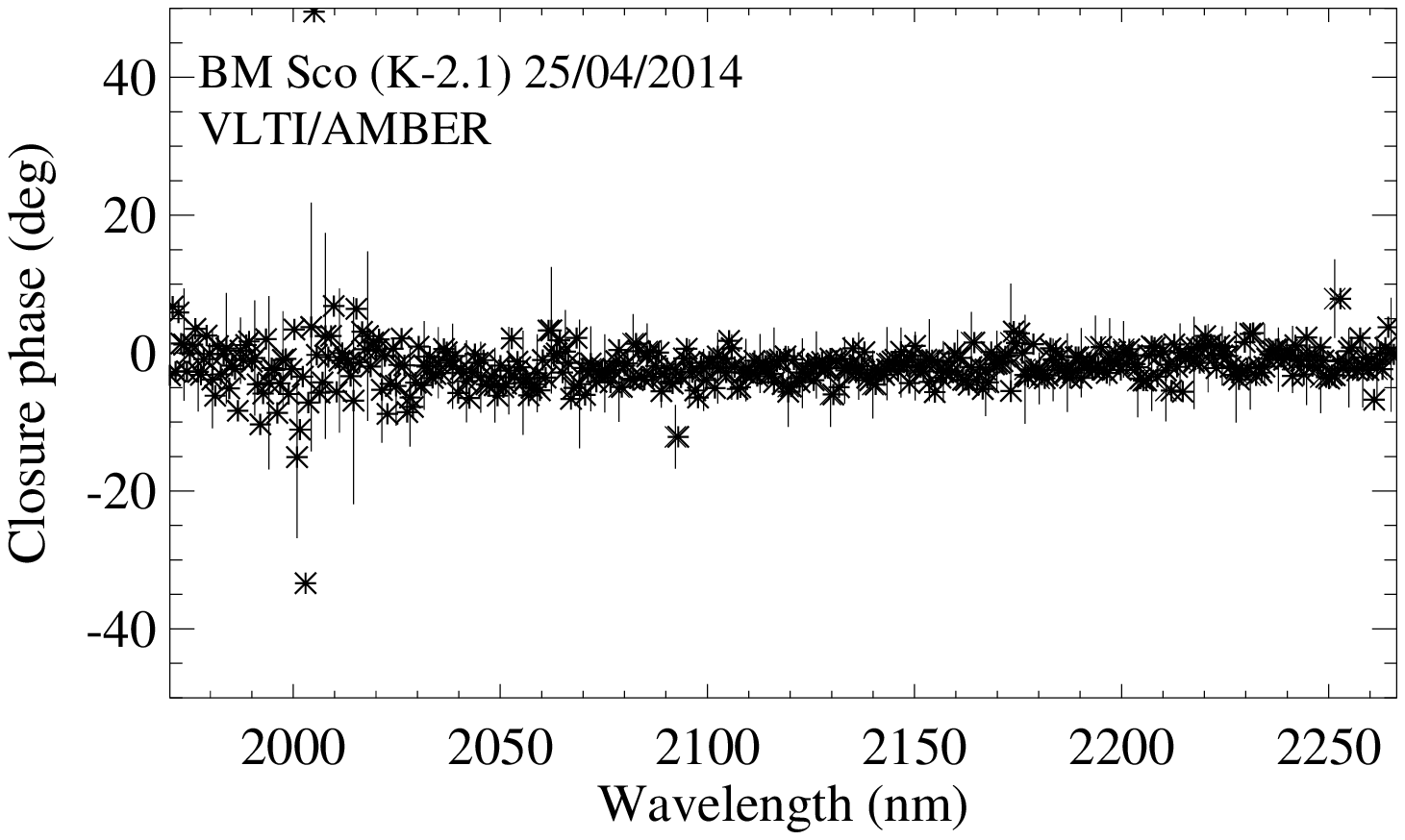}
\includegraphics[width=0.475\hsize]{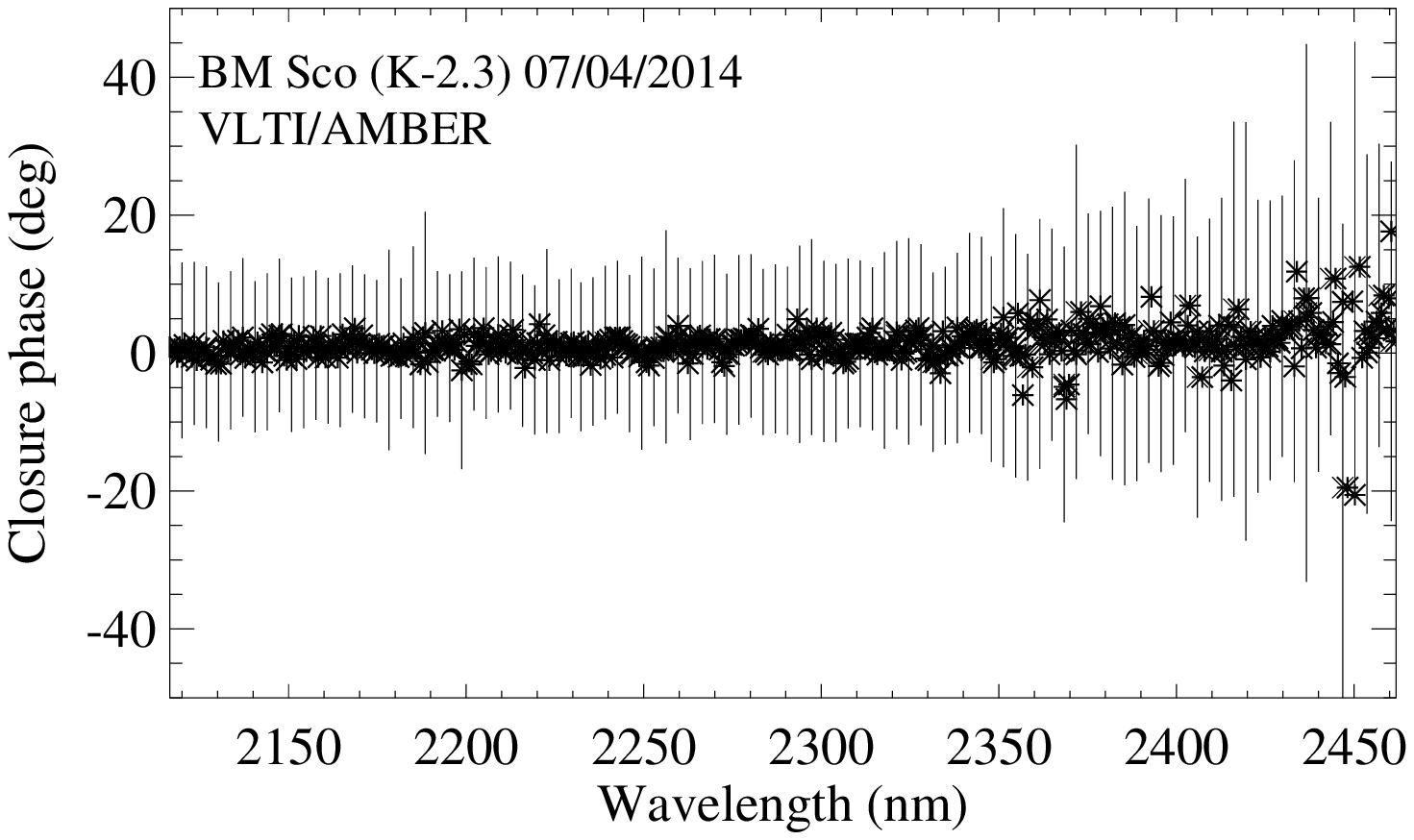}
\caption{As Fig. \ref{fig:resul_V766Cen}, but for data of BM~Sco obtained 
with the MR-K 2.1\,$\mu$m setting on 2014 April 25 (left) and with the 
MR-K 2.3\,$\mu$m setting on 2014 April 07 (right).}
\label{fig:resul_BMSco1}
\end{figure*}
\begin{figure}
\centering
\includegraphics[width=0.95\hsize]{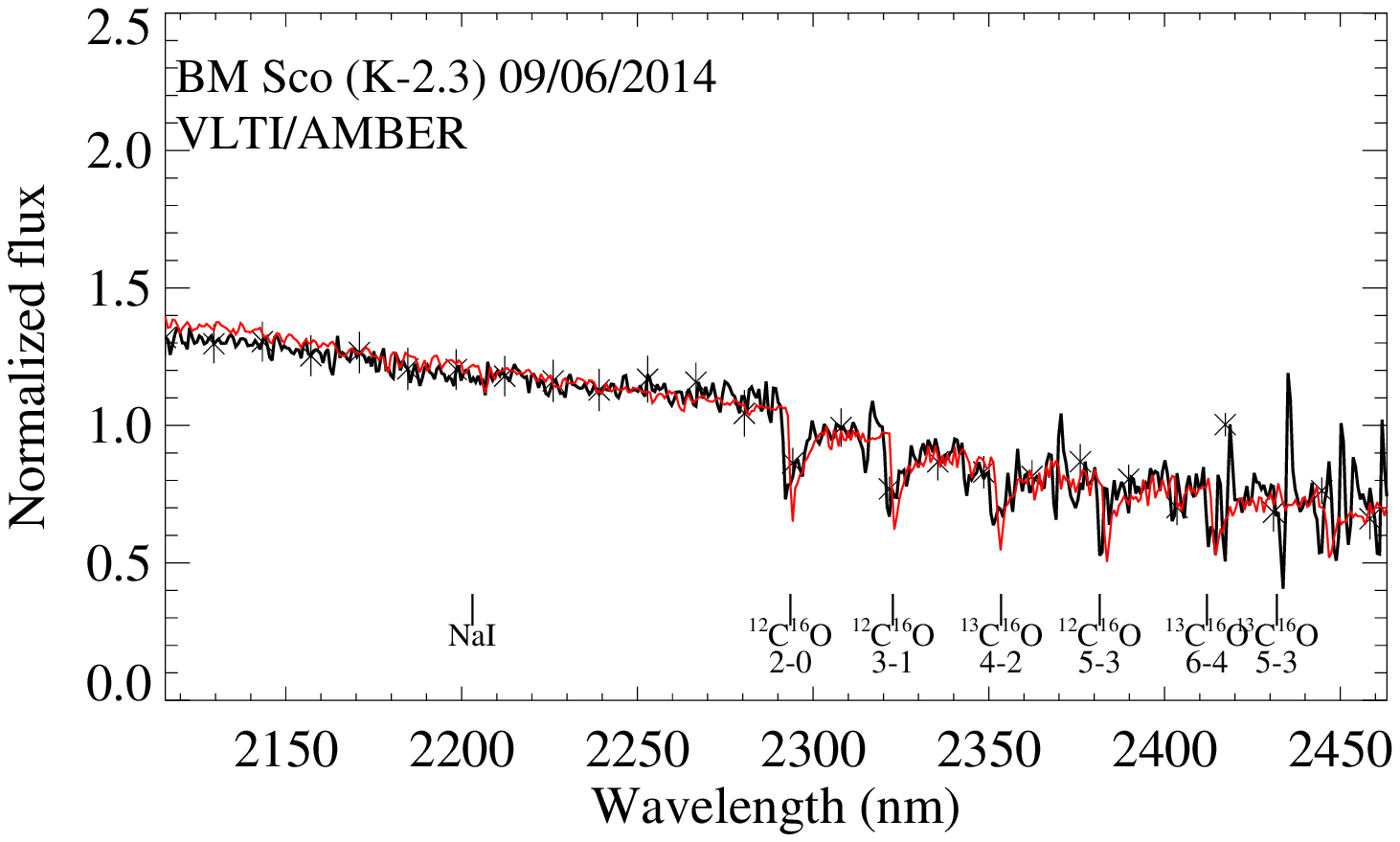}
\includegraphics[width=0.95\hsize]{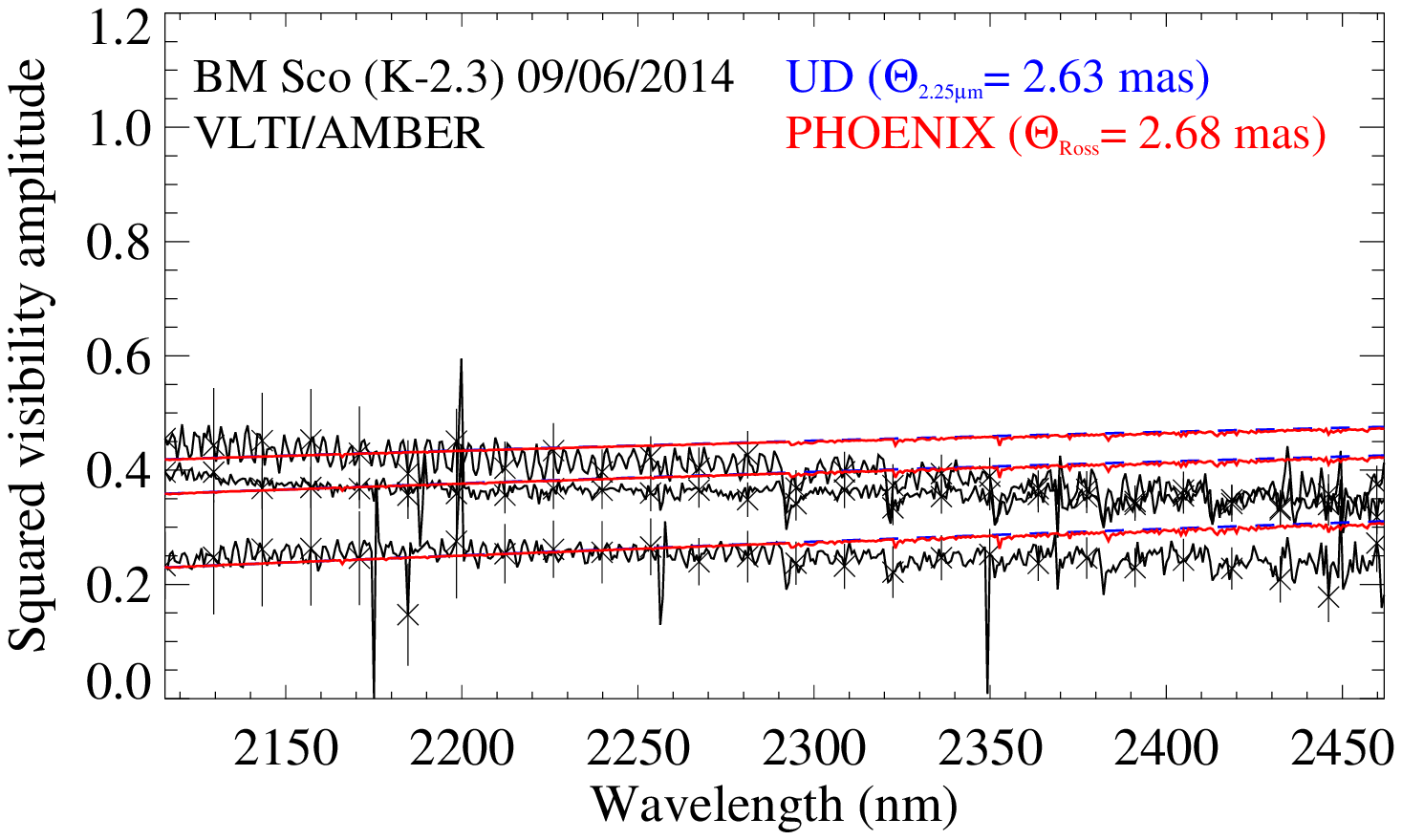}
\includegraphics[width=0.95\hsize]{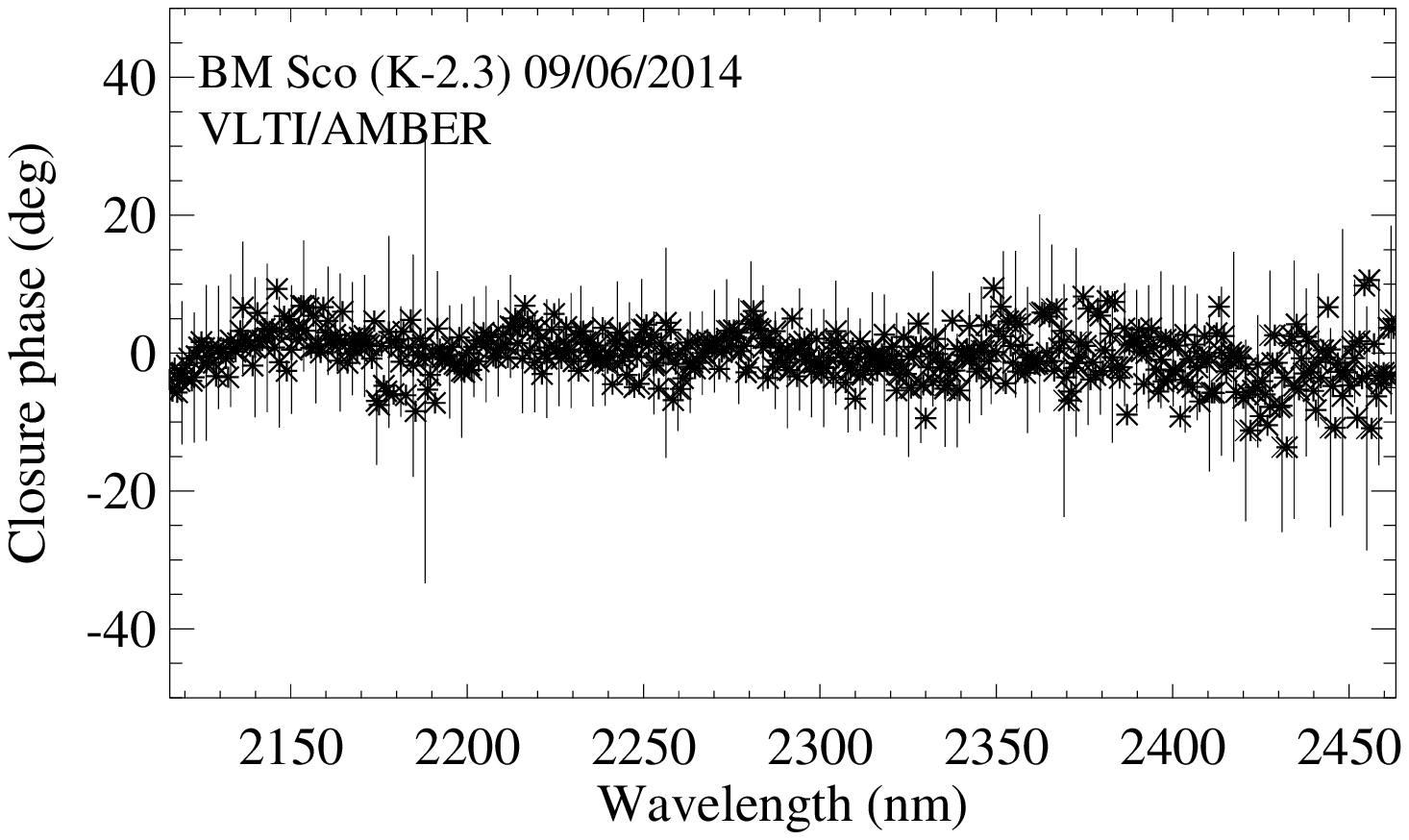}
\caption{As Fig. \ref{fig:resul_V766Cen}, but for data of BM~Sco obtained 
with the MR-K 2.3\,$\mu$m setting on 2014 June 09.
}
\label{fig:resul_BMSco2}
\end{figure}
\begin{figure*}
\centering
\includegraphics[width=0.475\hsize]{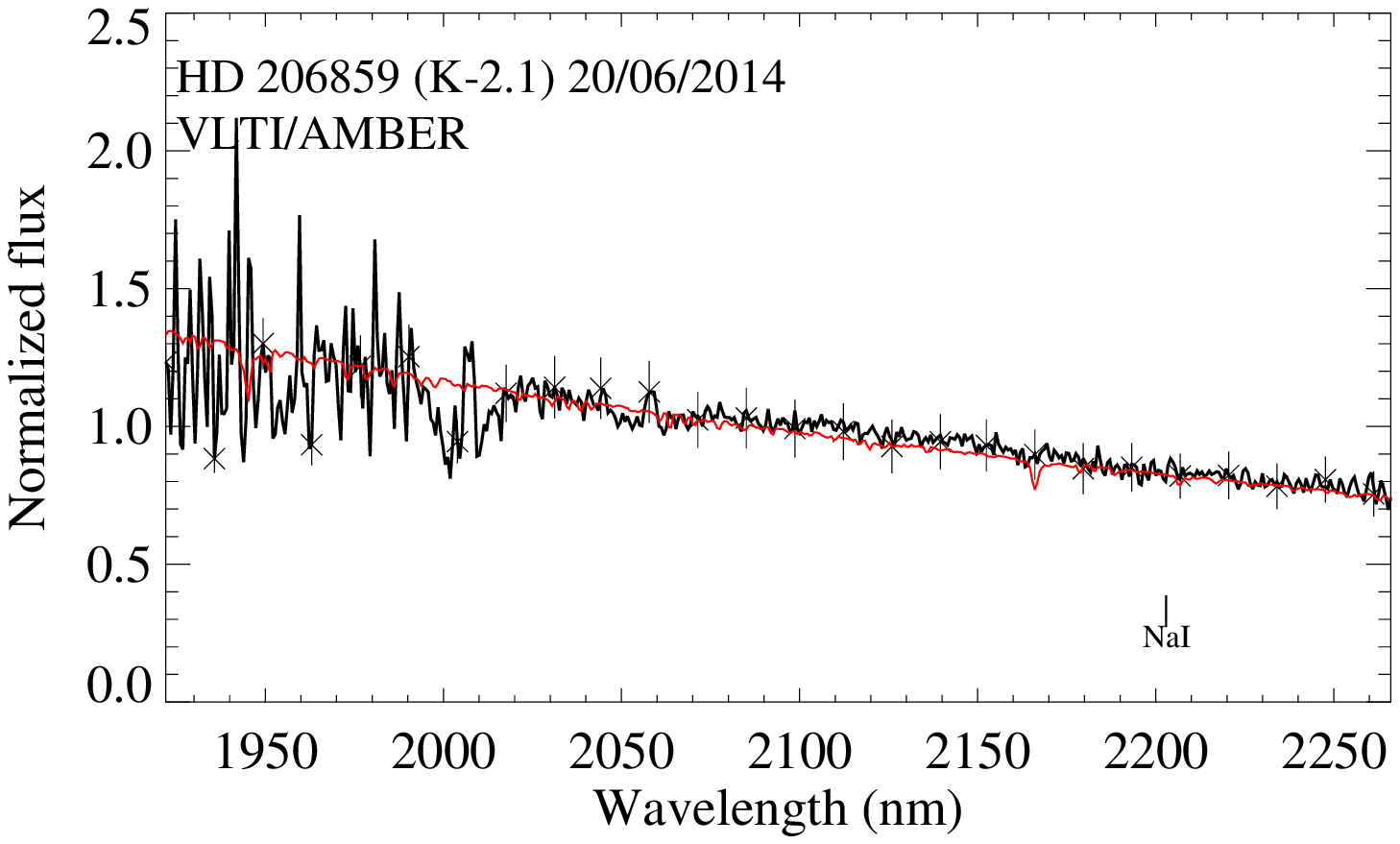}
\includegraphics[width=0.475\hsize]{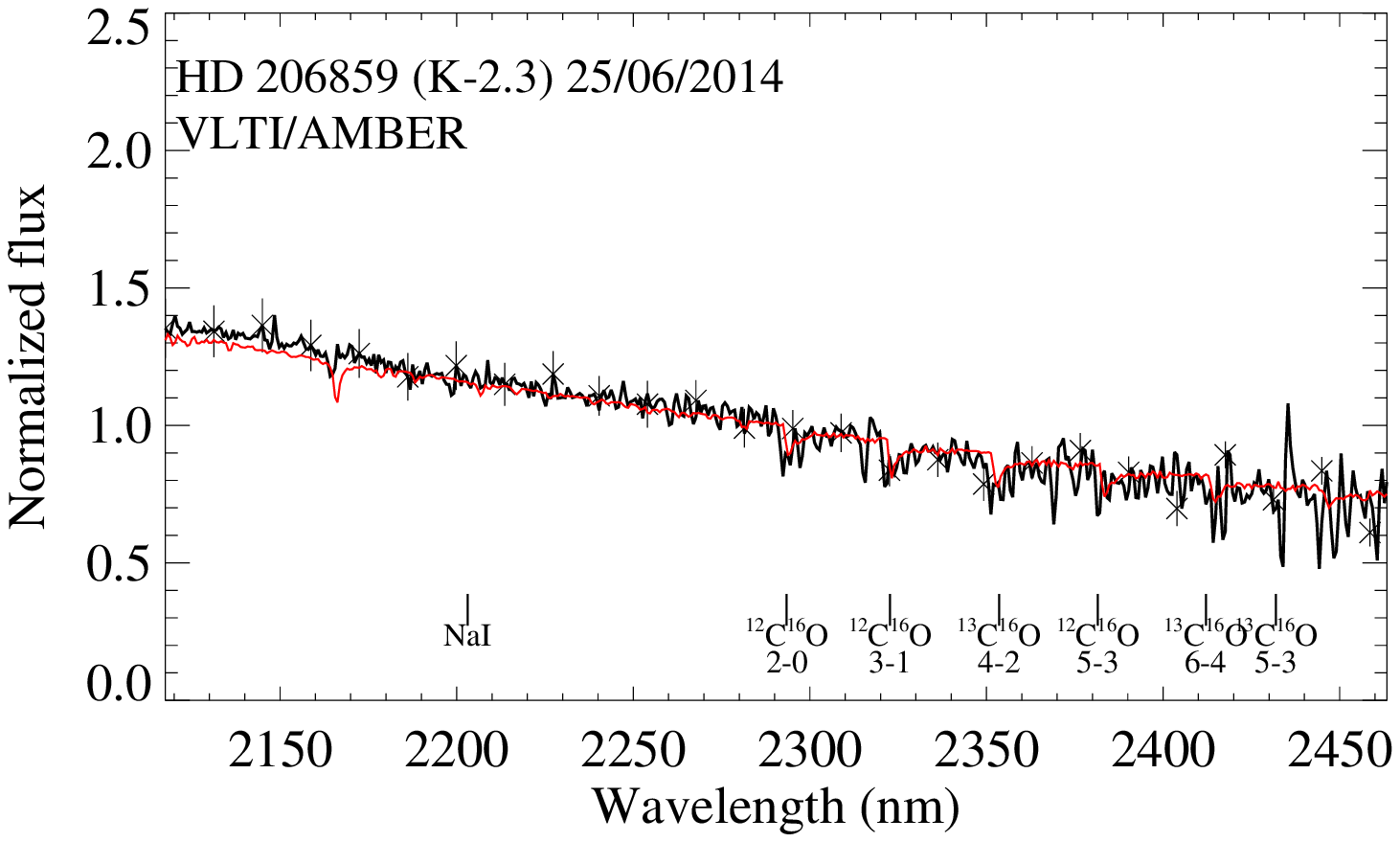}
\includegraphics[width=0.475\hsize]{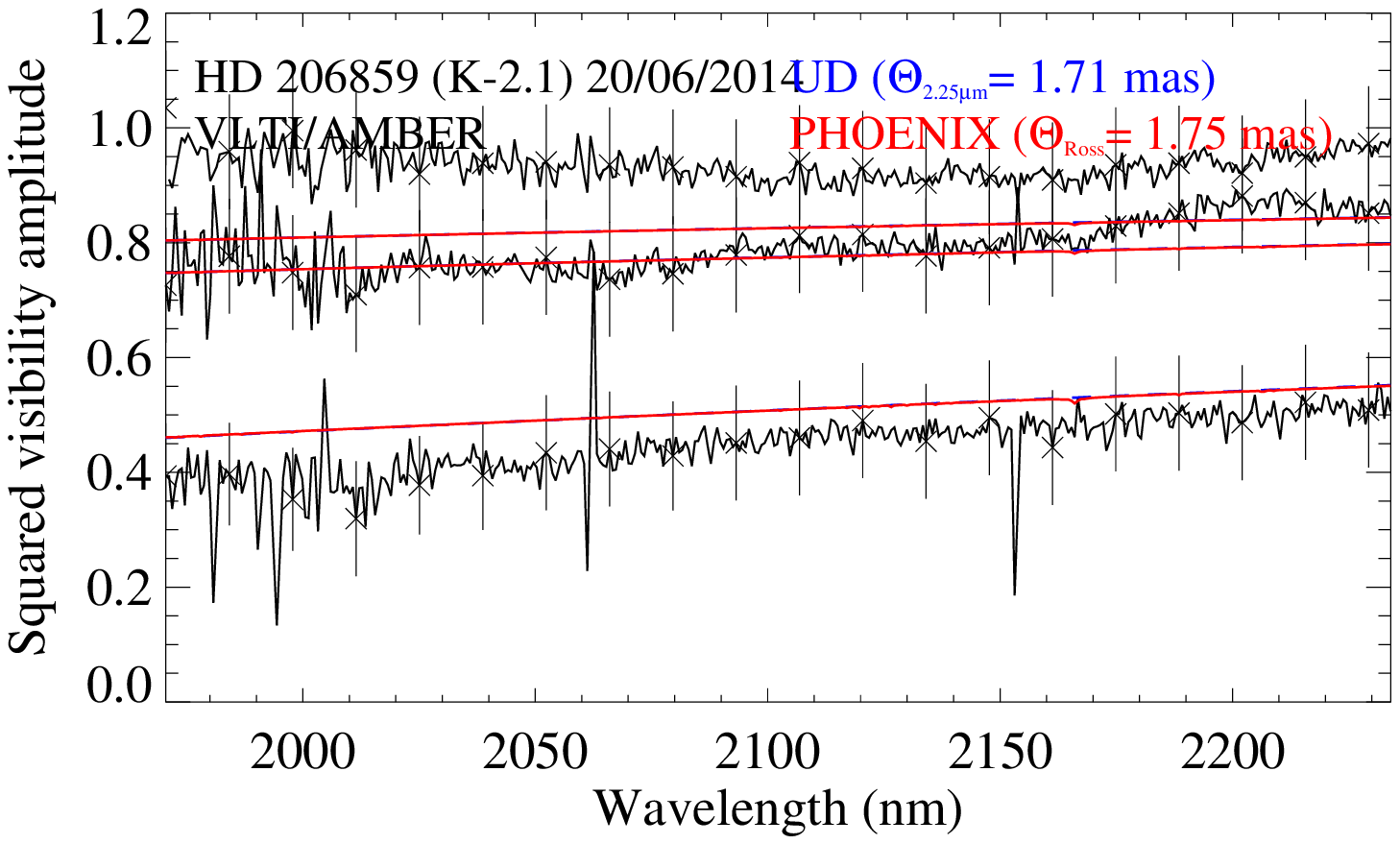}
\includegraphics[width=0.475\hsize]{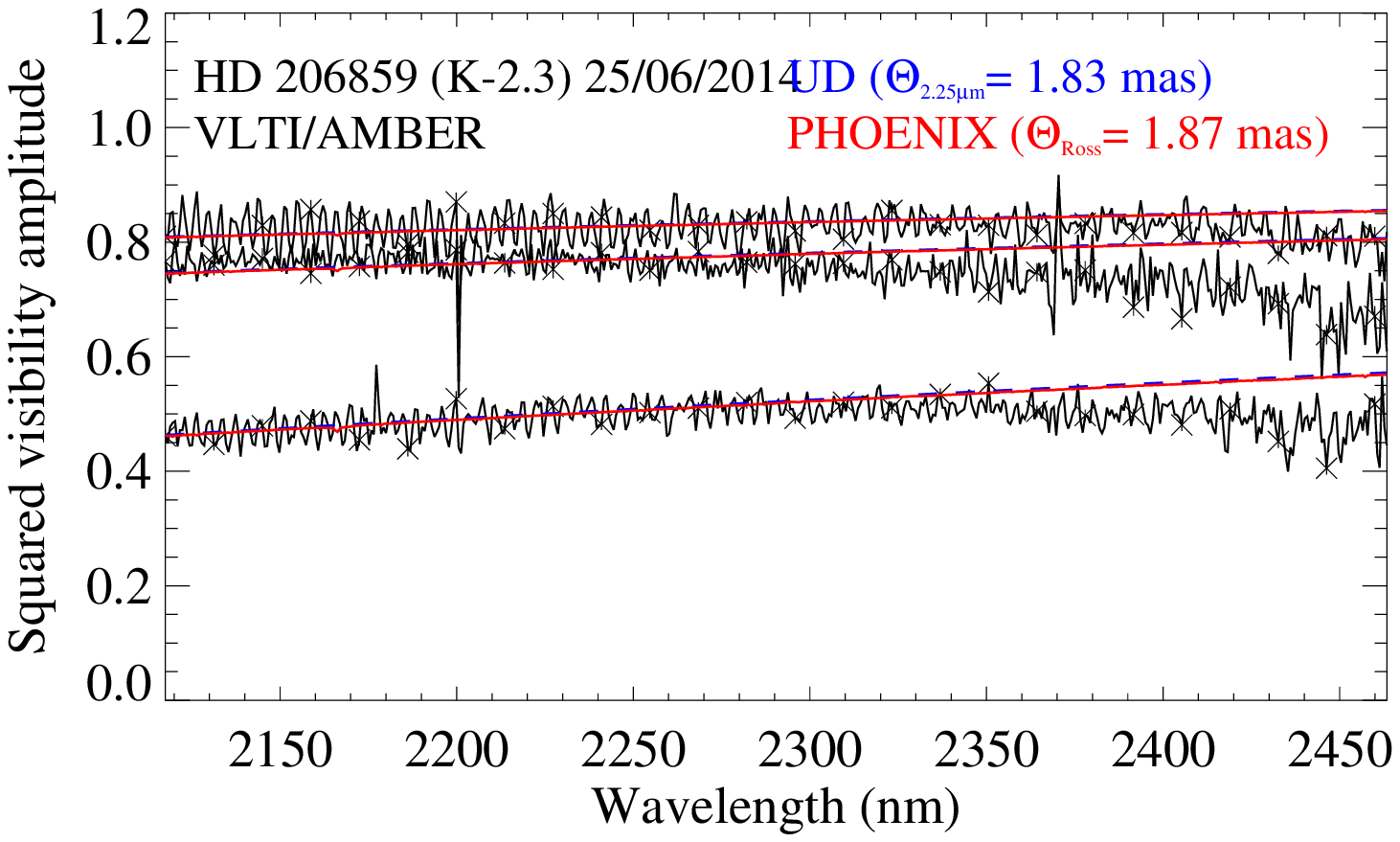}
\includegraphics[width=0.475\hsize]{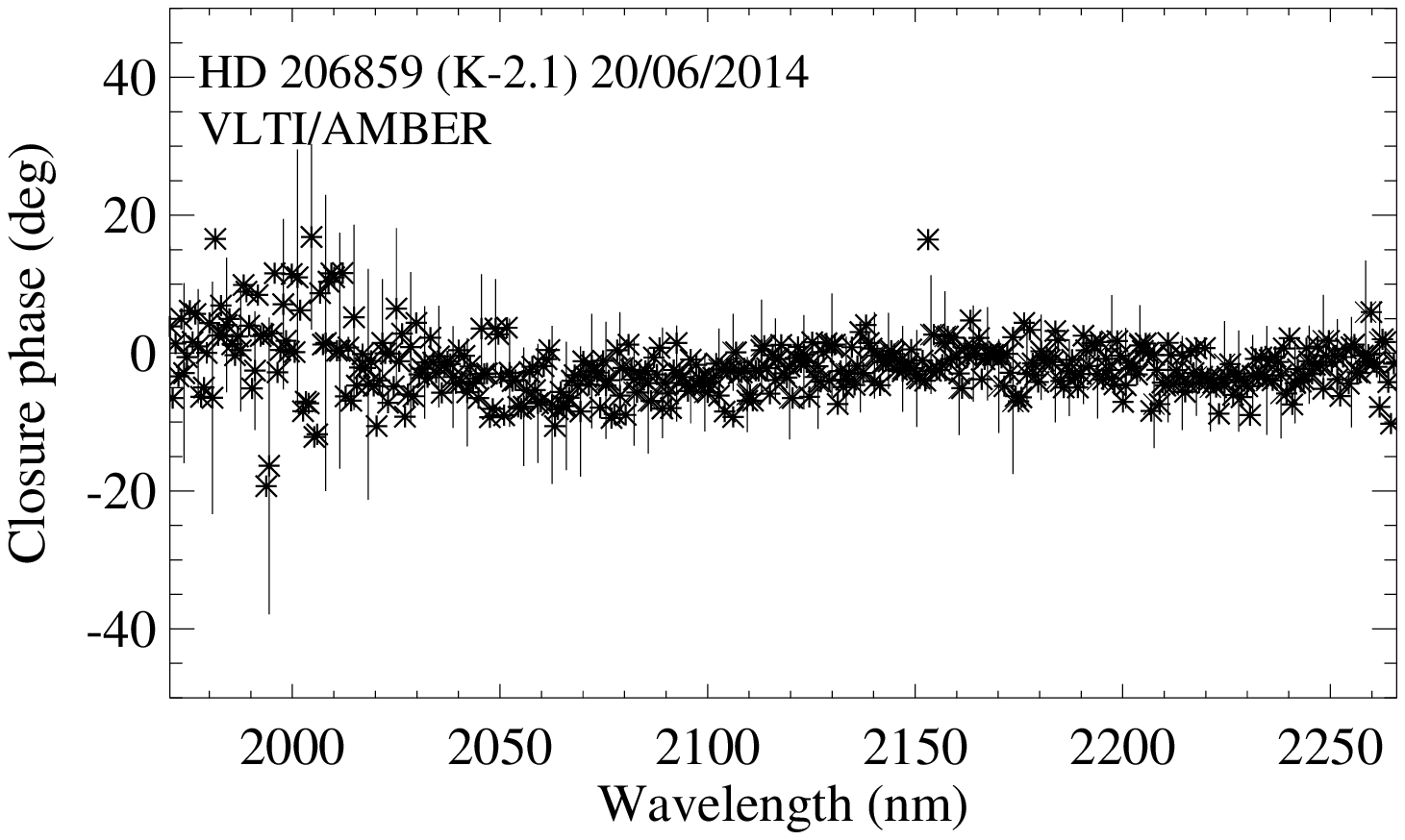}
\includegraphics[width=0.475\hsize]{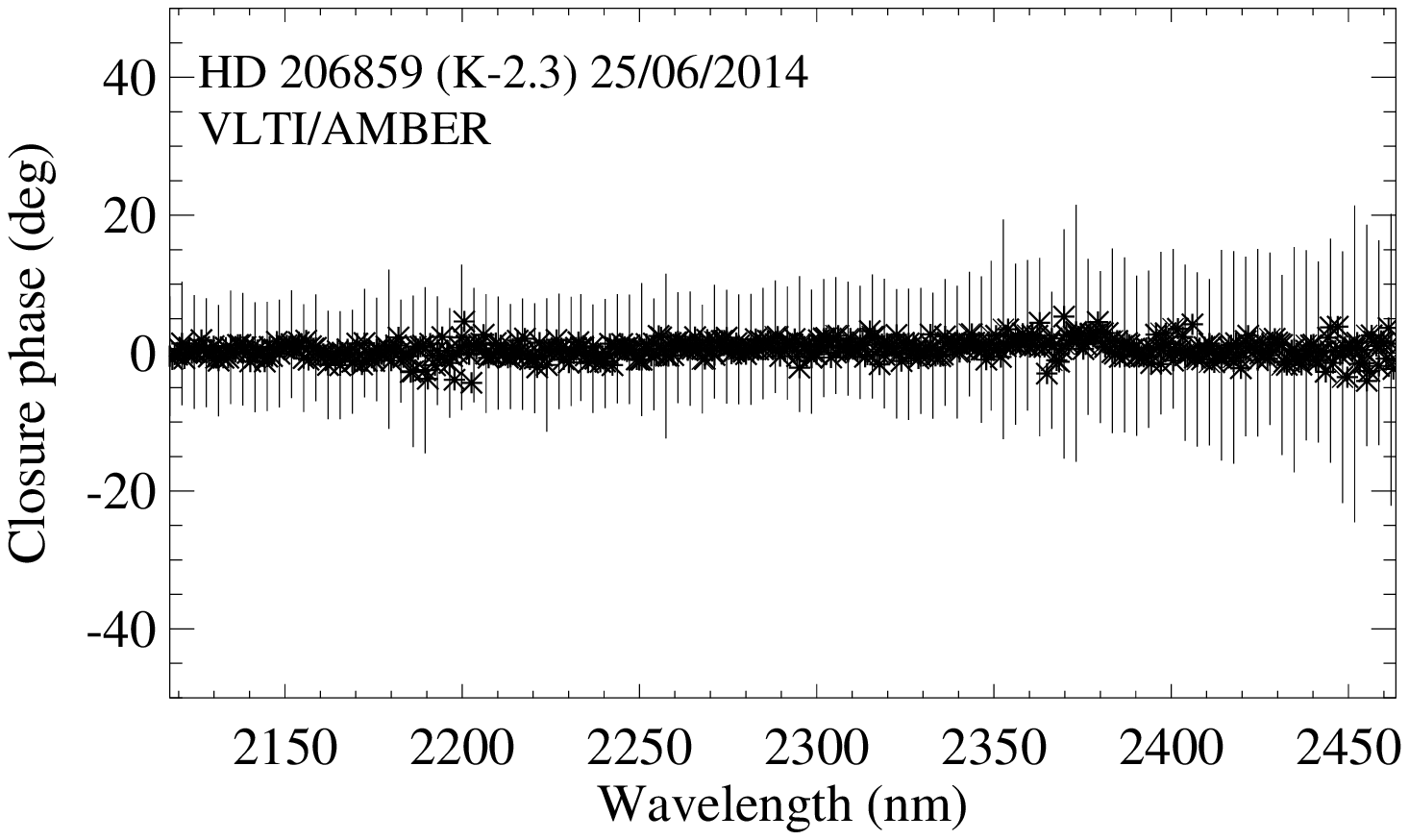}
\caption{As Fig. \ref{fig:resul_V766Cen}, but for data of HD~206859 
obtained with the MR-K 2.1\,$\mu$m setting on 2014 June 20 (left) and with 
the MR-K 2.3\,$\mu$m setting on 2014 June 25 (right).}
\label{fig:resul_HD2068591}
\end{figure*}
%
\section{Results}
\label{sec:results}
Figs.~\ref{fig:resul_V766Cen}--\ref{fig:resul_HD2068591} show the resulting 
reduced data for each of the data sets listed in Tab.~\ref{tab:Log_obs}.
The data include the normalized flux, the squared visibility amplitudes
for each baseline, and the closure phases. All quantities are plotted 
versus wavelength.
The figures also show the predictions by a uniform-disk (UD) model and by 
PHOENIX model atmospheres
as discussed below in Sect.~\ref{sec:modeling}.
All targets show relatively flat flux spectra, decreasing with wavelength, 
and with CO absorption features from 2.3\,$\mu$m onward, which is typical 
for late-type supergiants. The spectral features at the edges of the K band 
below about 2.0\,$\mu$m and above about 2.4\,$\mu$m are most likely from 
noise owing to the lower instrumental and atmospheric transmission at 
these wavelengths. Among our four sources, the CO features in the flux
spectra are relatively weak for V766~Cen
and HD~206859, and stronger
for $\sigma$~Oph and BM~Sco. This is consistent with the warmer spectral
types of G8/G5 for V766~Cen/HD~206859, compared to
K2 for $\sigma$~Oph and BM~Scou. However, the spectrum
of V766~Cen may be a composite spectrum including the suggested close
companion, cf. Sect.~\ref{sec:bolflux}.
However, in the visibility spectra,
the CO features are very strong for V766~Cen, indicating that CO is formed
at significantly higher layers in an extended atmosphere compared to the 
continuum. For the other targets, CO features are weak in the 
visibility spectra of BM~Sco, and are absent for $\sigma$~Oph and 
HD~206859, indicating that CO is formed only slightly above or at 
about the same layers as the continuum.

The closure phase of V766~Cen (bottom panels of Fig.~\ref{fig:resul_V766Cen}) 
shows significant variations at the positions of the CO bandheads 
between 2.3\,$\mu$m and 2.5\,$\mu$m, indicating deviations from 
point symmetry at the CO-forming extended layers. \citet{Chesneau2014} 
showed similar results in their Figs.~B.1 and B.2. 
The closure phase data of $\sigma$~Oph, BM~Sco, and HD~206859 
do not show any significant deviations from point symmetry at any 
wavelength. However, since our measurements lie in the first lobe of
the visibility,
we cannot exclude 
asymmetries on scales smaller than the photospheric stellar disk.

Our flux spectra of V766~Cen exhibit an emission feature at a 
wavelength of 2.205\,$\mu$m, which was also noticed by \citet{Chesneau2014} 
and identified as the \ion{Na}{i} doublet. It corresponds to significant drops 
in the visibility, indicating that \ion{Na}{i} is formed at higher 
atmospheric layers compared to the continuum. A more detailed study
of this feature follows below in Sect.~\ref{sec:NaI}. This feature is 
not visible for any of the other sources. There are no other significant
spectral features visible in the flux or visibility spectra of
our sources within our S/N.

\subsection{Estimates of bolometric fluxes and distances}
In order to derive fundamental parameters of our sources based
on the measured angular diameters derived below, we need to estimate 
bolometric flux values and distances as a prerequisite.

\subsubsection{Bolometric fluxes}
\label{sec:bolflux}
We estimated the bolometric flux of each of our sources by integrating 
broadband photometry available in the literature. 
The details
of our procedure are described in \citet[][Sect.~4.2]{Arroyo2014}.
We used $B$ and $V$ 
magnitudes from \citet{Kharchenko2001}, $J$, $H$, $K$ magnitudes from 
\citet{Cutri2003}, and the IRAS fluxes from \citet{Beichman1988} for 
V766~Cen, BM~Sco, and HD~206859 and from \citet{Moshir1990} for $\sigma$~Oph.
We dereddened the flux value using an estimated $E_{B-V}$ value. 
For V766~Cen and HD~206859, we used the $V$-$K$ color excess 
method with intrinsic colors from \citet{Ducati2001}.
For $\sigma$~Oph and BM~Sco, we used
the values estimated by \citet{Schlafly2011} and \citet{Levesque2005}, 
respectively. 
The adopted $E_{B-V}$ values are 0.12 for $\sigma$\,Oph, 
0.14 for BM\,Sco, 0.10 for HD\,206859, and 0.92 for V766\,Cen.
We assume an uncertainty of 15\% in the final bolometric
flux.

For V766~Cen, we estimate a total bolometric flux of 
$2.25\times10^{-9}$\,W/m$^{2}$.
This value includes the contribution by the close companion that
was suggested by \citet{Chesneau2014}. Given the effective temperature
of the companion of 4811\,K and its angular diameter of 1.8\,mas
from \citet{Chesneau2014},
we estimate a flux contribution of the
companion of $5.78\times10^{-10}$\,W/m$^{2}$ and conservatively 
adopt an error of 50\%. Subtracting this value from the total flux,
we derive a bolometric flux of the primary component alone of 
$1.68\times10^{-9}$\,W/m$^{2}$ $\pm$ $0.45\times10^{-9}$\,W/m$^{2}$.

\subsubsection{Distances}
V766~Cen belongs to the OB association R80, and we adopted its
distance by \citet{Humphreys1978}. For BM~Sco, we used the distance of 
the OB association M6 \citep{Mermilliod2003}. The stars $\sigma$~Oph and HD~206859 
are not known to belong to any OB association. We used the distance 
from \citet{Anderson2012}. The adopted distances are listed in
Tab.~\ref{tab:fundpar}.

\subsection{Model atmosphere fits}
\label{sec:modeling}
\begin{table}
\caption{Best-fit flux fractions and angular diameters}
\begin{center}
\begin{tabular}{lrrr}
\hline
Source       & $A$  & $\Theta_\mathrm{Ross}$ (mas) & $\Theta_\mathrm{UD}$ (mas) \\\hline
V766~Cen     & 0.94 & 3.86 $\pm$ 0.90 & 3.87 $\pm$ 0.90 \\
$\sigma$~Oph & 1.00 & 3.41 $\pm$ 0.90 & 3.35 $\pm$ 0.90 \\
BM~Sco       & 0.73 & 2.49 $\pm$ 0.35 & 2.45 $\pm$ 0.35 \\
HD~206859    & 1.00 & 1.86 $\pm$ 0.50 & 1.83 $\pm$ 0.50 \\\hline
\end{tabular}
\end{center}
\label{tab:ang_diameter}
\end{table}
We modeled our visibility data using a uniform disk (UD) model as 
well as PHOENIX model atmospheres following our previous 
papers \citep{Wittkowski2012,Arroyo2013,Arroyo2015}:
We used the grid of PHOENIX model atmospheres that we introduced in 
\citet{Arroyo2013}.
We used a scaled visibility function of the form
\begin{equation}
V(A,\Theta_\mathrm{Ross})=A\times V^\mathrm{PHOENIX}(\Theta_\mathrm{Ross}),
\end{equation}
where $A$ accounts for the flux fraction of the stellar disk, and where
the remaining flux is attributed to a possible over-resolved
circumstellar component within our field of view. 
The AMBER field of view using the ATs is about 270\,mas in size.
The Rosseland angular diameter $\Theta_\mathrm{Ross}$
of the respective PHOENIX model 
was the only fit parameter in addition to $A$. Here, $\Theta_\mathrm{Ross}$ 
corresponds to that model layer where the Rosseland optical depth 
equals $2/3$.
Likewise, we used a scaled UD model with fit parameters $A$ and
$\Theta_\mathrm{UD}$.
While we have shown
that none of the currently available model atmospheres predict observed
extensions of molecular layers (mainly in CO lines), our previous
data \citep{Wittkowski2012,Arroyo2013,Arroyo2015} as well as other 
studies \citep[e.g.,][]{Perrin2004}
showed that visibility 
curves of RSGs in near-continuum bandpasses can be well described 
by UD curves or hydrostatic model atmospheres.
As a result, we restricted the model fits to the near-continuum band 
at 2.05\,$\mu$m--2.20\,$\mu$m, which is least affected by contamination 
from molecular layers within the $K$ band. This bandpass
also avoids the \ion{Na}{i} doublet that has been detected for V766~Cen.  
We chose initial parameters of the PHOENIX model (mass $M$, effective 
temperature $T_\mathrm{eff}$, luminosity $L$) and 
iterated so that the model parameters were consistent with those
derived from the Rosseland angular diameters and the adopted bolometric 
fluxes and distances. 
Finally, for V766~Cen, $\sigma$~Oph, BM~Sco,
HD~206859, we used model atmospheres with effective temperatures
4300\,K, 4100\,K, 3900\,K, 5300\,K, and $\log g$ of values
-0.5, 1, 1, 2, respectively. We used models of masses 1\,M$_\odot$
and 20\,M$_\odot$ for all sources because our model grid includes only
these masses.
The structure of the atmosphere
is not very sensitive to mass \citep{Hauschildt1999} and the differences
in our final $\Theta_\mathrm{Ross}$ values are small compared to the
quoted errors. The $\Theta_\mathrm{Ross}$ values are the only
fit results that are used in the following.
Table~\ref{tab:ang_diameter} shows the resulting
best-fit angular diameters $\Theta_\mathrm{Ross}$/ $\Theta_\mathrm{UD}$
and flux fractions $A$.

\begin{figure}
\centering
\includegraphics[width=0.95\hsize]{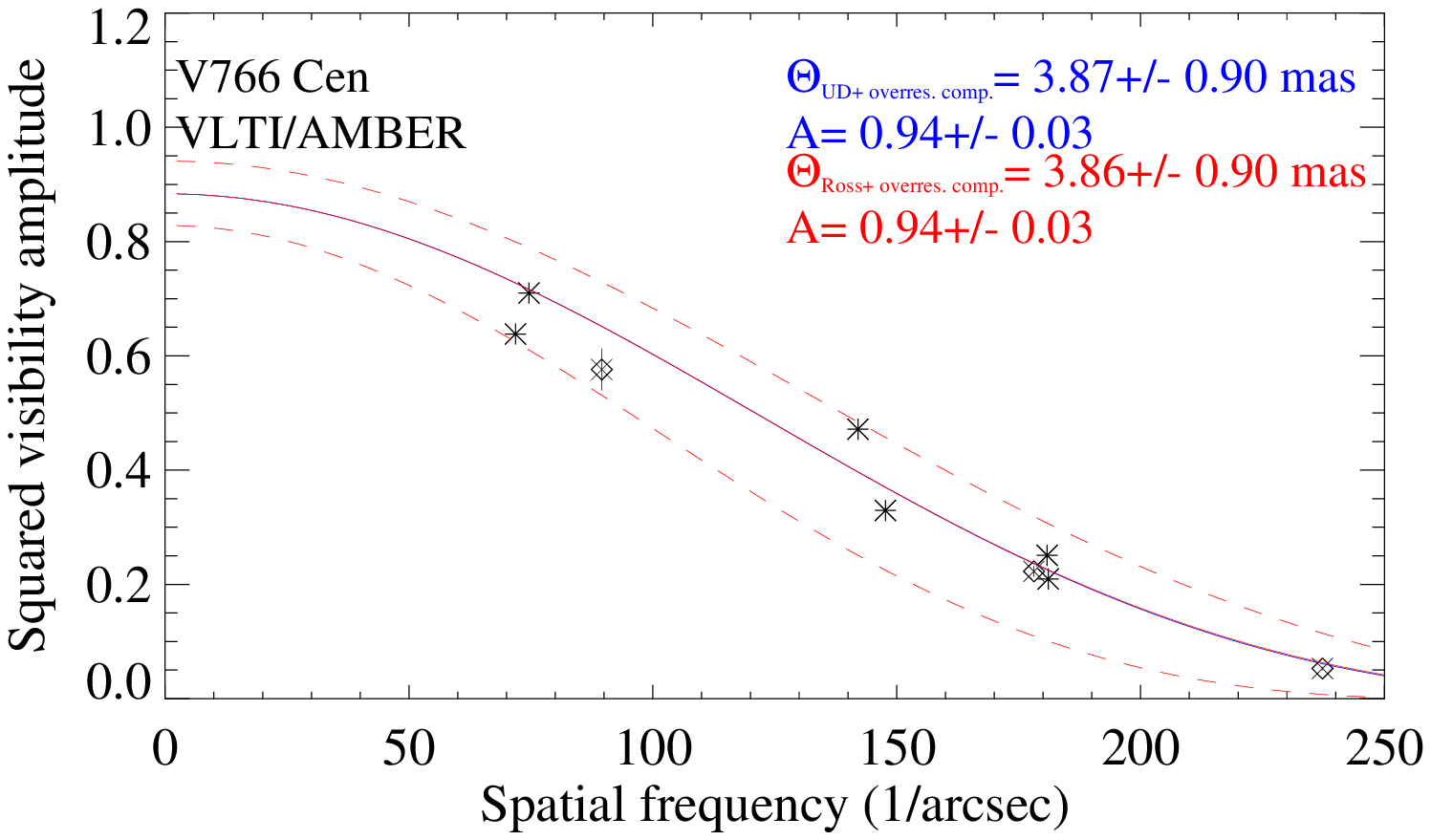}
\includegraphics[width=0.95\hsize]{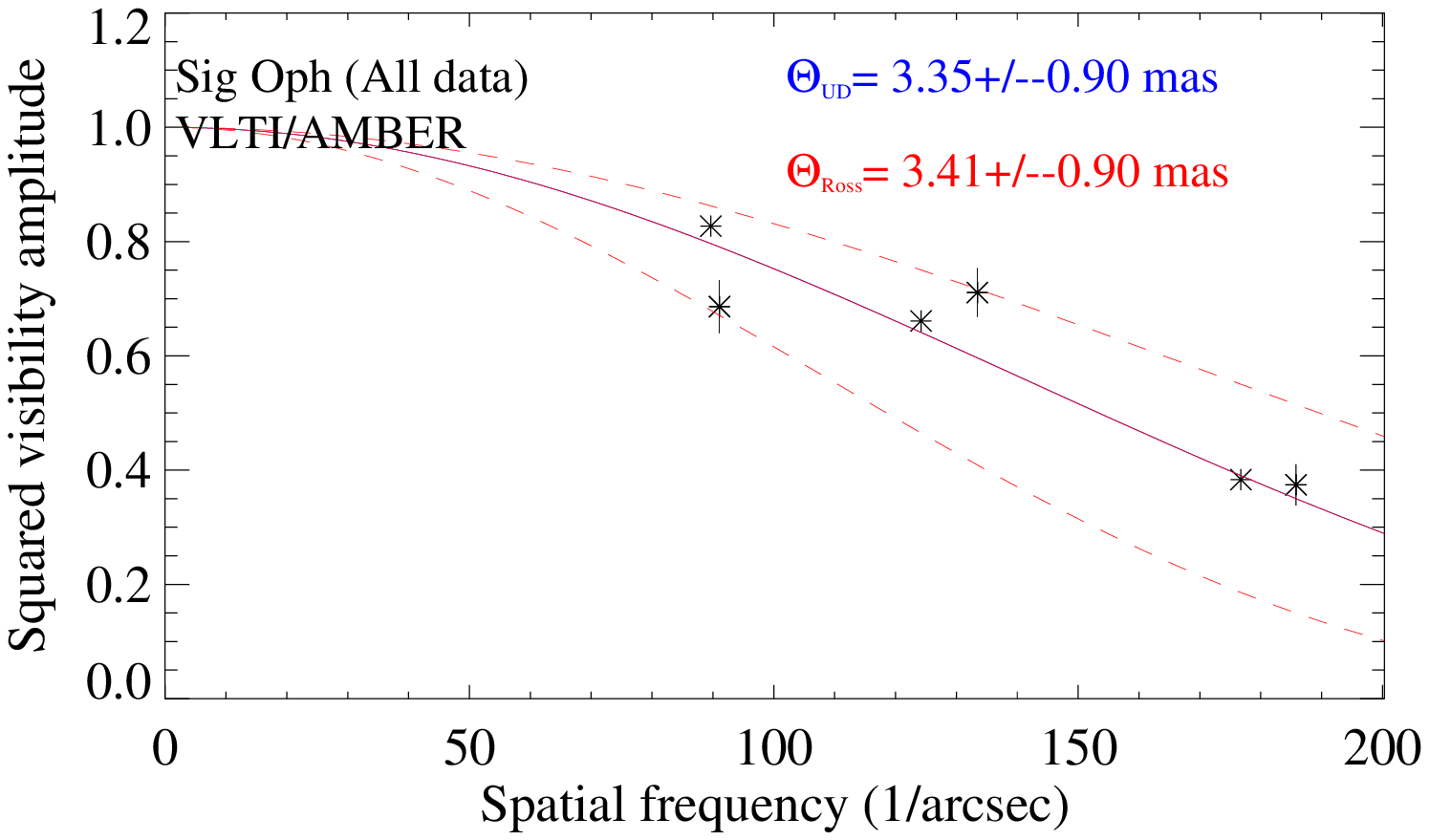}
\includegraphics[width=0.95\hsize]{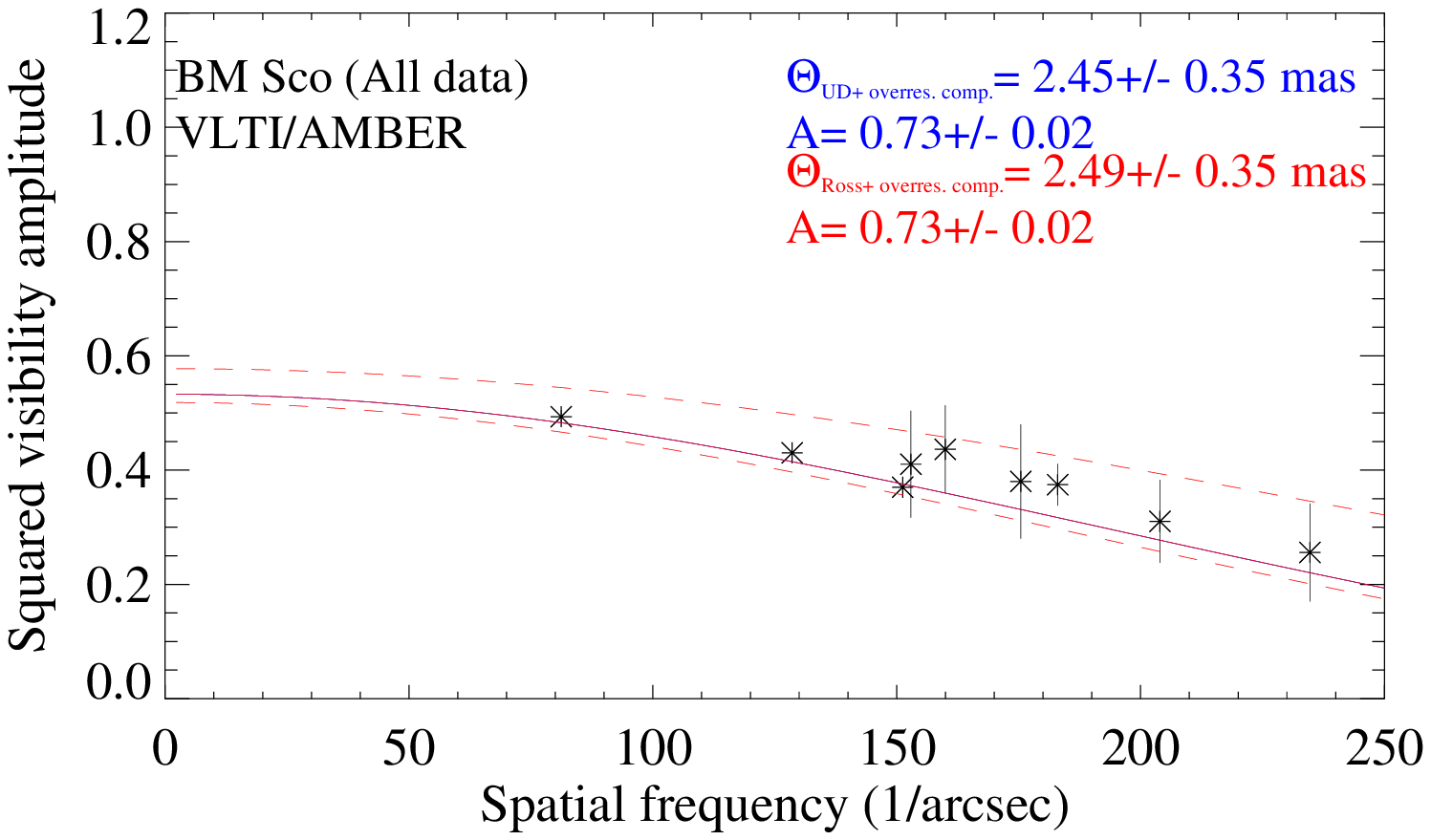}
\includegraphics[width=0.95\hsize]{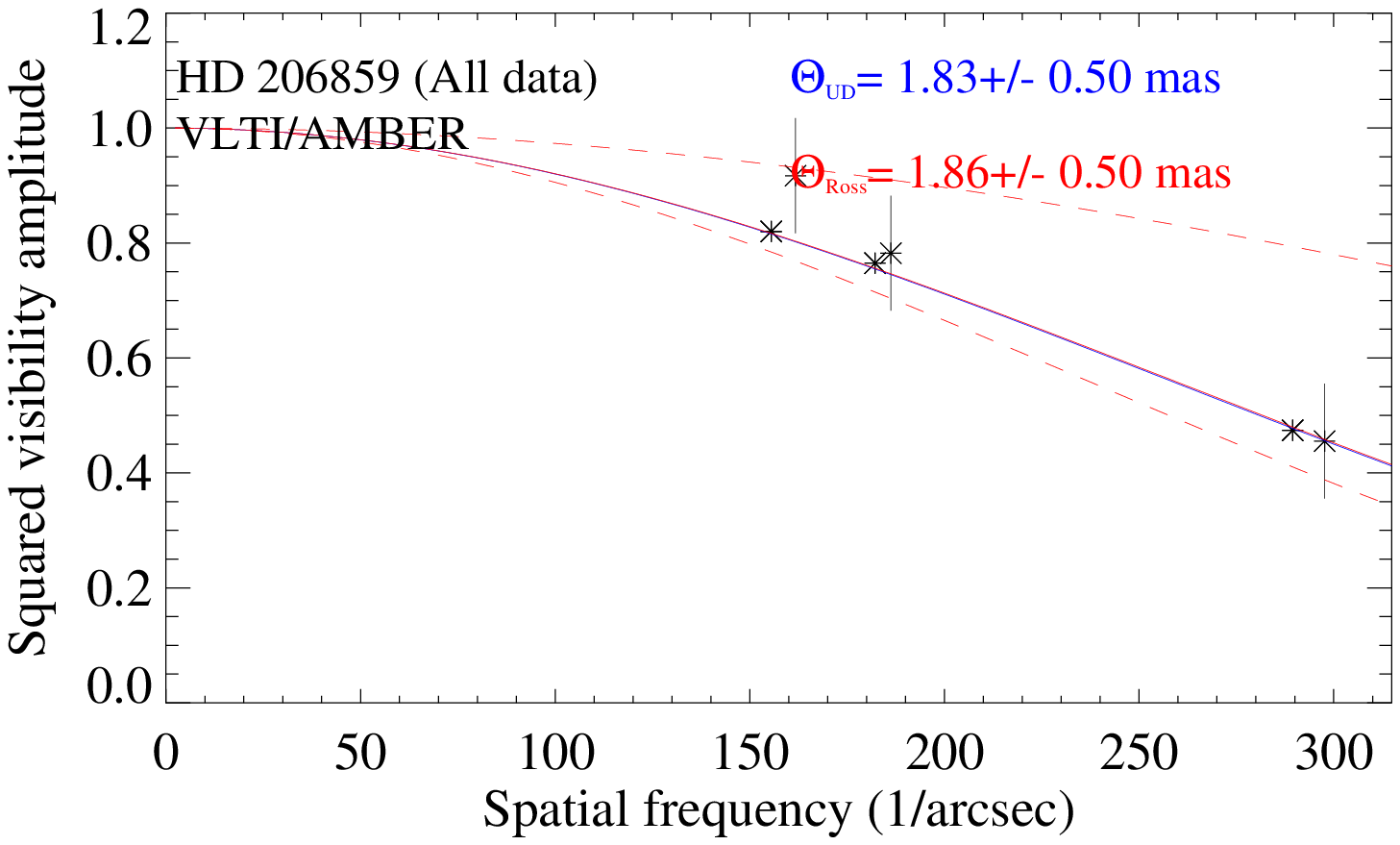}
\caption{Average of squared visibility amplitudes taken in the near-continuum
bandpass at 2.05--2.20\,$\mu$m
for (from top to bottom) V766~Cen, $\sigma$~Oph, BM~Sco,
and HD~206859 as a function of spatial frequency.
The red lines indicate the best-fit UD models and the blue lines the
best-fit PHOENIX models. The dashed lines are the maximum and minimum
visibility curves, from which we estimated the errors of the angular
diameters.}
\label{fig:Vis_spacialFrec}
\end{figure}

Fig.~\ref{fig:Vis_spacialFrec} shows for each of our sources 
the near-continuum (averaged over 2.05\,$\mu$m--2.20\,$\mu$m) squared 
visibility amplitudes as a function of spatial frequency together with the 
best-fit PHOENIX and UD models. 
V766~Cen and BM~Sco show $A$ values significantly below 1, indicating
the presence of an underlying over-resolved component within our
field of view, possibly due to dust.

To realistically estimate the error of the angular diameter, including statistical and systematic errors, 
we applied variations of the angular diameter so that the
model visibility curve lies below and on top of all measured points.
These maximum and minimum visibility curves are plotted as dashed lines
in Fig.~\ref{fig:Vis_spacialFrec}.
For V766~Cen (top panel of Fig.~\ref{fig:Vis_spacialFrec}), we also used the 
medium spectral resolution $K$-band data obtained 
by \citet{Chesneau2014} on 2012-03-09 in addition to our own data.
We processed them in the same way as our data. We did not use the low spectral resolution
data from \citet{Chesneau2014}
for consistency
with our strategy of basing our study on the medium spectral resolution
mode.

\begin{figure}
\centering
\includegraphics[width=0.95\hsize]{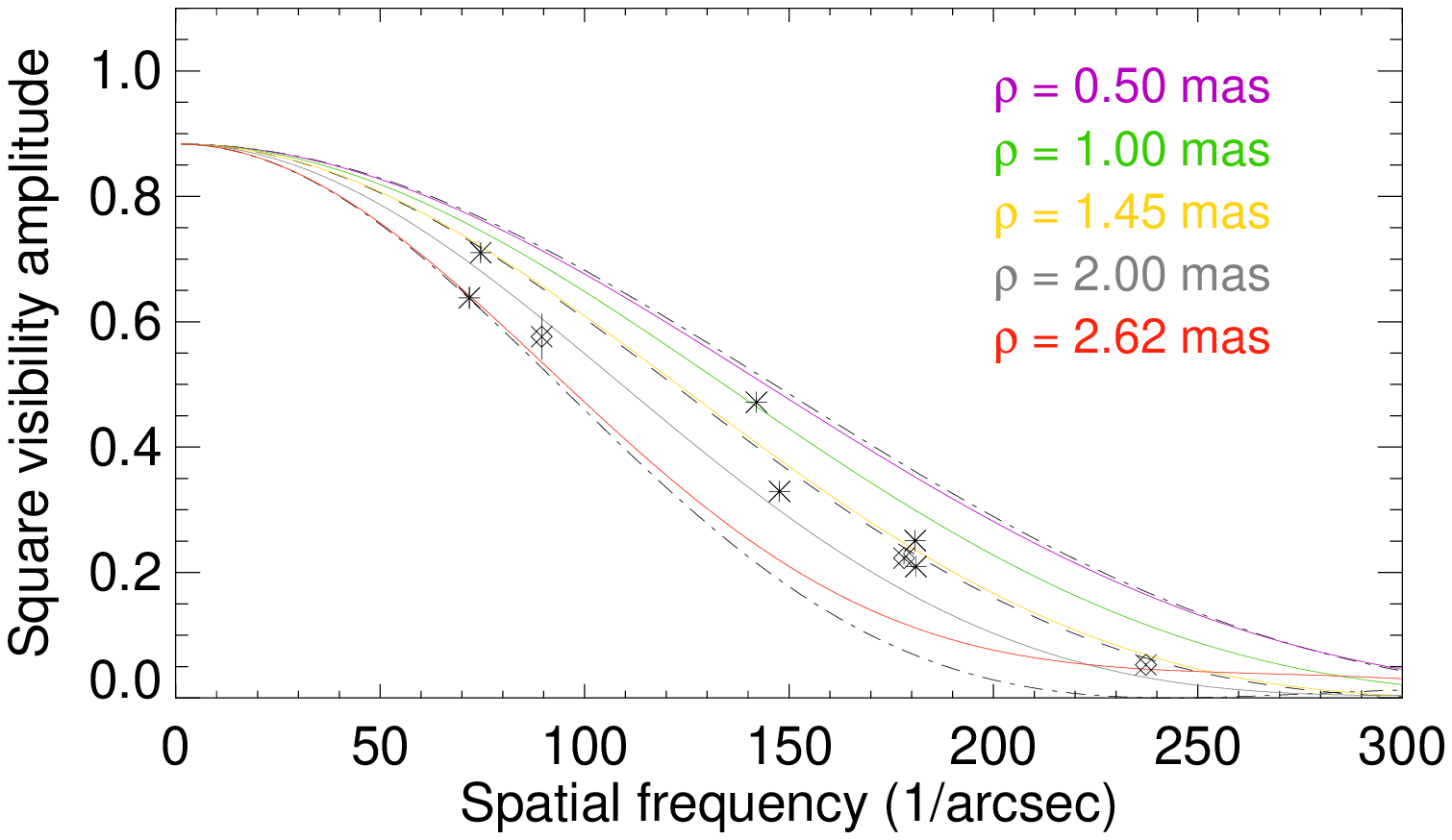}
\caption{Influence of the close companion of V766~Cen on the visibility curve.
The asterisk and diamond symbols denote our near-continuum
(2.05-2.20\,$\mu$m) data,
the central dashed line denotes the best-fit UD with over-resolved component,
and the dash-dotted lines the error of the UD diameter.
The color curves show the synthetic visibilities of a binary toy model
consisting of two UDs and an over-resolved component
with flux ratios as
derived by \citet{Chesneau2014}. 
The different colors indicate different
positions of the companion corresponding along the projected orbit.
The figure illustrates that our observations cannot separate between 
these curves and thus cannot fully characterize the companion. 
The range of the visibility curves of our binary toy model are within
our already adopted diameter error. 
}
\label{fig:V766Cen_binary_model}
\end{figure}
Moreover, for V766~Cen, we needed to investigate the effect that the
suggested close companion \citep{Chesneau2014} may have on our visibility 
data and on the determination of the angular diameter of the primary
(cf. the corresponding discussion for the bolometric flux in 
Sect.~\ref{sec:bolflux}.) 
We used a binary toy model,
consisting of two UDs, one UD for the primary and one offset UD for the
close companion, and an over-resolved component. We used the parameters of 
the orbit as obtained with the best NIGHTFALL model by 
\citet{Chesneau2014}\footnote{best-fit model 1 of their table 2}. Since we do 
not know the exact position along the orbit at the time of our observation, 
we computed synthetic visibility values for a range of such positions.
Fig.~\ref{fig:V766Cen_binary_model} shows the resulting synthetic
visibility values compared to the measured values along with the
uncertainties due to our UD analysis described above. As seen in the 
figure, the additional error introduced by the presence of the
close companion is comparable to that of our UD analysis.
Accordingly, we added in quadrature both contributions to the final
uncertainty and, in practice, we multiplied our error from the UD analysis
by a factor of $\sqrt{2}$.

Finally, we used the best-fit PHOENIX models 
to produce  synthetic
flux and visibility data as a function of wavelength.
These are shown in
Figs.~\ref{fig:resul_V766Cen}--\ref{fig:resul_HD2068591} together with the 
observed data.

For all four sources, the synthetic flux spectra (top panels) 
of the PHOENIX model are generally in good agreement with our observations, 
in particular at the locations of the CO bandheads. The PHOENIX model fluxes 
are consistent with the stronger CO bands for $\sigma$~Oph and BM~Sco, 
and the weaker CO bands for V766~Cen and HD~206859. For V766~Cen, the 
observed CO features may be weaker than predicted.
It is not clear whether this is due to noise or the composite 
spectrum with the close companion, which the model does not take into account. The observed presence of the \ion{Na}{i} doublet in emission for 
V766~Cen is not reproduced by the model atmosphere.

The visibility spectra (middle panels of Figs. 2--5) agree between 
observations and models
for $\sigma$~Oph, BM~Sco, and HD~206859; both the observed and model visibility spectra do not show significant 
CO features, indicating that CO is formed close to the photosphere.
BM~Sco shows weak CO features that slightly extend those of the best-fit 
PHOENIX model.  
It is not yet clear whether this indicates an extension of the CO layers 
beyond the predictions of the model for this source or whether,
for example, assumptions on the distance or bolometric flux and, thus
of model parameters, are erroneous.

However, the visibility spectra of V766~Cen (middle panels of 
Fig.~\ref{fig:resul_V766Cen}) show large visibility drops in the 
CO bandheads (2.3-2.5\,$\mu$m) and in the \ion{Na}{i} doublet (2.205\,$\mu$m), 
which are not reproduced by the PHOENIX model. The synthetic PHOENIX 
visibilities show CO features, but these features are much weaker. This indicates 
that the PHOENIX model structure is too compact compared to our observations.
We observed the same phenomenon previously for other RSGs as follows: 
VY~CMa \citep{Wittkowski2012}, AH~Sco, UY~Sct, KW~Sgr \citep{Arroyo2013},
and V602~Car, HD~95687 \citep{Arroyo2015}.

The panels showing the closure phases (bottom panels) 
do not include model atmosphere predictions because
spherical models cannot predict closure phase
signals other than zero in the first lobe.

We could not compare the data of V766~Cen to 3D radiation hydrodynamic (RHD)
simulations because such simulations do not currently exist for the
parameters of V 766 Cen. Moreover, we have already shown in \citet{Arroyo2015} that
the stratification of such models is generally close to hydrostatic models
and in particular cannot explain the extended molecular layers that are
observed for RSGs.
\subsection{\ion{Na}{i} emission toward V766~Cen}
\label{sec:NaI}
\begin{figure*}
\centering
\includegraphics[width=0.475\hsize]{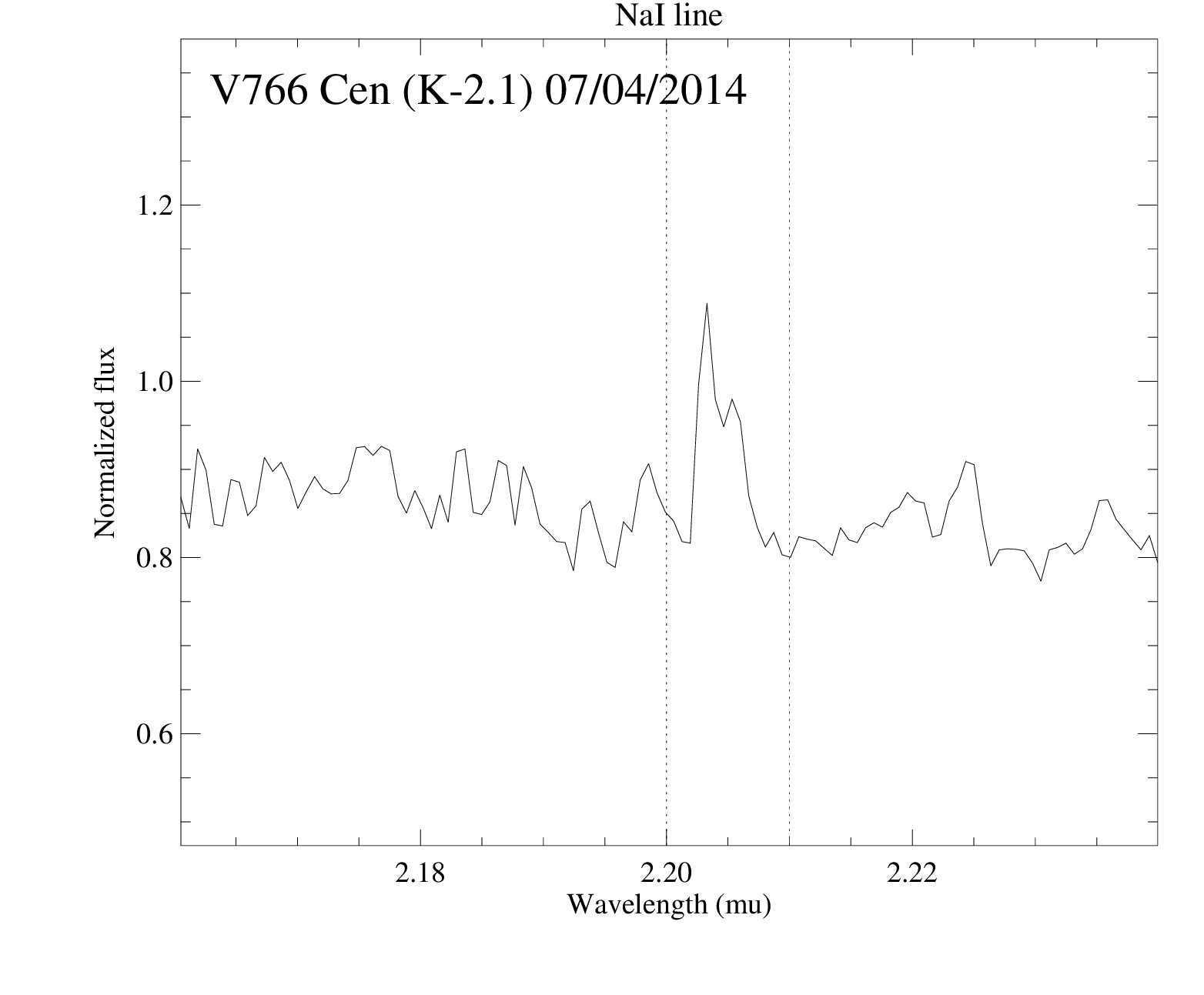}
\includegraphics[width=0.475\hsize]{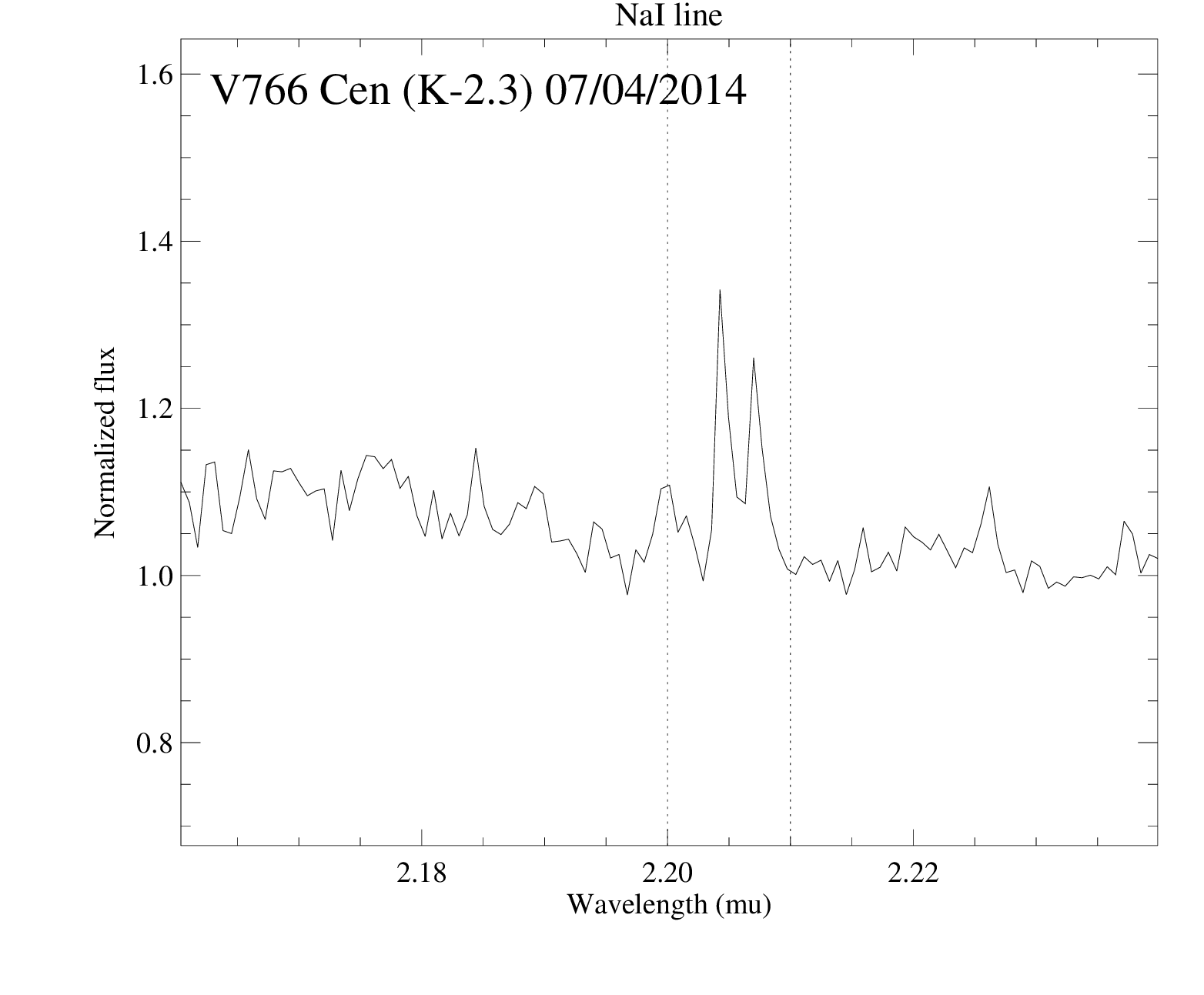}
\includegraphics[width=0.475\hsize]{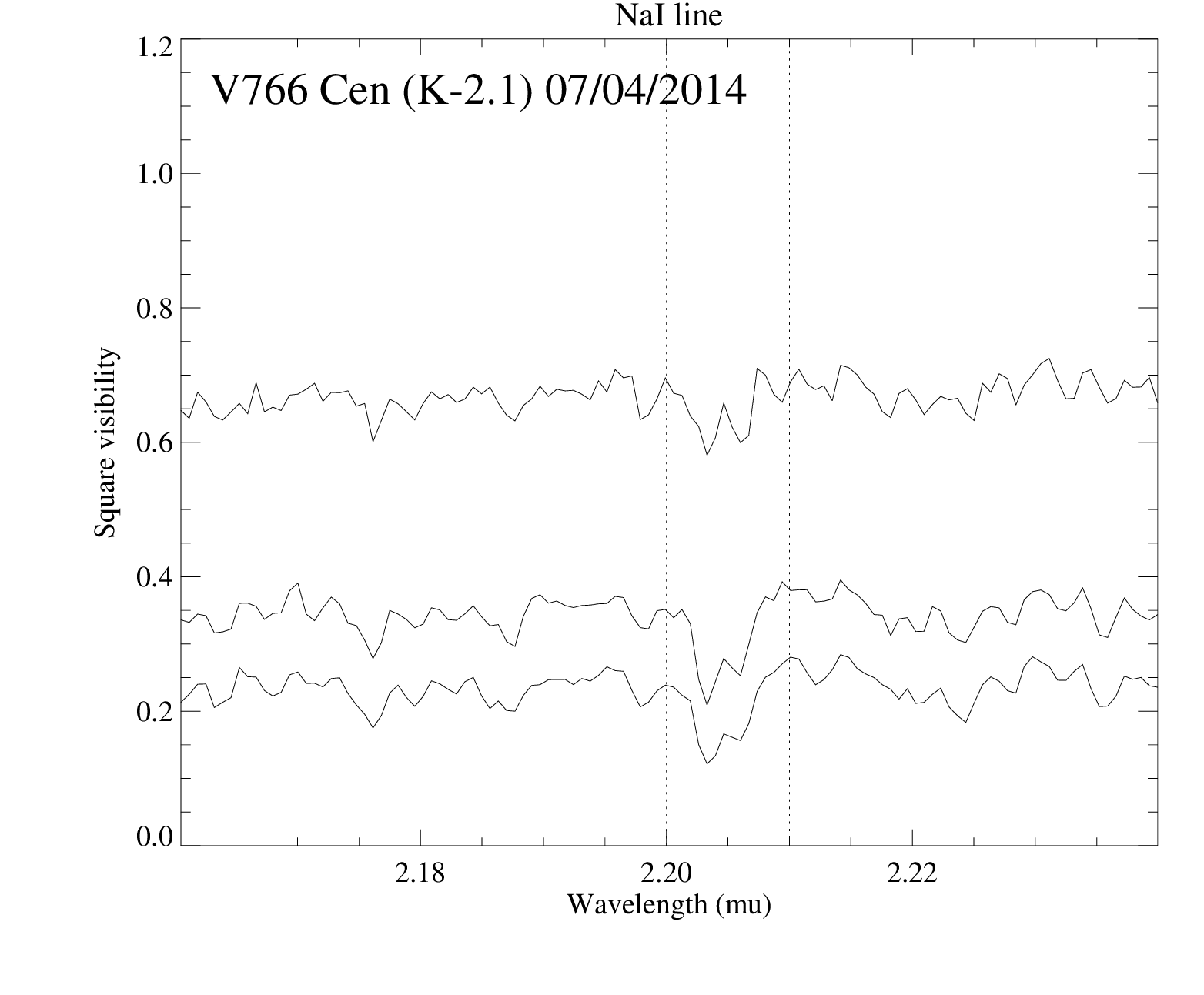}
\includegraphics[width=0.475\hsize]{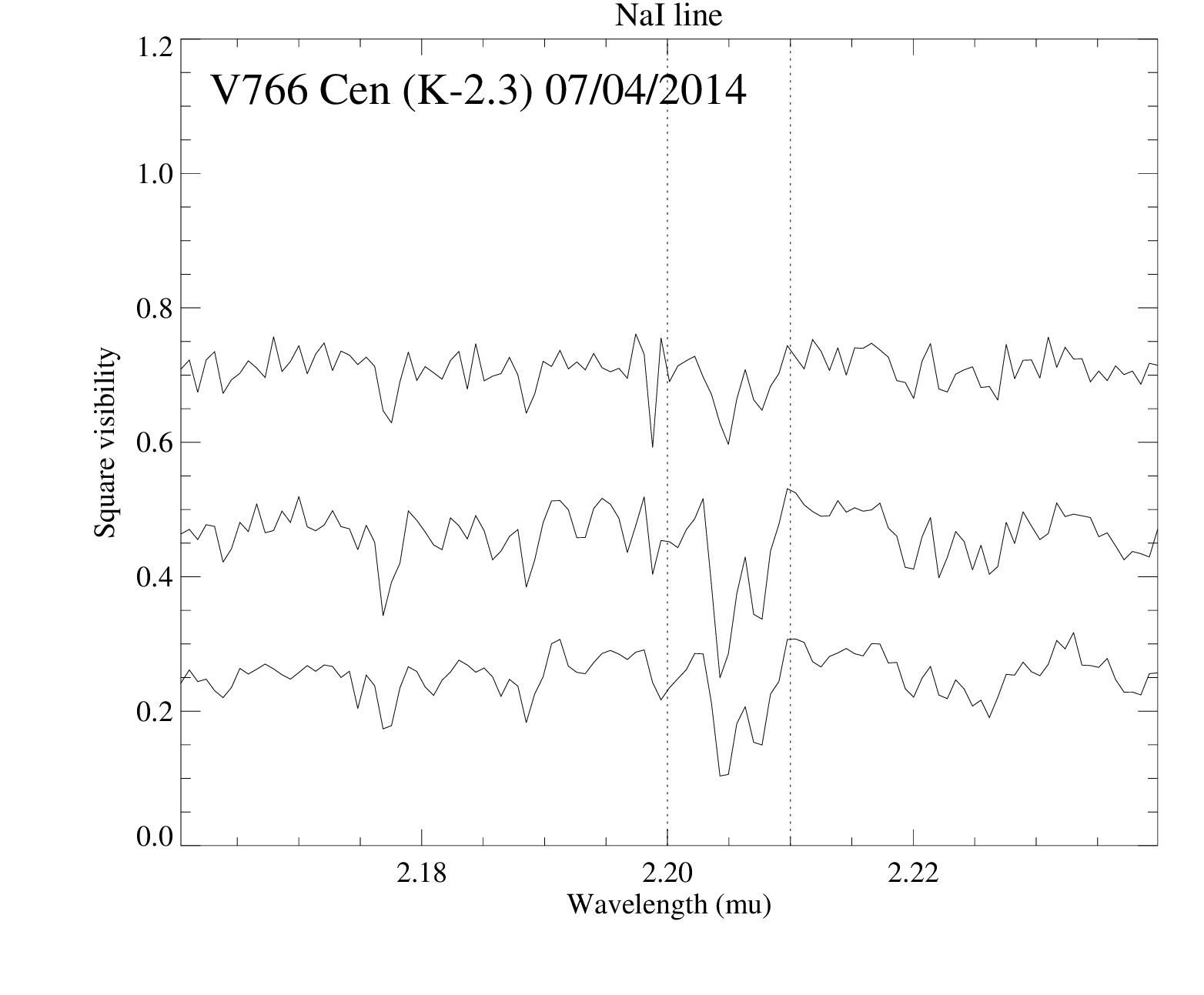}
\includegraphics[width=0.475\hsize]{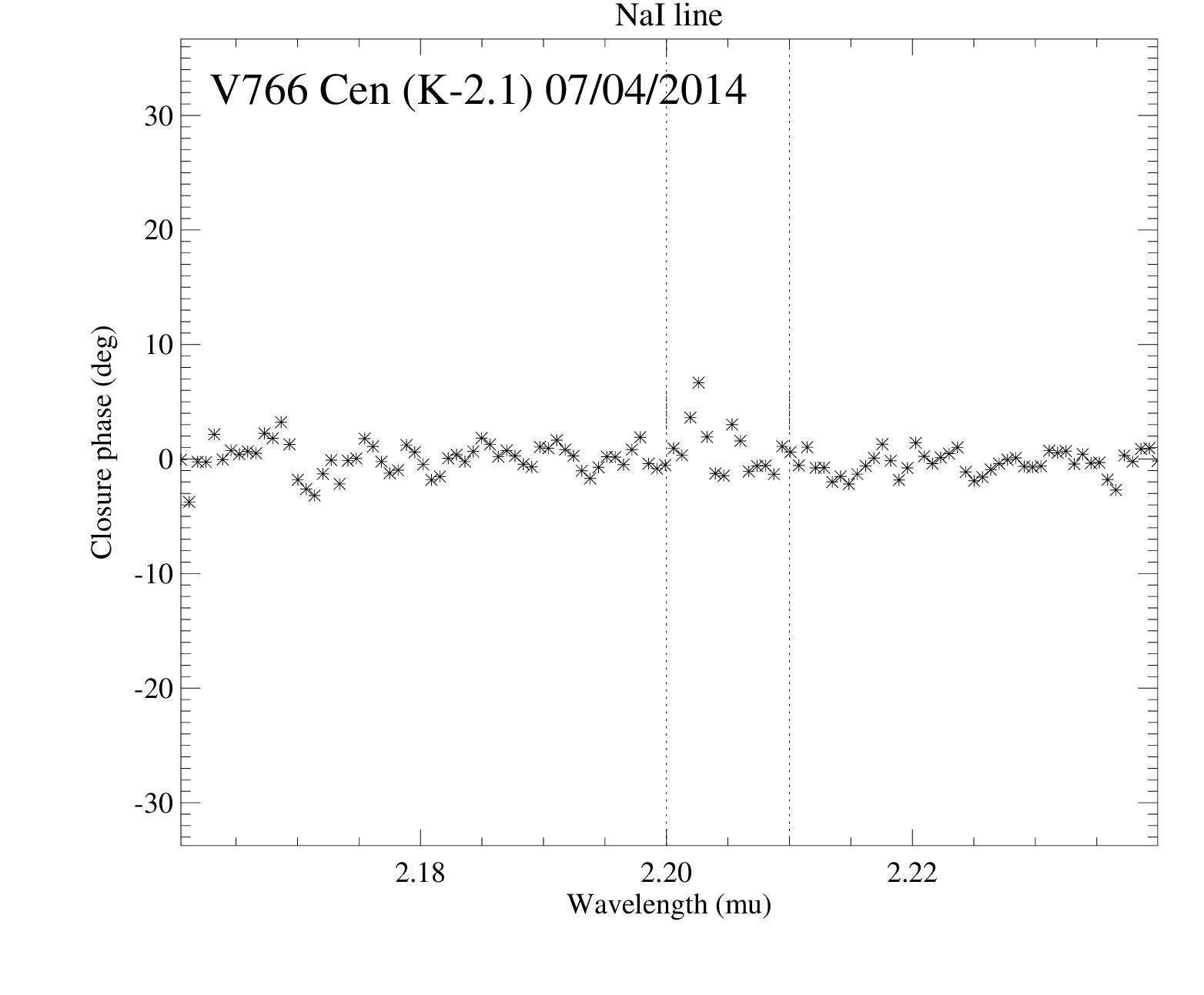}
\includegraphics[width=0.475\hsize]{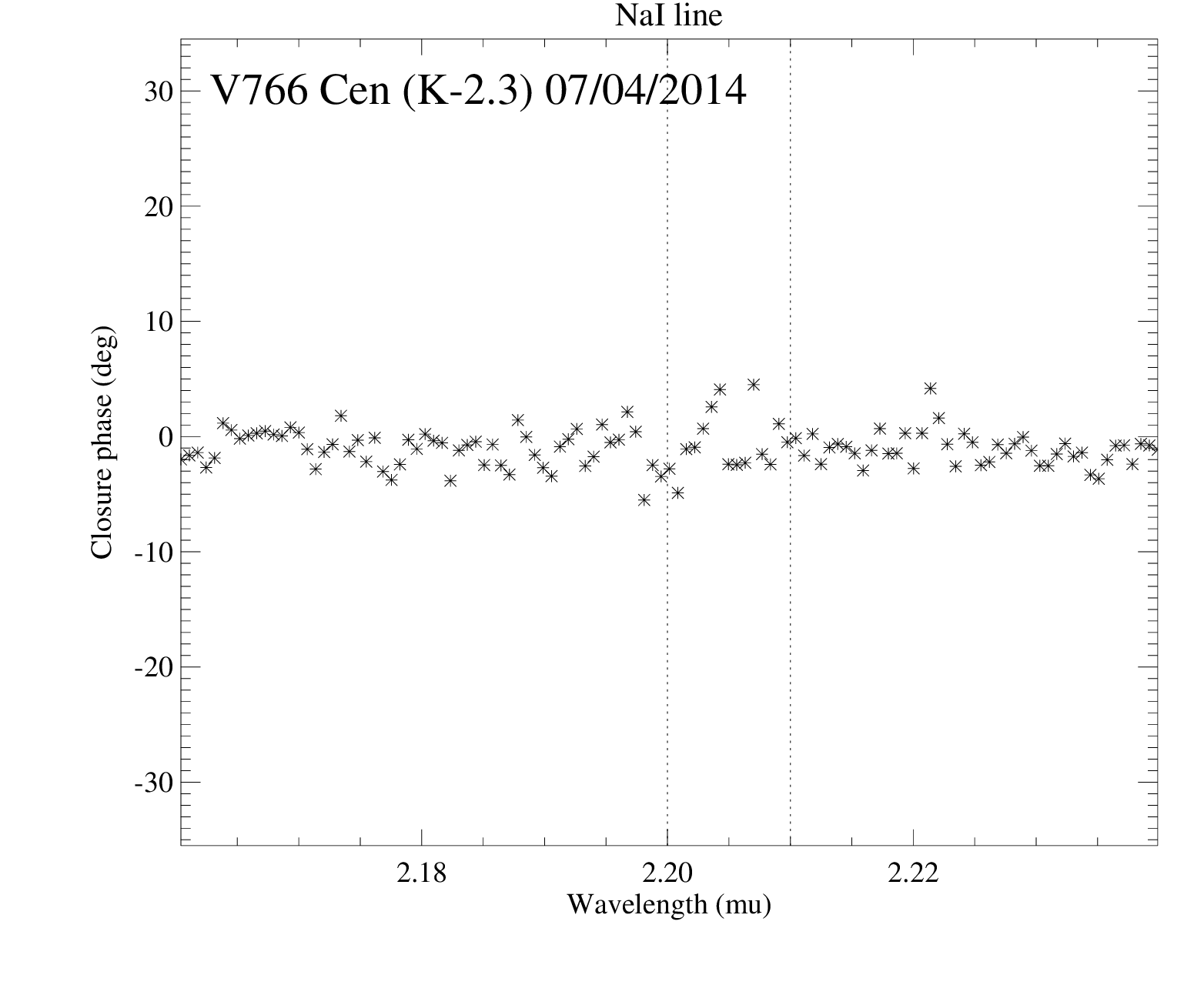}
\caption{Left: Observed normalized flux, squared visibility amplitudes,
and closure phases (from top to bottom) around the \protect\ion{Na}{i} line of
V766~Cen obtained with the MR-K 2.1\,$\mu$m setting. Right: Same as left
but obtained with the MR-K 2.3\ $\mu$m.}
\label{fig:zoom_NaI}
\end{figure*}

Our flux spectra of V766~Cen exhibit an emission feature at a
wavelength of 2.205\,$\mu$m, which \citet{Chesneau2014}
also noticed and identified as the \ion{Na}{i} doublet (see Sect.~\ref{sec:results} above). 
Fig.~\ref{fig:zoom_NaI} shows an enlargement
of the \ion{Na}{i} line extracted from Fig.~\ref{fig:resul_V766Cen}.
The emission line in the flux spectra corresponds to significant drops
in the visibility, indicating that \ion{Na}{i} is formed at higher
atmospheric layers compared to the continuum. The closure phase
suggests a small photocenter displacement of the \ion{Na}{i} emission with
respect to the nearby continuum emission.

A possible existence
of a high-temperature low-density shell around V766~Cen has been discussed
by \citet{Warren1973}. \citet{Gorlova2006} have discussed
shells around cool very luminous supergiants due to a quasi-chromosphere
or a steady shock wave at the interface of a fast expanding wind.
\citet{Oudmaijer2013} have discussed AMBER observations of the \ion{Na}{i} doublet
around the post-red supergiant IRC+10420. They have argued that the presence
of neutral sodium within the ionized region is unusual and requires
shielding from direct starlight. They have also discussed as an explanation
a dense chromosphere or an optically thick pseudo-photosphere related
to strong mass loss.

Assuming that \ion{Na}{i} is formed in a thin shell, we modeled the visibility
data in the \ion{Na}{i} line using a simple model of a UD, a thin ring, 
and an over-resolved, possibly dusty, component. Here, the UD represents 
the continuum-forming layers of the stellar photosphere, the ring represents 
the \ion{Na}{i} emission region, and the over-resolved component is the same
component as
indicated in our continuum fits in Sect.~\ref{sec:modeling}.
The continuum UD diameter
and the flux ratio between the UD and the over-resolved component
are known from our fits to the continuum visibilities; 
cf. Tab.~\ref{tab:ang_diameter}. Moreover, the flux contribution of the 
\ion{Na}{i} ring relative to the nearby continuum flux is given by the strength 
of the \ion{Na}{i} line in the flux spectra of Fig.~\ref{fig:zoom_NaI}.
We estimate a ratio of 12\% using a bandwidth of 2.202--2.207\,$\mu$m.
Altogether, this gives flux contributions at 2.2\,$\mu$m
of about 83\% from the continuum stellar component, 
12\% from the \ion{Na}{i} ring component, and 5\% from the 
over-resolved dusty component.
\begin{figure}
\centering
\includegraphics[width=0.95\hsize]{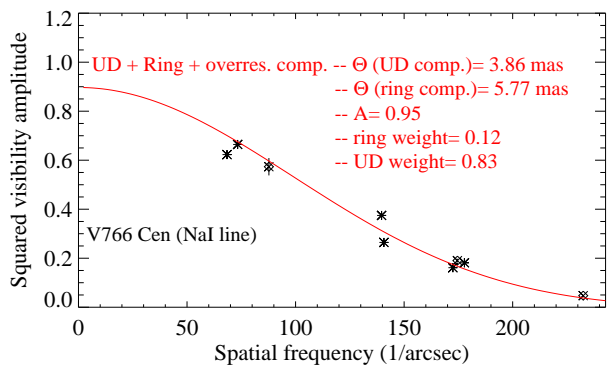}
\caption{Average of squared visibility amplitudes taken across the
\ion{Na}{i} line at 2.202--2.207\,$\mu$m 
for V766~Cen as a function of spatial frequency. The red lines indicate
the best-fit UD+ring+over-resolved component model (see text).}
\label{fig:V766Cen_UD_ring_model}
\end{figure}

Fig.~\ref{fig:V766Cen_UD_ring_model} shows the extracted average
visibilities in the \ion{Na}{i} line using the same bandwidth of 2.202--2.207\,$\mu$m.
Using the \ion{Na}{i} ring diameter as the only 
remaining free parameter, we obtained a ring diameter of 5.77\,mas, 
corresponding to about 1.5 times the photospheric continuum diameter. 
This model fit is shown together with the 
observed visibilities.

The closure phase signal of $\sim$\,5\,$\deg$ at the position of the 
\ion{Na}{i} doublet relative to the nearby continuum corresponds to a photocenter
displacement of the \ion{Na}{i} ring with respect to the stellar continuum
emission of $\sim$0.16\,mas. Here we 
use the formula for marginally
resolved targets $p=-\Phi/2\pi \times \lambda/B$
\citep{Lachaume2003,LeBouquin2009b}, 
where $p$ is the photocenter displacement and $\Phi$ is the phase signal.

\begin{figure}
\centering
\includegraphics[width=0.95\hsize]{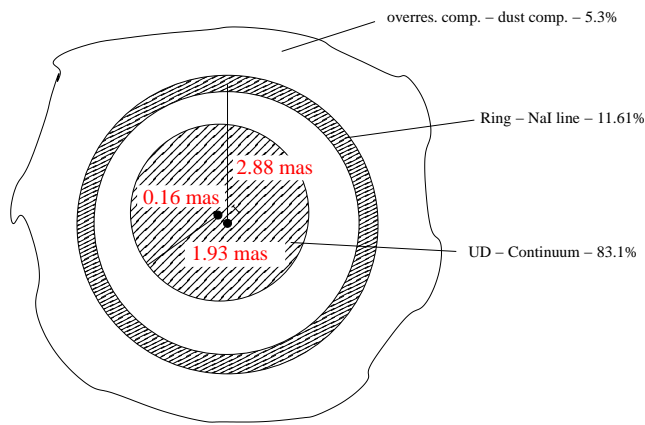}
\caption{
Sketch of V766~Cen's UD+ring+over-resolved component model for the emission
at the wavelength of the \ion{Na}{i} doublet.}
\label{fig:V766Cen_UD_ring_model_sketch}
\end{figure}

For illustration, Fig.~\ref{fig:V766Cen_UD_ring_model_sketch}
shows a sketch of the contributing components of our toy model at the 
wavelength of the \ion{Na}{i} doublet, including the continuum emission of the
central star (UD component), the \ion{Na}{i} emission from an offset ring-like
component, and continuum emission from an over-resolved dust component.
In addition, V766~Cen also shows extended CO emission, but that is
not visible at this wavelength.
The offset position of the \ion{Na}{i} ring in the sketch represents the photocenter
displacement between the \ion{Na}{i} emission and the nearby continuum emission.
It might be caused by effects of the nearby companion.
It may also be possible to explain this displacement via convection patterns
on the stellar surface that cause the photocenter of the stellar continuum to be 
different from the geometric center while the \ion{Na}{i} emission is symmetric with
respect to the outer radius of the star.
\section{Fundamental stellar parameters and statistical properties}
\begin{table*}
\caption{Fundamental properties of V766~Cen, $\sigma$~Oph, BM~Sco, 
and HD~206859.}
\begin{center}
\begin{tabular}{lccccccl}
\hline
\hline
Parameter  & V766~Cen & $\sigma$~Oph & BM~Sco & HD~206859 & Ref.   \\
\hline
$f_\mathrm{bol}$ ($10^{-9}$ Wm$^{-2}$)  &  1.68$\pm$0.45\tablefootmark{a} & 1.13$\pm$0.17 & 0.48$\pm$0.07 & 0.89$\pm$0.13 & 1  \\ 
$d$ (pc) &  3597$\pm$540 & 274$\pm$41 & 478$\pm$72 & 283$\pm$42  & 2 \\
$L$ ($10^{31}$ W) &  26.0$\pm$10.4 & 0.10$\pm$0.034 & 0.13$\pm$0.044 & 0.085$\pm$0.029  & 1,2 \\
$\log(L/L_{\odot}$) & 5.83$\pm$0.40 & 3.42$\pm$0.34 & 3.53$\pm$0.33 & 3.35$\pm$0.33 & - \\
$\theta _\mathrm{Ross}$ (mas) & 3.86$\pm$1.27\tablefootmark{b} & 3.41$\pm$0.90 & 2.50$\pm$0.35 & 1.86$\pm$0.50 & this work \\
$R$ ($R_\odot$) & 1492$\pm$540 & 100$\pm$30 & 129$\pm$26 & 57$\pm$17 & 2, this work \\
$T_\mathrm{eff}$ (K) & 4287$\pm$760 & 4129$\pm$566 & 3888$\pm$309 & 5274$\pm$736 & 1, this work \\
$\log(T_\mathrm{eff}$) & 3.63$\pm$0.18 & 3.62$\pm$0.14 & 3.59$\pm$0.07 & 3.72$\pm$0.14 & -  \\
$\log(g)$ & -0.5$\pm$0.6 & 1.3$\pm$0.5 & 1.1$\pm$0.4 & 1.8$\pm$0.5 & 4\\
Initial mass $M_i$ (M$_{\odot}$) & 25-40 & 5-9 & 5-9 & 5-9 & 3 \\
Current mass $M_c$ (M$_{\odot}$) & 13-36 & 5-9 & 5-9 & 5-9 & 3 \\
\hline
\end{tabular}
\tablefoot{1: \citet{Kharchenko2001}, \citet{Cutri2003}, 
\citet{Beichman1988} - V766~Cen, BM~Sco, HD~206859;
\citet{Moshir1990} - $\sigma$~Oph; 
2: \citet{Humphreys1978} - V766~Cen;
\citet{Mermilliod2003} - BM~Sco;
\citet{Anderson2012} - $\sigma$~Oph, HD~206859; 
3: Values obtained by the position of the stars in the HR diagram with the 
evolutionary tracks from \citet{Ekstrom2012}.
4: Calculated based on the current mass (see below) and radius.
We assumed a 15\% error each for the flux and distance. The errors 
in the luminosity, effective temperature, radius, and surface gravity
were estimated by error propagation.\\
\tablefoottext{a}{The bolometric flux for V766~Cen is that of the
primary component alone; it excludes the 
contribution by the close companion (see Sect.~\ref{sec:bolflux}).}\\
\tablefoottext{b}{The error includes the uncertainty due to the 
subtraction of the flux of the close companion.}}
\end{center}
\label{tab:fundpar}
\end{table*}
\begin{table*}
\caption{Properties of our sample of cool supergiants and giants}
\begin{center}
\begin{tabular}{llrrrrrll}
\hline\hline
Source & Luminosity Class & Lum.              & $T_\mathrm{eff}$ &  $R_\mathrm{Ross}$& $M_i$     &  $P$  & $A$    & Atm. Ext.\\
       & (Simbad)         & $\log(L/L_\odot)$ & K                &  R$_\odot$        & M$_\odot$ &  days &        & \\\hline
V766~Cen & G8 Ia          & 5.8 $\pm$ 0.4     & 4290 $\pm$ 760   &  1492 $\pm$ 540   & 25--40   & 1304  & 0.94   & yes \\  
AH~Sco   & M5 Ia-Iab      & 5.5 $\pm$ 0.3     & 3680 $\pm$ 190   &  1411 $\pm$ 124   & 25--40   &  725  & 0.81   & yes \\
UY~Sct   & M4 Ia          & 5.5 $\pm$ 0.3     & 3370 $\pm$ 180   &  1708 $\pm$ 192   & 25--40   &  705  & 0.94   & yes \\
VY~CMa   & M3-M4 II       & 5.4 $\pm$ 0.1     & 3490 $\pm$ 90    & 1420 $\pm$ 120    & 25--40   & 1600  & 0.50   & yes \\
KW~Sgr   & M4 Ia          & 5.2 $\pm$ 0.3     & 3720 $\pm$ 180   & 1009 $\pm$ 142    & 15--32   &  647  & 1      & yes \\
V602~Car & M3 Ia-Iab      & 5.1 $\pm$ 0.2     & 3430 $\pm$ 280   & 1050 $\pm$ 165    & 20--25   &  646  & 1      & yes \\
HD~95687 & M2 Iab         & 4.8 $\pm$ 0.2     & 3470 $\pm$ 300   & 674 $\pm$ 109     & 12--15   &       & 1      & weak\\[1ex]
HD~183589& K5 III         & 3.8 $\pm$ 0.3     & 3710 $\pm$ 340   & 197 $\pm$ 39      & 7--12    &       & 1      & no\\
NU~Pav   & M6 III         & 3.8 $\pm$ 0.2     & 3520 $\pm$ 280   & 204 $\pm$ 29      & 7--12    &  60   & 1      & no\\
BM~Sco   & K2 Ib          & 3.5 $\pm$ 0.3     & 3890 $\pm$ 310   & 129 $\pm$ 26      & 5--9     &       & 0.73   & weak\\
$\beta$~Peg& M2.5 II-III & 3.4 $\pm$ 0.2     & 3910 $\pm$ 190   & 109 $\pm$ 7       & 7--9     &  43   & 1      & no\\
$\sigma$~Oph& K2 III     & 3.4 $\pm$ 0.3     & 4130 $\pm$ 570   & 100 $\pm$ 30      & 5--9     &       & 1      & no\\
HD~206859& G5 Ib         & 3.4 $\pm$ 0.3     & 5270 $\pm$ 740   &  57 $\pm$ 17      & 5--9     &       & 1      & no \\[1ex]
$\epsilon$~Oct&M5 III    & 3.3 $\pm$ 0.2     & 3560 $\pm$ 260   & 112 $\pm$ 15      & 5--7     &  55   & 1      & no\\
$\psi$~Peg  & M3 III     & 3.2 $\pm$ 0.2     & 3710 $\pm$ 180   &  98 $\pm$ 6       & 5--7     &       & 1      & no\\
$\gamma$~Hya & G8 III     & 2.1 $\pm$ 0.2     & 4730 $\pm$ 440   &  16 $\pm$ 3       & 3        &       & 1      & no\\
\hline
\end{tabular}
\end{center}
\label{tab:comparison}
\end{table*}
\begin{figure}
\centering
\includegraphics[width=0.95\hsize]{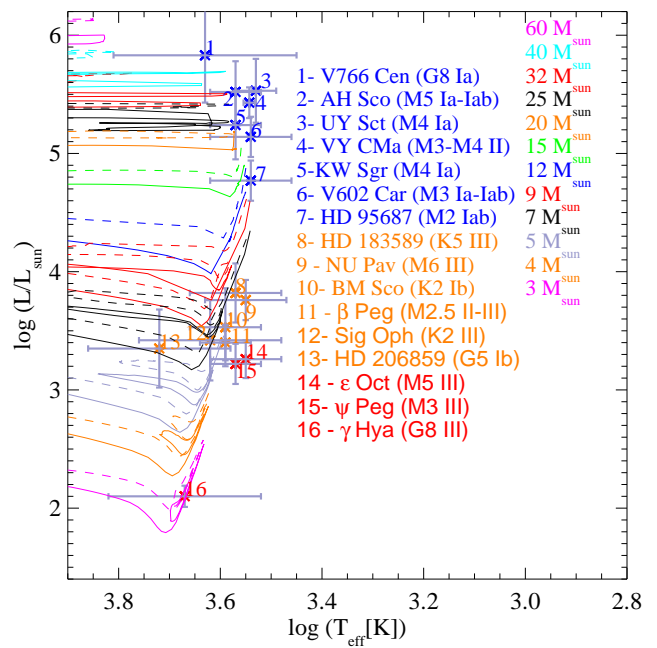}
\caption{Location of our updated sample of sources in the HR diagram 
(cf.~\protect\ref{tab:comparison}). Also shown 
are evolutionary tracks from \citet{Ekstrom2012} for masses 
of 3--60\,$M_{\odot}$.
The solid lines 
indicate models without rotation, and the dashed lines with rotation. 
Blue indicates the luminosity class I supergiants; orange indicates the transition region between 
giants and supergiants; and
red indicates the luminosity class III red giants.}
\label{fig:diagram_HR_all}
\end{figure} 
Table~\ref{tab:fundpar} lists the fundamental parameters of our new
sources based on our fits of the Rosseland angular diameters, the adopted
bolometric fluxes and distances, and the evolutionary tracks 
by \citet{Ekstrom2012}.
Fig.~\ref{fig:diagram_HR_all} shows the position of 
our new sources in the Hertzsprung-Russell (HR diagram) together with our
previously observed sources from \citet{Wittkowski2012}, \citet{Arroyo2013},
and \citet{Arroyo2014}, 
all of which were analyzed in the same way. 
This figure also shows the evolutionary tracks by \citet{Ekstrom2012} for initial masses
between 3\,M$_\odot$ and 60\,M$_\odot$ with and without rotation.
Our sample
includes supergiants and giants, where the giants are non-Mira semi-regular
or irregular variables.

\subsection{Fundamental parameters of our new sources}
The individual new sources are discussed in more detail in 
Sect.~\ref{sec:discussion} below. In short,
V766~Cen is found to be a high-luminosity star
of $\log L/L_\odot\sim$5.8$\pm$0.4, a radius of 
1490\,R$_\odot$\,$\pm$\,540\,R$_\odot$, and an
effective temperature of $\sim$4300\,K\,$\pm$\,760\,K. 
Its position in the HR diagram is close to the red edge of evolutionary
tracks of initial mass 25--40\,M$_\odot$. BM~Sco, $\sigma$~Oph, 
and HD~206859 show luminosities $\log L/L_\odot\sim 3.35-3.55$
corresponding to evolutionary tracks of initial masses
5--9\,M$_\odot$. The stars $\sigma$~Oph and
BM~Sco are close to the red edge of these tracks with effective
temperatures of 4130\,$\pm$\,566\,K and 3890\,$\pm$310\,K, respectively,
while HD~206859 is a warmer star of effective temperature
5250\,$\pm$\,740\,K.

\subsection{Statistical properties of our sample}
Table~\ref{tab:comparison} lists the fundamental
parameters of our full sample of 16 sources, i.e., those shown in
the HR diagram in Fig.~\ref{fig:diagram_HR_all}. It includes
the luminosity class from Simbad, the luminosity, effective temperature, 
and radius from our calculations, the initial mass from our position 
in the HR diagram and the evolutionary tracks by \citet{Ekstrom2012},
and the variability period from the GCVS 
\citep{Samus2009} and the AAVSO database. The period of VY CMa is 
from \citet{Kiss2006}. The table also includes the flux fraction $A$
of the stellar component, i.e., a value of $A<1$ indicates an additional
over-resolved, possibly dusty, background component of flux fraction.
Finally, we list whether or not we found an extended atmosphere
beyond that predicted by 1D and 3D model atmospheres; cf.
\citet{Arroyo2015}.

\begin{table}
\caption{Comparison to literature values}
\label{tab:levesquebelle}
\begin{center}
\begin{tabular}{lrr}
\hline\hline
                       & \citet{Levesque2005}\tablefootmark{a} & Our work       \\\hline

\multicolumn{3}{l}{\underline{HD\,95687}}      \\
$T_\mathrm{eff}$ (K)   &  3625                & 3470 $\pm$ 300 \\
$L$ ($\log L/\mathrm{L}_\odot$) &  4.95                & 4.8 $\pm$ 0.2  \\
$R$ (R$_\odot$)        &  760                 & 674 $\pm$ 109  \\[1ex]

\multicolumn{3}{l}{\underline{HD\,97671 (=V602 Car)}}      \\
$T_\mathrm{eff}$ (K)   &  3550                & 3430 $\pm$ 280 \\
$L$ ($\log L/\mathrm{L}_\odot$) &  5.02                & 5.1 $\pm$ 0.2  \\
$R$ (R$_\odot$)        &  860                 & 1050 $\pm$ 165  \\[1ex]

\multicolumn{3}{l}{\underline{HD\,160371 (=BM~Sco)}}      \\
$T_\mathrm{eff}$ (K)   &  3900                & 3890 $\pm$ 310 \\
$L$ ($\log {L/\mathrm{L}}_\odot$) &  3.35                & 3.5 $\pm$ 0.3  \\
$R$ (R$_\odot$)        &  100                 & 129  $\pm$  26  \\[1ex]

\multicolumn{3}{l}{\underline{KW Sgr}}      \\
$T_\mathrm{eff}$ (K)   &  3700                & 3720 $\pm$ 180 \\
$L$ ($\log L/\mathrm{L}_\odot$) &  5.56                & 5.2 $\pm$ 0.3  \\
$R$ (R$_\odot$)        &  1460                & 1009 $\pm$ 142  \\[1ex]
\hline
                       & \citet{Belle2009}\tablefootmark{b} & Our work       \\\hline

\multicolumn{3}{l}{\underline{HD 183589}}      \\
$T_\mathrm{eff}$ (K)   &  3955 $\pm$ 34       & 3710 $\pm$ 340 \\
$L$ ($\log L/\mathrm{L}_\odot$) &  3.74 $\pm$ 0.4      & 3.8 $\pm$ 0.3  \\
$R$ (R$_\odot$)        &  154 $\pm$ 46        & 197 $\pm$ 39   \\[1ex]

\multicolumn{3}{l}{\underline{HD~206859}}      \\
$T_\mathrm{eff}$ (K)   &  5072 $\pm$ 45       & 5270 $\pm$ 740 \\
$L$ ($\log L/\mathrm{L}_\odot$) &  3.31 $\pm$ 0.3      & 3.4 $\pm$ 0.3  \\
$R$ (R$_\odot$)        &  57 $\pm$ 12        & 57 $\pm$ 17   \\\hline
\end{tabular}
\tablefoottext{a}{\citet{Levesque2005} do not provide errors.}
\tablefoottext{b}{\citet{Belle2009} do not explicitly list the luminosity.
The values are calculated from their bolometric fluxes and their distances.}
\end{center}
\end{table}

In general, our results confirm the conclusion by \citet{Levesque2005}
that the positions of RSGs in the HR diagram are consistent 
with the red edges of the evolutionary tracks near the Hayashi line.
In particular, we have four sources in common 
with \citet{Levesque2005}: HD~95687, HD~97671 (=V602~Car),
HD~160371 (=BM~Sco), and KW~Sgr. We also have two sources in common
with \citet{Belle2009}: HD~183589 and HD~206859. 
Table~\ref{tab:levesquebelle} provides the comparisons
of the effective temperatures, luminosities, and absolute radii
with our values. They are all consistent within 1\,$\sigma$, except
for KW~Sgr, for which \citet{Levesque2005} found a higher luminosity
and a larger radius at the 2--3\,$\sigma$ level. 
Our effective temperatures and luminosities of $\sigma$~Oph,
BM~Sco, and HD~206859 are consistent with the estimates
by \citet{McDonald2012} as well, except for the luminosity of BM~Sco, where
\citet{McDonald2012} found a lower value of $\log L/\mathrm{L}_\odot=3.0$
compared to our value of $\log L/\mathrm{L}_\odot=3.5\pm 0.3$.

Tab.~\ref{tab:comparison} is ordered by decreasing
luminosity. Our values confirm decreasing radius, initial mass,
and pulsation period with decreasing luminosity,
giving additional confidence in the consistency
of our values. The outliers are VY~CMa, which has an unusually long period,
and UY~Sct, which has  an unusually large radius. The reasons for these
outliers are not yet clear.

Our sources can be separated into three groups: 
(I) Sources of luminosity $\log L/L_\odot$ between about 5.8 and 4.8,
which are consistently classified as luminosity class I sources.
They correspond to initial masses between about 60\,M$_\odot$ and 
12\,M$_\odot$. These are clearly supergiants. (II) Sources of 
$\log L/L_\odot$ between about 3.8 and 3.4, of which some are classified 
as luminosity class Ib and some as (II-)III. Some
of these sources have conflicting classifications in the literature.
They correspond to initial masses between 12\,M$_\odot$ and 5\,M$_\odot$.
These sources may represent a transitional region between 
giants and supergiants.
(III) Sources of luminosity $\log L/L_\odot$ below about 3.4, which
are consistently classified as luminosity class III sources, corresponding
to initial masses below about 7\,M$_\odot$. These are clearly giant stars.
The groups are indicated by different colors in 
Fig.~\ref{fig:diagram_HR_all} (HR diagram) and 
Fig.~\ref{fig:Teff_sp_R} (effective temperature scale, discussed
below).

An extended atmosphere, beyond the predictions by 1D and 3D model
atmospheres, is seen only for the clear RSG sources (group I above)
including the warmer source V766~Cen.
Similarly, a significant contribution of a very extended over-resolved
(dust) component is only seen for the most luminous RSG sources.
BM~Sco as a transitional source represents an exception and possibly shows
 a weak extension of the atmosphere and a strong contribution
by an over-resolved component. The latter is consistent with
\citet{McDonald2012}, who list BM~Sco as one of a few
giants with an infrared flux excess indicating circumstellar material.
This is unusual for a non-Mira giant of this luminosity.
The cause of these effects is not clear; they might
might possibly be due to an unseen binary companion.

Our sampling of luminosities presents an unsampled locus 
in the HR diagram around $\log L/L_\odot$ 3.8--4.8, corresponding to 
initial masses around 10--13\,M$_\odot$. Other well-known RSGs,
including Betelgeuse, VX~Sgr \citep[cf.][]{Arroyo2013}, Antares \citep{Ohnaka2013}, and
$\mu$\,Cep \citep{Perrin2005}, which are not a part of our sample,
have luminosities above $\log L/L_\odot\sim$4.8 as well.
This gap appeared very naturally
in our sampling,
which was based on sources from the GCVS \citep{Samus2009}, which were
classified as luminosity classes I--III and cool spectral types of 
K--M (with an emphasis in Lum. class I) and well observable with the 
VLTI/AMBER instrument in terms of brightness 
and expected angular size. It is not yet clear whether this gap is caused
by a selection bias, or whether it may indicate a shorter lifetime of
sources of this mass near the red edge of the evolutionary tracks.
Possibly, stars below this
mass range are (super)-AGB stars and return to warmer effective temperatures
toward becoming planetary nebulae and stars above this mass range return
to warmer effective temperatures toward becoming yellow supergiants
and WR stars, while stars of this mass range explode as core-collapse
SNe at the RSG stage. This may explain a shorter lifetime
of these sources near the red edge of the evolutionary tracks. Such a scheme
is roughly consistent with the evolutionary tracks of \citet{Ekstrom2012},
where RSGs with initial masses larger than about 15\,$_\odot$
start to return to warmer effective temperatures.
Interestingly this range of masses corresponds to about that of some
recently confirmed RSG progenitors of SNe of type II-P
\citep{Maund2014a,Maund2014b,Maund2015}, while for instance the higher mass
(40\,M$_\odot$) progenitor of SN type Ic was found to be a yellow
supergiant, i.e., it had not already exploded at the RSG stage
\citep{Maund2016}. Likewise, \citet{Smartt2015}
recently pointed out the deficit of high-mass stars 
(larger than 18\,$M_\odot$) among the progenitors of mostly 
II-P, II-L, or IIb core collapse SNe.

\subsection{Effective temperature scale}
\begin{figure}
\centering
\includegraphics[width=0.95\hsize]{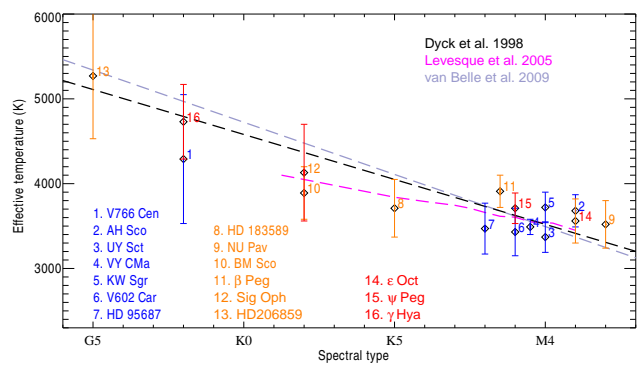}
\caption{Effective temperature vs. spectral type of our sources together 
with calibrations of the effective temperature scale by \citet{Dyck1998}, 
\citet{Levesque2005}, and \citet{Belle2009}.  
Blue indicates the luminosity class I supergiants;
orange indicates the  transition region between giants and supergiants;
and red represents the luminosity class III red giants.}
\label{fig:Teff_sp_R}
\end{figure} 
Fig.~\ref{fig:Teff_sp_R} shows the effective temperature versus 
spectral type for our sample, together with empirical
calibrations by \citet{Dyck1998} 
for cool giants stars, by \citet{Belle2009} for cool giants stars 
and RSG stars, and by \citet{Levesque2005} for RSGs. 
Our data points are consistent with all of these calibrations 
within the error bars. In particular we are not yet able to 
prove or disprove a different effective temperature scale
for supergiants and giants as suggested by \citet{Levesque2005}.

\subsection{Atmospheric extension}
\begin{figure}
\centering
\includegraphics[width=0.95\hsize]{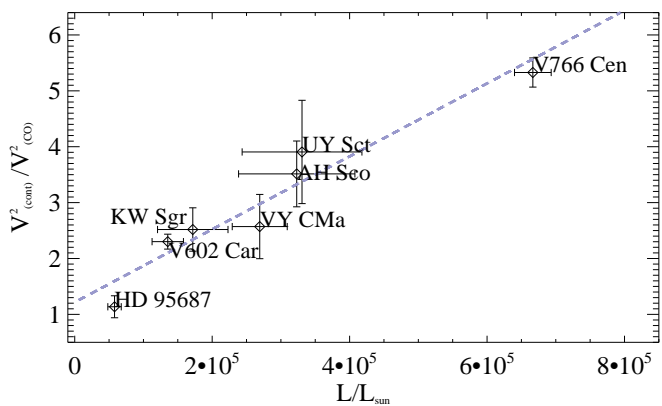}
\caption{Atmospheric extension vs. luminosity of our RSGs.
The 
contribution by extended atmospheric CO layers
is measured by
ratio between the squared visibility 
amplitude in the continuum (average between 2.27 and 2.28\,$\mu$m) and the 
squared visibility amplitude in the CO (2-0) line at 2.29\,$\mu$m.
The dashed gray line shows a linear fit to the data points. 
}
\label{fig:visratio_RSGs}
\end{figure}
\citet{Arroyo2015} found a linear relation for RSGs
(i.e., group I above) between the 
contribution by extended atmospheric CO layers,
as measured by the 
squared visibility in the CO(2-0) bandhead relative to the nearby 
continuum, and the luminosity. This may support a scenario of 
radiative acceleration on molecules as a missing driver for the atmospheric
extension of RSGs as suggested by \citet{Josselin2007}.
Fig.~\ref{fig:visratio_RSGs} shows the same relation as in
\citet{Arroyo2015}, but now also including our new supergiant source
V766~Cen. 
As in \citet{Arroyo2015}, we limited the calculation
to continuum squared visibilities between 0.2 and 0.4, which is a range where
the source is resolved and the visibility function is nearly linear.
Interestingly, this relation is confirmed up to the location
of V766~Cen, now extending to twice the previous luminosity range.
This supports once again a similar mass loss mechanism of all RSGs
over about an order of magnitude in luminosity.
We note that V766~Cen has a luminosity ($\log L/L_\odot\sim$5.8)
close to the Eddington luminosity ($\log L/L_\odot\sim$5.6--6.1 for
masses 13--36\,M$_\odot$), above which
stars become unstable owing to strong radiative winds. This may
further support a scenario of steadily increasing radiative winds,
and thus increasing magnitude of extended molecular layers
with increasing luminosity.

The visibility drops discussed above indicate 
that CO is formed above the continuum-forming layers.
The visibility ratios provide an estimate of the total 
contribution of the extended layers compared to the stellar 
continuum layers
and not necessarily the ratio of the radii of these layers.
We are not able to directly constrain the radii of the CO layers by
our few snapshot observations owing to the unknown complex
intensity profiles of CO \citep[cf.][Fig. 3]{Wittkowski2016}
and the lack of a satisfactory geometrical model to describe
them.
However, from comparisons of model atmospheres of AGB stars
that predict extended layers and
show similar visibility drops \citep[e.g.,][]{Wittkowski2016},
we can estimate that CO layers of RSG stars have comparable 
typical extensions of about 1.5--2 stellar radii. 
This puts the CO emission of V766~Cen at a similar radius 
as the \ion{Na}{i} emission.

\section{Discussion of individual sources}
\label{sec:discussion}
\subsection{V766~Cen}
\begin{figure}
\centering
\includegraphics[width=0.95\hsize]{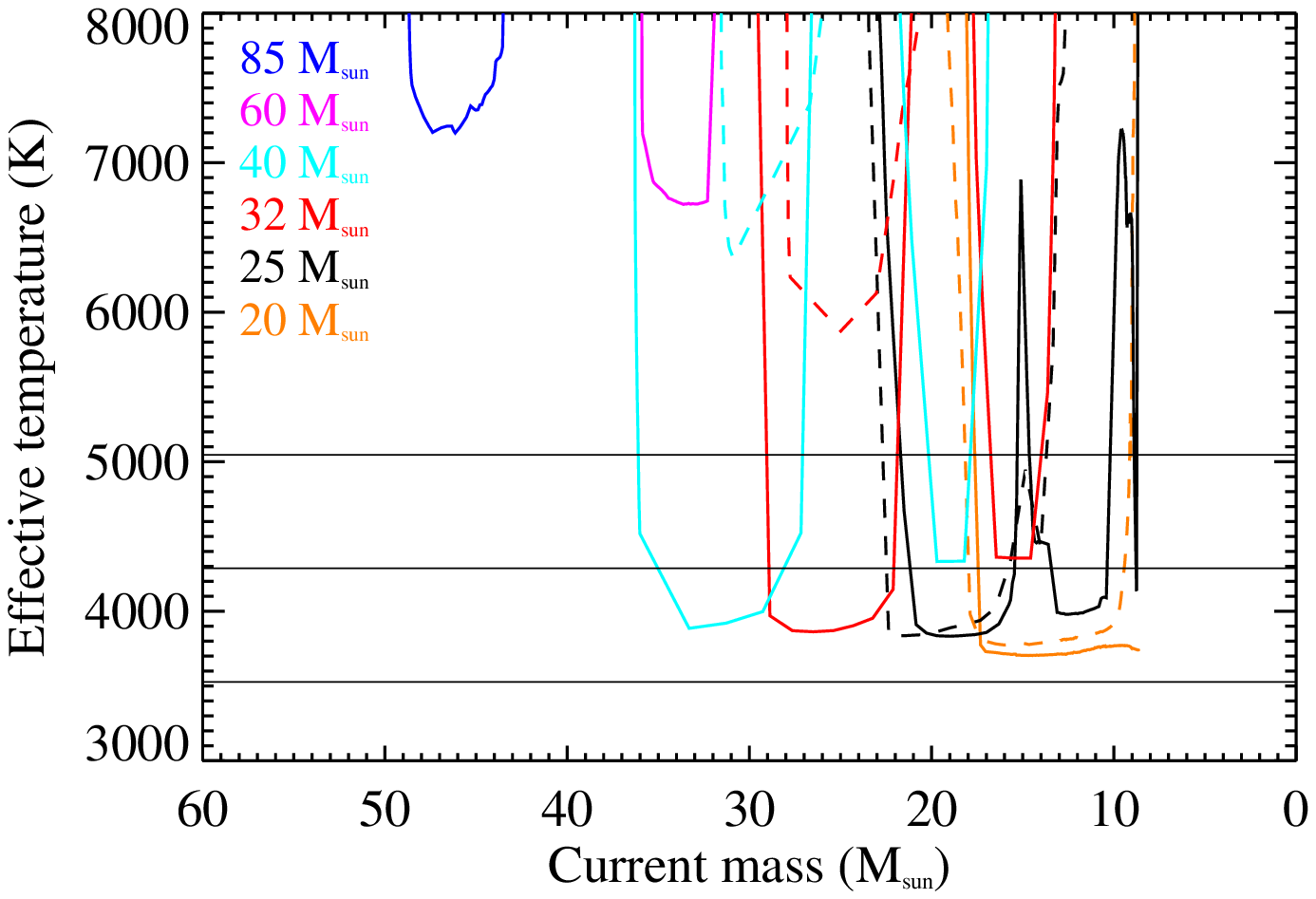}
\includegraphics[width=0.95\hsize]{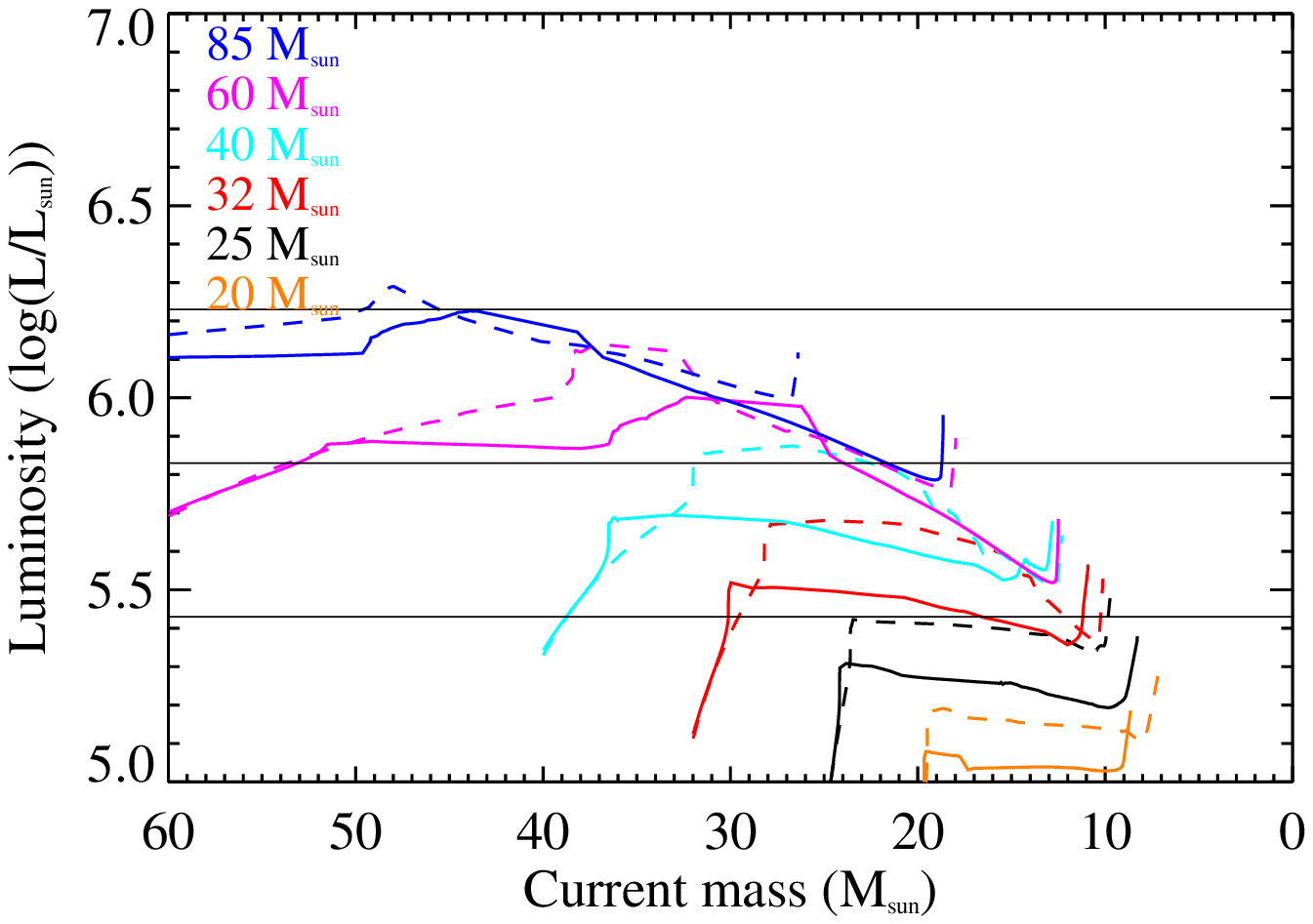}
\caption{Evolutionary status of V766~Cen. Effective temperature (top) and 
luminosity (bottom) as a function of mass during the 
evolution of 25\,M$_\odot$ to 80\,M$_\odot$ ZAMS evolutionary tracks 
from \citet{Ekstrom2012}.
The solid lines denote tracks without rotation and the dashed lines
tracks with rotation.
The horizontal lines denote our measurement of the effective temperature
and luminosity of V766~Cen together with the error ranges.
}
\label{fig:ekstrom_v766cen}
\end{figure}
V766~Cen (=HR~5171\,A) was classified and discussed as a yellow
hypergiant (YHG) by, for example, \citet{Humphreys1971}, \citet{vanGenderen1992},
\citet{deJager1998}, and \citet{vanGenderen2015}.
\citet{Chesneau2014} suggested that it has a very
close, possibly common-envelope companion, which we took into account in
our analysis.
This classification implies that the
star has a high luminosity and is located in the HR diagram between 
the blue and red parts of its evolutionary tracks, either 
evolving from the RSG phase back toward higher effective temperatures, 
or --less likely-- from the main sequence toward the RSG phase.

Our results confirm that V766~Cen is a high-luminosity star.
However, its effective temperature 
compared to the evolutionary tracks by
\citet{Ekstrom2012} indicate that it is located
close to the red edge of its evolutionary track and close to the Hayashi line,
thus resembling a luminous RSG rather than a typical YHG. It may have
a higher effective temperature compared to a more typical RSG, such as
our other RSGs, simply because the Hayashi line shifts 
slightly to the left in the HR diagram for increasing mass
\citep[$\partial \log T_\mathrm{eff}/\partial \log M \sim 0.2$; 
see][]{Kippenhahn1990}. For example, this would move the Hayashi line
from an effective temperature of 3500\,K to 4300\,K between stars of current mass
10\,$M_\odot$ and 30\,$M_\odot$.
Details depend on metallicity and rotation.
Figure~\ref{fig:ekstrom_v766cen} shows the effective temperature (top)
and luminosity (bottom) as a function of current mass
based on the tracks by \citet{Ekstrom2012} for solar metallicity
with and without rotation. Also indicated are our values
for V766~Cen with their errors. It illustrates that only those tracks
of initial mass 40\,M$_\odot$ without rotation, 32\,M$_\odot$ without
rotation, and (marginally) 25\,M$_\odot$ with rotation are consistent
with our observations, corresponding to current masses of 
about 27--36\,M$_\odot$, 22--29\,M$_\odot$, 13--23\,M$_\odot$, respectively.
The other tracks either do not decrease to our effective temperature
or do not increase to our luminosity.
Here, the track of 40\,M$_\odot$ without rotation is consistent
with our values.
A current mass of 27--36\,M$_\odot$ is also consistent with
the current system mass estimated by \citet{Chesneau2014} of
$39^{+40}_{-22}$\,M$_\odot$ based on their NIGHTFALL models.
The upper panel of Fig.~\ref{fig:ekstrom_v766cen} also illustrates nicely
that V766~Cen is located close to the coolest part of its evolutionary
track. The comparison of the position of  V766~Cen in the HR diagram
may be complicated by effects of the companion on its stellar evolution.
Nevertheless, the position in the HR diagram itself is independent 
of the comparison to evolutionary tracks.

\citet{Chesneau2014} already noted that the radius of V766~Cen
is unusually large for a typical YHG. Indeed, it is more consistent
with an evolutionary status of a RSG with 
a fully convective envelope that is close to the Hayashi limit, as discussed above.

Our data of V766~Cen indicate a shell of \ion{Na}{i} with a radius of 
1.5 photospheric radii. Our closure phases
indicate a small photocenter displacement of the \ion{Na}{i} shell with
respect to the stellar continuum of $\sim$0.16\,mas (8\% of the 
photospheric radius). This displacement might possibly be caused
by the presence of the close companion.
\citet{Gorlova2006} and \citet{Oudmaijer2013} have described such shells around
cool very luminous supergiants  as due to a 
quasi-chromosphere or a steady shock wave at the interface of a fast 
expanding wind shielding the gas from direct starlight.
The luminosity of $\log L/L\odot\sim$5.8$\pm$0.4 is close
to the Eddington luminosity of $\log L/L_\odot \sim$\,5.6--6.1 
for current masses of 13--36\,M$_\odot$, so that indeed strong radiative
winds can be expected.  The high luminosity compared
to the Eddington luminosity also points toward the upper current mass 
range of V766~Cen discussed above.

V766~Cen shows extended molecular CO layers similar to the other
RSGs of our sample. \citet{Arroyo2015} showed that the extension of these
layers cannot be explained by current pulsation or convection models
and that additional physical mechanisms are needed to explain them.
We confirmed that the relation of increasing contribution by extended 
CO layers with increasing luminosity \citep{Arroyo2015} extends to
the luminosity of V766~Cen, which is close to the Eddington limit.

Finally, our data show evidence of an over-resolved (i.e., uncorrelated)
background flux, which is consistent with a known silicate 
dust shell \citep{Humphreys1971} that vastly exceeds our interferometric 
field of view, which is about 270\,mas.

In summary, V766 is located close to both the Hayashi limit toward cooler
effective temperatures and the Eddington limit toward higher 
luminosities. This source shows properties of red supergiants, such as a relatively
large radius and an extended molecular layer. It also shows properties
of high-luminosity stars, such as a spatially extended shell of neutral
sodium at about 1.5 stellar radii. The inner dust shell radius is not
yet constrained. \citet{Verhoelst2006} suggested an inner dust shell of Al$_2$O$_3$ as close as 1.5 stellar radii for the RSG $\alpha$\,Ori, which is consistent
with most recent models \citep{Gobrecht2016,Hoefner2016} and observations 
\citep{Ireland2005,Wittkowski2007,Norris2012,Karovicova2013} of AGB stars.
If this is confirmed, it would put the \ion{Na}{i} emission, the extended
molecular (CO) layer, and inner dust formation zone at a similar radius
of about 1.5 stellar radii, forming an optically thick region 
(aka pseudo-photosphere) at the onset of the wind.

\subsection{BM~Sco}
BM~Sco is generally classified as a K1-2.5 Ib supergiant 
(Simbad adopted K2\,Ib). 
However, our
luminosity of $\log L/L_\odot =$\,3.5\,$\pm$0.3 places this source into a 
transitional zone between red giants and red supergiants corresponding
to evolutionary tracks of initial mass 5--9\,M$_\odot$.
\citet{McDonald2012} placed this souces at an even lower luminosity of 
$\log L/\mathrm{L}_\odot =3.0$.
This position in the HR diagram is more consistent with a 
classification as K3\,III by \citet{Houk1982}. 
The measured effective temperature of 3890$\pm$310\,K is consistent
with the spectral classification. BM~Sco is also among the source
of \citet{Levesque2005} and our fundamental properties are 
consistent with theirs.
We have found a significant uncorrelated flux fraction of $\sim$25\%
from an over-resolved component, most likely pointing to a 
large background dust shell. This is consistent with \citet{McDonald2012}
who list BM\ Sco as one of relatively few giants with infrared excess and
evidence for circumstellar material.
BM~Sco also shows extended CO features in 
the visibility that may be slightly beyond model predictions.
These features are not typical for a star of this mass range,
but would be more typical for a higher luminosity and thus higher mass
star. This makes BM~Sco an interesting target for follow-up observations.
These features might
be caused by an interaction with a close unseen companion.
We can also not exclude that adopted values for the distance or 
for the calculation of the bolometric flux may be erroneous,
and that BM~Sco in fact is a higher luminosity and higher mass
star.

\subsection{$\sigma$~Oph}
The star $\sigma$~Oph is classified as spectral type K2-5 and luminosity class II-III
(Simbad adopted K2\,III).
We find a luminosity of $\log L/L_\odot =$\,3.4\,$\pm$0.3 and an 
effective temperature of 4130\,$\pm$\,570\,K, placing it  at a similar place as BM~Sco in the 
HR diagram within the errors, again
in a transitional zone between red giants and red supergiants,
corresponding to initial masses of 5--9\,M$_\odot$.
Our luminosity and effective temperature are consistent with those
of \citet{McDonald2012}.

\subsection{HD~206859}
HD~206859 is generally classified as a G5\,Ib supergiant.
However, similar to BM~Sco and $\sigma$~Oph, we find that it lies
in the transitional zone between giants and supergiants 
at a luminosity of $\log L/L_\odot =$\,3.4\,$\pm$0.3
corresponding to tracks of initial mass 5--9\,M$_\odot$.
We find that it has the highest effective temperature of our 
sample of 5270\,$\pm$740\,K, which is consistent with having the warmest
spectral type of our sample of G5. Our fundamental parameters
of HD~206859 are consistent with those by \citet{Belle2009} and
\citet{McDonald2012}.

\section{Summary and conclusions}
We observed four late-type supergiants of spectral types 
G5 to K2.5 with the AMBER instrument in order to derive their
fundamental parameters and study their atmospheric structure.
These sources include V766~Cen (=HR~5171\,A), $\sigma$~Oph,
BM~Sco, and HD~206859. The new observations complement our 
previous observations of cooler, mostly M-type,  giants and 
supergiants, increasing our full sample to 16 sources.
We derived effective temperatures, luminosities, and 
absolute radii. The values are generally consistent with other
estimates available in the literature. 

V766~Cen was classified as a yellow hypergiant in the literature, 
and is thought to have most likely evolved from the red supergiant
stage back toward larger effective temperatures. 
\citet{Chesneau2014} suggested a close, possibly common-envelope
companion, which we took into account in our analysis.
We derived a luminosity of $\log L/\mathrm{L}_\odot=$\,5.8$\pm$0.4,
an effective temperature of 4290\,$\pm$\,760\,K, a radius
of 1490\,$\pm$540\,R$_\odot$, and a surface gravity $\log g\sim$ -0.3.
Contrary to the classification as a yellow hypergiant, our 
observations indicate that V766~Cen is located at the
red edge of a high-mass, most likely 40\,M$_\odot$ track,
thus representing a high-luminosity fully convective red supergiant rather
than a yellow hypergiant.
This is supported by a relatively large radius, which would be unusual
for a typical YHG, and by the observed presence of extended CO layers,
which are typical for RSGs.
V766~Cen is located close to both the Hayashi limit toward
cooler effective temperatures and the Eddington limit 
toward higher luminosities. 
V766~Cen shows a $K$-band flux spectrum that exhibits CO lines
and the \ion{Na}{i} doublet. The CO lines are weaker in the flux spectrum
compared to our previously observed red supergiants, which is
consistent with the earlier spectral type. 
However, the visibility spectra indicate a relatively strong extension 
of the CO forming layers. The closure phases indicate deviations 
from symmetry at the CO bandheads. Extended clumpy
molecular layers, most importantly water and CO in the near-infrared, 
are typical for
red supergiants and were previously observed for
other red supergiants. \citet{Arroyo2015} showed that the observed 
molecular extensions of red supergiants cannot be reproduced by 
1D and 3D pulsation and convection models, a missing physical process may explain the inability to
reproduce these observed molecular extensions.
The \ion{Na}{i} doublet toward V766~Cen is also spatially resolved. We estimated
a shell radius of about 1.5 stellar photospheric radii. The closure and differential phases indicate a photocenter displacement of the \ion{Na}{i}
shell compared to the nearby continuum of about 0.1\,R$_\mathrm{Phot}$. 
Lines of neutral sodium are typical for high-luminosity red supergiants
and hypergiants as discussed, for instance, by \citet{Gorlova2006} 
and \citet{Oudmaijer2013}. These studies suggest that a dense quasi-chromosphere
located just above the photosphere or an optically thick pseudo-photosphere
formed by the wind shields the gas from direct starlight and prevents
it from being ionized. 
Our observations suggest that the extended molecular 
atmosphere and the \ion{Na}{i} shell may coincide and represent an optically
thick region at the onset of the wind.
The photocenter displacement between the \ion{Na}{i} line and the continuum
may be caused by the effects of the close companion on the nearby 
circumstellar environment or by asymmetric convection patterns on the 
continuum-forming surface of the star.

We found luminosities $\log L/\mathrm{L}_\odot$ for BM~Sco, $\sigma$~Oph, 
and HD~206859 of 3.5\,$\pm$\,0.3, 3.4\,$\pm$0.3, 3.4\,$\pm$0.3, 
effective temperatures of 3890\,$\pm$\,310\,K, 4130\,$\pm$\,570\,K,
5270\,$\pm$\,740\,K, Rosseland radii of 129\,$\pm$\,26\,R$_\odot$,
100\,$\pm$\,30\,R$_\odot$, 57\,$\pm$\,17\,R$_\odot$, and surface
gravities $\log g$ of about 1.1, 1.3, and 1.8, respectively.
These sources show CO features in the flux spectra, but with visibility
spectra indicating that CO is formed at photospheric layers. There
is no indication of extended molecular layers for these sources.
Compared to evolutionary tracks, they 
correspond to initial masses of 5\,M$_\odot$ and 9\,M$_\odot$.
Altogether, despite their classification as luminosity class Ib sources,
they more likely represent higher mass giants than supergiants.
BM~Sco exhibits a significant flux fraction of about 25\% from
an underlying over-resolved dusty background component, which is consistent
with an earlier indication of circumstellar material \citep{McDonald2012}.
This is unusual among our sources of this mass, which makes BM~Sco
an interesting target for follow-up observations.

The relatively low luminosities of the luminosity class Ib sources of 
our sample leaves us with an unsampled locus in the HR diagram corresponding
to luminosities $\log L/\mathrm{L}_\odot\sim$3.8--4.8 or masses
10--13\,M$_\odot$. This region might correspond to evolutionary tracks
where red supergiants explode as core-collapse (type II-P) SN,
while stars of lower and higher masses return to higher effective
temperatures. In fact, this mass range corresponds to about the mass range
of recently confirmed red supergiant progenitors of type II-P SN. 
It will be of interest for upcoming observations to find red supergiants 
of this luminosity range to determine whether the gap
is due to a selection effect or a lower probability of finding
red supergiants of these luminosities.

The previously found relation of increasing strength
of extended molecular layers with increasing luminosity \citep{Arroyo2015}
was now confirmed to extend to the luminosity of V766~Cen. 
This represents twice the previous luminosity range and now reaches a level near the Eddington limit.
This might further point to steadily increasing
radiative winds with increasing luminosity as a possible explanation for
the observed extensions of atmospheric molecular layers toward
red supergiants.

\begin{acknowledgements}
We thank Antxon Alberdi for scientific discussions during the data reduction.
This research has made use of the \texttt{AMBER data reduction package} 
of the Jean-Marie Mariotti Center.
This research has made use of the SIMBAD database,
operated at CDS, France. This research has made use of NASA's
Astrophysics Data System. We acknowledge with thanks the variable star 
observations from the AAVSO International Database contributed by 
observers worldwide and used in this research.
\end{acknowledgements}

\bibliographystyle{aa}
\bibliography{RSG-MIRA.bib}
\end{document}